\documentclass[%
aps,
pra,
amsmath,amssymb,
floatfix,
twocolumn,
10pt,
]{revtex4-1}

\usepackage{braket}  
\usepackage{graphicx}
\usepackage{color}

\usepackage{numprint} 
\usepackage{multirow}

\usepackage{tabularx}
\newcommand{\ab}[0]{\emph{ab initio} }
\newcolumntype{C}[1]{>{\setlength\hsize{#1\textwidth}\centering}X}
\newcolumntype{L}[1]{>{\setlength\hsize{#1\textwidth}\raggedright}X}

\begin{document}

\title{Dynamic dipole polarizabilities of heteronuclear alkali dimers: optical response, trapping and control of ultracold molecules  }	
\author{R. Vexiau, D. Borsalino, M. Lepers, A. Orb{\'a}n, M. Aymar, O. Dulieu and N. Bouloufa-Maafa
\email{Corresponding author: nadia.bouloufa@u-psud.fr}}

\affiliation{ Laboratoire Aim\'e Cotton, CNRS, Universit\'e Paris-Sud, ENS Cachan, Universit\'e Paris-Saclay, 
91405 Orsay Cedex, France 
}%

\begin{abstract}
In this article we address the general approach for calculating dynamical dipole polarizabilities of small quantum systems, based on a sum-over-states formula involving in principle the entire energy spectrum of the system. We complement this method by a few-parameter model involving a limited number of effective transitions, allowing for a compact and accurate representation of both the isotropic and anisotropic components of the polarizability. We apply the method to the series of ten heteronuclear molecules composed of two of ($^7$Li,$^{23}$Na,$^{39}$K,$^{87}$Rb,$^{133}$Cs) alkali-metal atoms. We rely on both up-to-date spectroscopically-determined potential energy curves for the lowest electronic states, and on our systematic studies of these systems performed during the last decade for higher excited states and for permanent and transition dipole moments. Such a compilation is timely for the continuously growing researches on ultracold polar molecules. Indeed the knowledge of the dynamic dipole polarizabilities is crucial to model the optical response of molecules when trapped in optical lattices, and to determine optimal lattice frequencies ensuring optimal transfer to the  absolute ground state of initially weakly-bound molecules. When they exist, we determine the so-called ``magic frequencies'' where the ac-Stark shift and thus the viewed trap depth, is the same for both weakly-bound and ground-state molecules.
\end{abstract}

\vspace{2pc}

\maketitle


\section{Introduction}
\label{sec:intro}

The interaction of a neutral system composed of charged particles with an electromagnetic field is governed by its so-called dynamic dipole polarizability (DDP), which expresses the deformation of the charge distribution under the influence of the oscillating electric field. Knowing this quantity is of crucial importance in the context of the ongoing development of experiments aiming at trapping atoms and molecules with laser fields \cite{Grimm2000}. For instance, spatially-periodic configurations of light are created by using  standing wave  laser fields. These optical lattices act as periodic potentials allowing for trapping quantum gases of ultracold atoms and molecules, and thus for controlling their internal and external degrees of freedom \cite{Bloch2005b, Morsch2006, Bloch2008, Rom2004}. Such lattice-based ultracold systems are ideal  to test fundamental theoretical concepts \cite{Anderson1998b, Steck2001, Greiner2002, Paredes2004, Bloch2005}, to achieve and control ultracold chemical reactivity  \cite{Ni2010, DeMiranda2011} as well as for applications in quantum optics and quantum computation \cite{Bloch2004}. 

The DDP can be obtained both through spectroscopic measurements of light-induced shifts of energy levels, and advanced theoretical calculations based on accurate modeling of the electronic structure of the systems under study. In the present article we address the general approach for calculating DDPs of small quantum systems, based on a sum-over-states formula involving in principle the entire energy spectrum of the system. We complement this method by a few-parameter model involving a limited number of effective transitions, allowing for a compact and accurate representation of both the isotropic and anisotropic components of the polarizability. 

We apply both approaches to the series of heteronuclear alkali-metal diatomic molecules composed of $^7$Li, $^{23}$Na, $^{39}$K, $^{87}$Rb and $^{133}$Cs atoms, using state-of-the-art molecular potential energy curves (PECs) and electronic permanent and transition dipole moments (PDMs and TDMs, respectively). A combination of up-to-date spectroscopically-determined PECs for the lowest electronic states, and of results from our own systematic studies of these systems performed during the last decade for higher excited states, PDMs and TDMs is employed for the DDP computation of both real and imaginary parts of the DDP as a function of the field frequency.

Such a compilation is timely for the continuously growing researches on ultracold polar molecules, namely, the achievement of a quantum degenerate dipolar gas. Heteronuclear alkali-metal dimers possess a permanent dipole moment in their own frame, --which varies from 0.57 to 5.59 D--, making them ideal candidates for manipulation by external electric fields. The observation of anisotropic effects in the long-range interaction between ultracold molecules is foreseen, with exciting prospects in terms of novel quantum phases, quantum simulation, quantum information \cite{Baranov2008,Lahaye2009}. Even if generally those molecules do not exhibit an appropriate closed radiative cycle, which prevents them from being directly laser-cooled, the alkali atoms which compose them are nowadays efficiently cooled down to ultralow temperatures. Recently, a method relying on the presence of Feshbach resonances (FR) in the collision of alkali-metal atom pairs has been proved very efficient to create ultracold molecules. FRs have been recorded for most of the heteronuclear pairs, \textit{i.e.} LiNa \cite{Stan2004}, KRb \cite{Inouye2004}, RbCs \cite{Pilch2009}, LiCs \cite{Repp2013}, LiK \cite{Wille2008}, NaK \cite{Wu2012}, NaRb \cite{Wang2013}, LiRb \cite{Deh2008,Marzok2009}. Exposing these pairs to a suitable ramp of magnetic field allowed to stabilize these resonances as weakly-bound molecules, which can then be transformed into molecules lying in the lowest energy level of their ground state. Over the past few years, this has been achieved through the Stimulated Raman Adiabatic Passage (STIRAP) optical technique \cite{Gaubatz1990} for several homonuclear and heteronuclear species : Cs$_2$\cite{Danzl2008}, Rb$_2$ \cite{Lang2008a}, KRb \cite{Ni2008,Chotia2012}, RbCs \cite{Takekoshi2014,Shimasaki2015,Molony2014}, NaK \cite{Park2015}, NaRb \cite{Guo2016}. 

It is worthwhile to note that in a couple of cases where the transfer was successful, it was performed in the presence of an optical lattice in order to prevent inelastic and reactive molecular collisions before the STIRAP process, which can  greatly limit the lifetime of the trapped particles as discussed in \cite{Danzl2008, Chotia2012}, thus reducing the density of ground state molecular samples. With one molecule per site of the optical lattice, the molecules are shielded from collisional losses during their preparation and manipulation. The wavelength of the standing wave which constitutes the lattice has to be conveniently chosen so that the initial Feshbach level and the final ground level involved in the STIRAP sequence {}``feel'' the same trap depth in the lattice. At this wavelength called {}``magic wavelength''\cite{Katori1999}, the ac Stark shifts and so the DDP of the two molecular levels are equal. Only few studies of the DDP have been done on specific molecules like KRb \cite{Kotochigova2006}, RbCs \cite{Kotochigova2006},  Cs$_2$  \cite{Vexiau2011} and Rb$_2$ \cite{Deiss2015}. However the large range of species currently considered in experiments requires an extensive study. 

This article is organized as follows: In section \ref{sec:formalism}, we present the general expressions of the DDP. In particular we illustrate the pending controversy on its imaginary part with a simple two-level model (subsection \ref{sec:formalism-2lev}), and we apply the general definitions to the case of diatomic molecules (subsection \ref{sec:formalism-diatom}). In section \ref{sec:PEC} we detail the up-to-date molecular structure data used in our calculations; we also recall the basics of our quantum chemistry and vibrational level calculations. Sections \ref{sec:low-freq} and \ref{sec:opt-freq} present our results in the low-frequency (microwave) and optical frequency (visible and near-infrared) regimes. In section \ref{sec:lifetime} we discuss the influence of excited levels lifetimes on the calculated DDPs. In section \ref{sec:compar} we compare our results to the corresponding theoretical and experimental values when they exist in literature. In section \ref{sec:Feshbach} we present our results for Feshbach molecules before discussing in section \ref{sec:magic} the existence of magic frequencies for these systems. In section \ref{sec:effective} we give a simple expression of the DDP for all vibrational levels of the electronic ground state, as a function of a limited number of parameters. A table with the corresponding parameters can be found in the Supplementary Material, as well as files containing the PECs, PDMs and TDMs necessary for the calculation of DDPs.

Except otherwise stated, atomic units (a.u.) will be used for distances (1 a.u. = 0.052917721092 nm), for energies (1 a.u. = 219474.63137 cm$^{-1}$), for dipole moments (1 a.u. = 2.54158059 D) and for polarizabilities ($1 \mathrm{a.u.} = 4.6883572 \times 10^{-8}$~MHz/[W/cm$^2$]) throughout the paper.

\section{Formalism}
\label{sec:formalism}

\subsection{Expression of the DDP}
\label{sec:formalism-2lev}

The dynamic dipole polarizability is the microscopic counterpart of the relative permitivity $\varepsilon_r(\omega)$ of a material. For a dilute gas, the two quantities are related by  \cite{Jackson1999}
\begin{equation}
  \varepsilon_r(\omega) = 1 + \frac{N_v}{\varepsilon_0}
    \alpha_n(\omega)
\end{equation}
where $N_v$ is the volume density of the gas, $\varepsilon_0$ the vacuum permitivity, and $\alpha_n(\omega)$ the DDP of one particle of the gas in the quantum level $|n\rangle$. In the general case, $\varepsilon_r(\omega)$ and $\alpha_n(\omega)$ are complex quantities, whose imaginary part characterizes the absorption of the incoming electromagnetic field by the gas. If the expression of the real part of $\alpha_n(\omega)$ is well-established, the imaginary part is the subject of a long-standing controversy, that we illustrate in this section.

Indeed, we calculate the DDP of a quantum particle submitted to a classical electromagnetic field. First we follow the approach of \cite{Novotny2012} for a two-level particle; but we point out that the method of \cite{Novotny2012} contradicts the principles of expectation-value calculations introduced by N.~Moiseyev for non-Hermitian Hamiltonians \cite{Moiseyev1998, Moiseyev2011}. When we follow these rules, we obtain the same real part, but a different imaginary part from \cite{Novotny2012}. This discrepancy is an illustration of the dispute between the ``constant-sign'' rule and the ``opposite-sign'' rule (see for instance Refs.~\cite{Andrews2003, Milonni2004}).

We consider a quantum particle (atom or molecule), with a spectrum containing two non-degenerate levels, $|1\rangle$ and $|2\rangle$, of energies $E_1$ and $E_2$. The upper level $|2\rangle$ is also characterized by its finite lifetime $\tau_2 = 1/\gamma_2$, due to spontaneous emission toward $|1\rangle$, or to non-radiative processes. The system is submitted to an oscillating electric field of angular frequency $\omega$, amplitude $\mathcal{E}$, and linearly polarized in the $Z$-direction. In the basis $\{|1\rangle, |2\rangle\}$, the particle-field system can be represented by the Hamiltonian
\begin{equation}
  \hat{H}(t) = \left( \begin{array}{cc}
    E_1 & -d_{12} \mathcal{E} \cos(\omega t) \\
    -d_{12} \mathcal{E} \cos(\omega t) & E_2 - i\gamma_2/2
  \end{array} \right) \,,
  \label{eq:hamit22}
\end{equation}
where we used the electric-dipole approximation and introduced $d_{12} = d_{21} = \langle 1| \hat{d}_Z |2\rangle$ the transition dipole moment which is assumed to be real. It is important to note that the Hamiltonian \eqref{eq:hamit22} \textit{is not Hermitian}, since the level $|2\rangle$ is characterized by a complex energy.

Introducing the state vector
\begin{equation}
  |\Psi(t)\rangle = \left( \begin{array}{c}
    c_1(t) \\ c_2(t)
  \end{array} \right) \,,
\end{equation}
we can write the time-dependent Scr\"odinger equation as a system of ordinary differential equations
\begin{align}
  i\hbar \frac{dc_1}{dt} & = E_1c_1 
    - d_{12}\mathcal{E} \cos(\omega t) c_2 \nonumber \\
  i\hbar \frac{dc_2}{dt} & = \left(E_2-i\frac{\gamma_2}{2}\right) 
    c_2 - d_{12}\mathcal{E} \cos(\omega t) c_1 .
  \label{eq:ode-cn}
\end{align}
In the weak-field $|d_{12}\mathcal{E}| \ll (E_2-E_1)$ and off-resonant regime $|E_2-E_1-\hbar\omega| \gg \gamma_2/2$, the system of equations \eqref{eq:ode-cn} can be solved using time-dependent perturbation theory \cite{Langhoff1972}, \textit{i.e.}~setting $c_l(t) = \sum_k c_l^{(k)}(t)$ for $l=1,2$. Assuming the initial conditions $c_1^{(k)} (t\to -\infty) = \delta_{k0}$ ($\delta_{k0}$ being Kronecker's symbol), and $c_2^{(k)} (t\to -\infty) = 0\, \,\forall k$, we write for the zeroth order
\begin{equation}
  c_1^{(0)}(t) = \exp\left( -i\frac{E_1t}{\hbar}\right),
   \quad c_2^{(0)}(t) = 0 .
\end{equation}
Note that considering initial conditions starting at $t\to -\infty$ rather than $t=0$ allows for considering that the field is switched on sufficiently slowly to avoid the so-called secular terms which diverge with time \cite{Langhoff1972}. At the first order of perturbation, Eq.~\eqref{eq:ode-cn} becomes
\begin{align}
  i\hbar \frac{dc_1^{(1)}}{dt} & = E_1c_1^{(1)} \nonumber \\
  i\hbar \frac{dc_2^{(1)}}{dt} & = \left( E_2 
    - i\frac{\hbar\gamma_2}{2} \right) c_2^{(1)} \nonumber \\
   & - \frac{d_{12}\mathcal{E}}{2}
    \left( e^{-i\left(\frac{E_1}{\hbar}-\omega\right)t}
         + e^{-i\left(\frac{E_1}{\hbar}+\omega\right)t} \right).
  \label{eq:ode-1-ord}
\end{align}
Because $c_1^{(1)}(t\to -\infty) = 0$, then $c_1^{(1)}(t) = 0 \, \forall t$, and
\begin{align}
  c_2^{(1)}(t) & = C e^{-\frac{\gamma_2t}{2}}
    e^{-i\frac{E_2t}{\hbar}} \nonumber \\
   & + \frac{d_{12}\mathcal{E}}{2\hbar}
    \left( \frac{e^{-i\left(\frac{E_1}{\hbar}+\omega\right)t}}
           {\omega_{12}-i\frac{\gamma_2}{2}-\omega}
         + \frac{e^{-i\left(\frac{E_1}{\hbar}-\omega\right)t}}
           {\omega_{12}-i\frac{\gamma_2}{2}+\omega} \right)
\end{align}
with $\hbar\omega_{12} = E_2-E_1$, and $C=0$ to satisfy the initial conditions $c_2^{(1)} (t\to-\infty) = 0$.

Now we want to calculate the leading contribution to the induced dipole moment $\langle d_1^{(1)}(t)\rangle$; however the way to calculate the mean value is controversial. Using the standard definition of an operator mean value, Novotny and Hecht obtained in Eq.~(A.22) of \cite{Novotny2012},
\begin{align}
  \left\langle d_{1,-}^{(1)}(t) \right \rangle & 
    = \left( c_1^{(0)*}(t) c_2^{(1)}(t) 
    + c_2^{(1)*}(t) c_1^{(0)}(t) \right) d_{12} \notag \\
   & = \frac{d_{12}^2\mathcal{E}}{2\hbar}
   \left( \frac{e^{-i\omega t}}
           {\omega_{12}-i\frac{\gamma_2}{2}-\omega}
        + \frac{e^{i\omega t}}
           {\omega_{12}-i\frac{\gamma_2}{2}+\omega} \right.
  \nonumber \\
   & \quad \quad \left. + \frac{e^{-i\omega t}}
           {\omega_{12}+i\frac{\gamma_2}{2}+\omega}
        + \frac{e^{i\omega t}}
           {\omega_{12}+i\frac{\gamma_2}{2}-\omega} \right)
  \label{eq:dip-1}
\end{align}
The subscript ``-'' comes from the change of sign in front of $i\gamma_2/2$ inside each pair of terms $e^{\pm i\omega t}$. Note that in \cite{Novotny2012}, the $c_l$ are defined up to a phase factor $\exp(-iE_lt/\hbar)$. Using the relation $\langle d_{1,-}^{(1)}(t)\rangle = \alpha_1^-(\omega) \mathcal{E} e^{-i\omega t}/2+c.c.$, Novotny and Hecht extracts the DDP as
\begin{equation}
  \alpha_1^-(\omega) = \frac{d_{12}^2}{\hbar}
    \left( \frac{1}{\omega_{12}-i\frac{\gamma_2}{2}-\omega}
         + \frac{1}{\omega_{12}+i\frac{\gamma_2}{2}+\omega}
    \right)
  \label{eq:alpha-herm-2lev}
\end{equation}
the real and imaginary parts,
\begin{align}
  \mathrm{Re}(\alpha^-_1(\omega)) & = \frac{d_{12}^2}{\hbar}
    \left( \frac{\omega_{12}-\omega}
             {(\omega_{12}-\omega)^2 + \frac{\gamma_2^2}{4}}
         + \frac{\omega_{12}+\omega}
             {(\omega_{12}+\omega)^2 + \frac{\gamma_2^2}{4}}
    \right)
  \label{eq:alpha-re-2lev} \\
  \mathrm{Im}(\alpha^-_1(\omega)) & = 
    \frac{d_{12}^2\,\gamma_2}{2\hbar}
    \left( 
      \frac{1}{(\omega_{12}-\omega)^2 + \frac{\gamma_2^2}{4}}
    - \frac{1}{(\omega_{12}+\omega)^2 + \frac{\gamma_2^2}{4}}
    \right)
  \label{eq:alpha-im-herm-2lev}
\end{align}
Regarding the real part, we obtain the well-established expression (see for instance \cite{Manakov1986}). As for the imaginary part, as it agrees with the opposite-sign rule (hence the label ``-'' in Eqs.~\eqref{eq:dip-1}--\eqref{eq:alpha-im-herm-2lev}), it is odd with respect to $\omega$ \cite{Andrews2003, Milonni2004}. Moreover, $\mathrm{Im}(\alpha_1^-(\omega))$ scales as $\omega$ and not $\omega^3$ for vanishing frequencies \cite{Jentschura2015}, especially because the width $\gamma_2$ is $\omega$-independent.

Nevertheless the way of calculating the mean value in Eq.~\eqref{eq:dip-1} does not seem appropriate for non-Hermitian Hamiltonians as Eq.~\eqref{eq:hamit22}. According to N.~Moiseyev \cite{Moiseyev1998, Moiseyev2011}, one needs to define a left state vector $|\Psi_L(t)\rangle$ which must be constructed in such a way that one takes the complex conjugate of the right vector $|\Psi(t)\rangle$ for the terms resulting from the Hermitian dynamics, and the right eigenvector itself for the terms resulting from the non-Hermitian dynamics (for a complex-symmetric Hamiltonian as Eq.~\eqref{eq:hamit22}). Namely, setting $|\Psi_L(t)\rangle = \left( \begin{array}{c} c_{1,L}(t) \\ c_{2,L}(t) \end{array} \right)$, one has, up to the first order,
\begin{align}
  c_{1,L}^{(0)}(t) & = e^{i\frac{E_1t}{\hbar}} \notag \\
  c_{2,L}^{(1)}(t) & = \frac{d_{12}\mathcal{E}}{2\hbar}
    \left( \frac{e^{i\left(\frac{E_1}{\hbar}+\omega\right)t}}
           {\omega_{12}-i\frac{\gamma_2}{2}-\omega}
         + \frac{e^{i\left(\frac{E_1}{\hbar}-\omega\right)t}}
           {\omega_{12}-i\frac{\gamma_2}{2}+\omega} \right) ,
\end{align}
and $c_{1,L}^{(1)}(t) = c_{2,L}^{(0)}(t) = 0$. Calculating the mean dipole moment as $\langle d_{1,+}^{(1)} \rangle = ( c_{1,L} c_{2} + c_{2,L} c_{1} ) \times d_{12}$, which is now complex, we can extract the DDP using the relation $\langle d_{1,+}^{(1)}(t)\rangle = ( \alpha_1^+(\omega) e^{-i\omega t} + \alpha_1^+(-\omega) e^{i\omega t} ) \times \mathcal{E}/2$,
\begin{equation}
  \alpha_1^+(\omega) = \frac{d_{12}^2}{\hbar}
    \left( \frac{1}{\omega_{12}-i\frac{\gamma_2}{2}-\omega}
         + \frac{1}{\omega_{12}-i\frac{\gamma_2}{2}+\omega}
    \right) .
  \label{eq:alpha-non-herm-2lev}
\end{equation}
Its real part is identical to Eq.~\eqref{eq:alpha-re-2lev}, while the imaginary part becomes
\begin{align}
  \mathrm{Im}(\alpha_1^+(\omega)) & = 
    \frac{d_{12}^2\,\gamma_2}{2\hbar}
    \left( 
      \frac{1}{(\omega_{12}-\omega)^2 + \frac{\gamma_2^2}{4}}
    + \frac{1}{(\omega_{12}+\omega)^2 + \frac{\gamma_2^2}{4}}
    \right)
  \label{eq:alpha-im-non-herm-2lev}
\end{align}
Equation \eqref{eq:alpha-im-non-herm-2lev} is in agreement with the constant-sign rule \cite{Andrews2003, Milonni2004} (hence the ``+'' label in Eqs.~\eqref{eq:alpha-non-herm-2lev} and \eqref{eq:alpha-im-non-herm-2lev}), obtained by considering complex energies in the usual expression of the DDP (see our previous works on other diatomics \cite{Vexiau2011, Deiss2015} and lanthanide atoms \cite{Lepers2014, Li2017, Li2017b}). Equation \eqref{eq:alpha-im-non-herm-2lev} is even with respect to $\omega$, and non-zero for $\omega=0$. 

By consequence, the above calculations tend to support the constant-sign rule against the opposite-sign one. However, Eq.~\eqref{eq:alpha-im-non-herm-2lev} is not similar to imaginary parts obtained in second quantization \cite{Milonni2004, Milonni2008, Wang2009, Jentschura2015} (which do not solve the controversy either). The main reason is the following. In Eqs.~\eqref{eq:alpha-im-herm-2lev} and \eqref{eq:alpha-im-non-herm-2lev}, the width $\gamma_2$ is taken as an intrinsic property of the upper level, and so it is $\omega$-independent. By contrast, in the second-quantization formalism, the analog of $\gamma_2$ turns out to depend on $\omega$. One possible way to sort out this paradox is to consider that Eq.~\eqref{eq:alpha-im-non-herm-2lev} does not exactly account for the absorption of the incoming field, but rather the loss of coherence of the quantum particle, which is crucial limitation in ultracold experiments, coming from spontaneous emission or from non-radiative processes like molecular predissociation.

We can straightforwardly extend Eqs.~\eqref{eq:alpha-re-2lev}, \eqref{eq:alpha-im-herm-2lev} and \eqref{eq:alpha-im-non-herm-2lev} to a multilevel particle in a non-degenerate level $|n\rangle$,
\begin{align}
  \mathrm{Re}(\alpha_n(\omega)) & = \sum_{n'\neq n}
    \frac{d_{nn'}^2}{\hbar}
    \left( \frac{\omega_{nn'}-\omega}
             {(\omega_{nn'}-\omega)^2 + \frac{\gamma_{n'}^2}{4}}
    \right. \notag \\
    & \quad \left. + \frac{\omega_{nn'}+\omega}
             {(\omega_{nn'}+\omega)^2 + \frac{\gamma_{n'}^2}{4}}
    \right)
  \label{eq:alpha-re} \\
  \mathrm{Im}(\alpha_n^-(\omega)) & = \sum_{n'\neq n} 
    \frac{d_{nn'}^2\,\gamma_{n'}}{2\hbar}
    \left( 
     \frac{1}{(\omega_{nn'}-\omega)^2 + \frac{\gamma_{n'}^2}{4}}
    \right. \notag \\
   & \quad \left.
    -\frac{1}{(\omega_{nn'}+\omega)^2 + \frac{\gamma_{n'}^2}{4}}
    \right)
  \label{eq:alpha-im-herm} \\
  \mathrm{Im}(\alpha_n^+(\omega)) & = \sum_{n'\neq n} 
    \frac{d_{nn'}^2\,\gamma_{n'}}{2\hbar}
    \left( 
     \frac{1}{(\omega_{nn'}-\omega)^2 + \frac{\gamma_{n'}^2}{4}}
    \right. \notag \\
   & \quad \left.
    +\frac{1}{(\omega_{nn'}+\omega)^2 + \frac{\gamma_{n'}^2}{4}}
    \right) ,
  \label{eq:alpha-im-non-herm}
\end{align}
where $\mathrm{Re}(\alpha_n(\omega)) = \mathrm{Re}(\alpha^\pm_n(\omega))$, see Eq.~\eqref{eq:alpha-herm-2lev} and \eqref{eq:alpha-non-herm-2lev}. In those expressions we set $\gamma_n \ll \gamma_{n'} \, \forall n'$, which means that the level $|n\rangle$ has a sufficiently long lifetime to be trapped efficiently. Because the expression of the imaginary part is questionable, in Sec.~\ref{sec:opt-freq}, we will give results for the two equations \eqref{eq:alpha-im-herm} and \eqref{eq:alpha-im-non-herm}. We will see that significant differences only arise at very low frequencies, at which optical trapping is not achieved.

\subsection{DDP of a diatomic molecule}
\label{sec:formalism-diatom}

In the case of a diatomic molecule, the level $|n\rangle$ is a rovibrational level supported by an electronic state. Even if the sums in Eqs.~\eqref{eq:alpha-re}--\eqref{eq:alpha-im-non-herm} run \textit{a priori} over all bound and free molecular levels $|n'\rangle$, they can be reduced by taking into account selection rules. The matrix element $d_{nn'}$ of the dipole moment operator can be expressed in a more explicit way, assuming that the $\left | n \right \rangle$ level is labeled using the electronic state $e=X$, $A$, $a$, ..., the vibrational quantum number $v$, as well as $J$, $M$ and $\Lambda$ associated with the total angular momentum and its projections on the $Z$ axis of the laboratory frame and on the molecular axis $z$, respectively. The level $|n'\rangle$ is labeled with the corresponding primed quantum numbers. One finds for a linearly polarized field along $Z$
\begin{align}
\label{eq:tdm}
  d_{nn'} & = \displaystyle {\sum_{q=0,\pm1}} 
   \left\langle v'\left| d_q(e',e;R)
   \right| v \right\rangle
  \notag \\
   & \times \left( \begin{array}{ccc} J' & 1 & J \\
      -M' & 0 & M\end{array} \right)
     \left( \begin{array}{ccc} J' & 1 & J \\
      -\Lambda' & q & \Lambda \end{array} \right)
\end{align}
where $d_q(e',e;R)$ is the matrix element evaluated in the molecular frame ($xyz$) of the electronic PDM (if $e=e'$) or TDM (if $e\neq e'$) at the internuclear distance $R$, and the quantities $(:::)$ are Wigner 3-j symbols. The index $q$ characterizes the component of the dipole moment, namely $d_0=d_z$ and $d_{\pm 1}=\mp\frac{1}{\sqrt{2}}(d_x\pm id_y)$. The 3-j symbol of Eq.~\eqref{eq:tdm} imposes $q=\Lambda'- \Lambda$. The angled brackets refer to the integration over the internuclear distance $R$. The wave function of the vibrational levels $\psi_{v}(R)=\langle R|v\rangle$ and $\psi_{v'}(R)=\langle R|v'\rangle$ are assumed independent from the rotational quantum numbers $J$ and $J'$. This assumption is valid as we will limit our study to the lowest values of $J$.

For alkali-metal diatomic molecules in their electronic ground state $X^1\Sigma^+$ ($\Lambda=0$) and in an arbitrary rovibrational level ($v,J,M$), the accessible levels according to electric dipole selection rules are of $^1\Sigma^+$ and $^1\Pi$ symmetries. The $\Sigma-\Sigma$ electronic transitions involve the $q=0$ component of the dipole moment $d_z$, \textit{i.e.} the electronic TDM along the molecular axis; hence these transitions are referred to as parallel transitions. Similarly, $\Sigma-\Pi$ electronic transitions, which involve $d_x$ and $d_y$, are referred to as perpendicular transitions. After evaluating the 3-j symbols and gathering all the parallel and perpendicular contributions in $\alpha_{ev\parallel}^\pm$ and  $\alpha_{ev\perp}^\pm$, respectively, we have for $\alpha_n^\pm(\omega) = \alpha_{evJM}^\pm(\omega)$:
\begin{align}
  \alpha_{evJM}^\pm(\omega) & = \frac{2J^2+2J-1-2M^2}
    {(2J+3)(2J-1)} \alpha_{ev\parallel}^\pm(\omega) 
  \nonumber \\
   & + \frac{2J^2+2J-2+2M^2}{(2J+3)(2J-1)} 
       \alpha_{ev\perp}^\pm(\omega)
  \label{eq:pola-JM}
\end{align}
\begin{align}
  \alpha_{ev\parallel}^\pm(\omega) & 
    = \displaystyle {\sum_{e'v' \in \Sigma}} \left(
     \frac{1}{\omega_{ev,e'v'}-i\gamma_{e'v'}/2-\omega} \right.
  \nonumber \\
    & \left. \quad \quad +
    \frac{1}{\omega_{ev,e'v'}\mp i\gamma_{e'v'}/2+\omega} \right)
  \nonumber \\
    & \quad \times \left|\left\langle v'\left|
    d_z(e'e;R) \right|v\right\rangle\right|^{2} 
  \label{eq:pola-para} \\
  \alpha_{ev\perp}^\pm(\omega) & 
    = \displaystyle {\sum_{e'v' \in \Pi}} \left(
     \frac{1}{\omega_{ev,e'v'}-i\gamma_{e'v'}/2-\omega} \right.
  \nonumber \\
    & \left. \quad \quad +
    \frac{1}{\omega_{ev,e'v'}\mp i\gamma_{e'v'}/2+\omega} \right)
  \nonumber \\
   & \quad \times \left|\left\langle v'\left|d_x(e'e;R)
    \right|v\right\rangle\right|^{2} \,.
  \label{eq:pola-perp}
\end{align}
Note that as $e$ is the electronic ground state $X^1\Sigma^+$,  the different indexes $e'$ in Eq.~\eqref{eq:pola-para} and in Eq.~\eqref{eq:pola-perp} refer to electronic excited states of $^1\Sigma^+$ and $^1\Pi$ symmetry, respectively. The parallel and perpendicular polarizabilities depend on the electronic $e$ and vibrational $v$ quantum numbers, but not on the rotational ones, which only appear as prefactors in Eq.~\eqref{eq:pola-JM}. We can rewrite the DDP in terms of the isotropic DDP $\tilde{\alpha}(\omega)$ and the anisotropy of the DDP $\gamma(\omega)$ :
\begin{align}
\label{eq:pola-iso}
\tilde{\alpha}_{ev}(\omega) &= \frac{1}{3} \alpha_{ev\parallel}(\omega) + \frac{2}{3} \alpha_{ev\perp}(\omega) \\
\gamma_{ev}(\omega) &= \alpha_{ev\parallel}(\omega) - \alpha_{ev\perp}(\omega) \\
\label{eq:pola:JM-2}
  \alpha_{evJM}(\omega)& =  \tilde{\alpha}_{ev}(\omega) + \frac{2J(J+1)-6M^2}{3(2J+3)(2J-1)} \gamma_{ev}(\omega)
\end{align}
where we drop the $\pm$ superscripts to highlight the generality of those relations. In the present study, we focus on the polarizability $\alpha_{ev00}(\omega) \equiv \alpha(\omega)$ of the lowest rotational level $J=M=0$ of a vibrational level v of the ground state X$^1\Sigma^+$, which is in fact the isotropic polarizability $\tilde{\alpha}(\omega)$, independent from $J$, $M$ and thus from the direction of the oscillating field in the laboratory frame. The relation $\alpha(\omega) = \tilde{\alpha}(\omega)$ still holds for an excited rotational $J$ level when all the sublevels $M$ are statistically populated. We will however discuss the variations of the anisotropy $\gamma$ with the frequency in Section \ref{sec:effective}. A point we need to emphasize on is the clear distinction between the anisotropy of the molecules evaluated in the molecular frame and the isotropy of the polarizability relevant in the laboratory frame. The two notions being defined in different frames, we can have an isotropic behavior despite the anisotropy of the molecule due to its cylindrical geometry.

\section{Molecular structure data}
\label{sec:PEC}

The calculation of $\alpha(\omega)$ requires a detailed knowledge of the structure of all the molecular states -- PECs and TDMs -- involved in the sums of Eqs.~\eqref{eq:pola-JM}--\eqref{eq:pola-perp} according to the selection rules. First of all, it is crucial to have an accurate description of the $X^1\Sigma^+$ ground-state PEC. To that end we rely on the extensive work done by several groups (in particular in Hannover and Riga), who extracted accurate potential energy curves (PECs) of the ground state from spectroscopic measurements, for all bialkaline heteronuclear molecules. 

In addition, we need the PECs of excited states of $^1\Sigma^+$ and $^1\Pi$ symmetries. We used PECs deduced from spectroscopic measurements when available and PECs computed in our group by quantum chemistry otherwise. For each molecule we typically include in our DDP calculations four excited $^1\Sigma^+$ and three $^1\Pi$ states, insuring the convergence of the sums in Eqs.~\eqref{eq:pola-JM}--\eqref{eq:pola-perp}. Electronic TDMs towards higher excited states are usually weaker; combined with the energy dependence of the DDP, this leads to smaller contributions as the excited states get higher. We will illustrate this pattern on the RbCs molecule in the next section. Indeed we find that the main contribution to the sum, up to $90~\%$, is due to the $A^1\Sigma^+$ and the $B^1\Pi$ low excited states.

Even if we calculate the polarizability on a broad range (0-20000 cm$^{-1}$) of frequencies, we focus our analysis on optical frequencies in the near-infrared and visible domains used in experiments. Such frequencies can induce transitions from the $X^1\Sigma^+$ ground state towards levels of the $A^1\Sigma^+$, accentuating the need to have an accurate knowledge of this state. The $A^1\Sigma^+$ is coupled by the spin-orbit (SO) interaction to the $b^3\Pi$ state. Therefore, we also need the $b^3\Pi$ state PECs as well as its SO coupling to the $A^1\Sigma^+$ state, especially for heavy molecules. Note that in what follows, the ground and these excited electronic states will be often labeled with the first letter of the above notations, namely $X, A, B, b$...

In table \ref{table:PEC} we give for each molecule the source of all the PECs used in our calculations. To get a good description of weakly bound levels as well as the continuum of electronic states, one needs an accurate determination of the long-range part of the electronic potential curves. To that end, we smoothly connect the short-range part of the PEC to an analytical long-range expansion described by the $C_n$ coefficients given in Ref. \cite{Marinescu1999a}, unless the long-range part is already provided in a complete experimental PEC.

Table \ref{table:PEC} also presents the SO-coupling matrix elements between $b^3\Pi$ and $A^1\Sigma^+$ states, as well as the PEC of the $b^3\Pi$ state. For six molecules (KCs, RbCs, NaK, NaRb, NaCs and LiCs), we used $R$-dependent SO coupling functions determined by an extensive deperturbation work of the $A$/$b$ system. In contrast, for the other molecules, we used SO coupling functions from another molecule, rescaled to give the correct asymptotic limits, \textit{i.e.}~the atomic spin-orbit splittings at infinite internuclear distances. For example for KRb we used rescaled NaRb SO couplings (see \cite{Borsalino2014} for more details).

\begingroup
\squeezetable
\begin{table*}[!p]
\caption{\label{table:PEC}References for all the electronic PECs used in the calculations (SOCME $\equiv$ Spin-Orbit Coupling Matrix Elements). For heavy molecules (RbCs, KRb, KCs) we added the short-range core-core repulsion term \cite{Jeung1997} to the quantum chemistry  PECs. This term is less important for lighter molecules. For the quantum chemistry PECs of LiRb and NaRb we extrapolate the short-range potentials by comparing our ground state quantum chemistry  PEC to the RKR potential \cite{Borsalino2014}. For the other molecules this term is not included.}	
\begin{ruledtabular}
        \begin{tabularx}{\textwidth}{C{0.08}C{0.08}C{0.16}C{0.08}C{0.16}C{0.08}C{0.08}L{0.24}}
        \hline
		Molecule & {$X^1\Sigma^+$\linebreak experimental state} & Experimental excited states & Long Range & \ab PECs (tw) & { SOCME} & 
		{ PDMs\linebreak TDMs} & 
		{\centering Notes}\tabularnewline
		\hline
		KCs & \cite{Ferber2009} & $A^1\Sigma^+$, $b^3\Pi$ \cite{Kruzins2010}, $B^1\Pi$ \cite{Birzniece2012}, 
			$E^1\Sigma^+$ \cite{Busevica2011} & \cite{Kruzins2010,Busevica2011,Marinescu1999a} & 
			$(3,5,6)^1\Sigma^+$, $(2,3)^1\Pi$ & 
			$(b/A)$ \cite{Kruzins2010} & 
			\cite{Aymar2005}\linebreak  (tw) & 
			$B^1\Pi$ : RKR + \ab + long range.\linebreak
			$E^1\Sigma^+$ : modified empirical PEC to converge toward the unperturbed atomic limit (4s+5d)\tabularnewline
		\hline
		KRb & \cite{Pashov2007} & $A^1\Sigma^+$ \cite{Kim2012}, $b^3\Pi$ \cite{Kim2011}, 
			$C^1\Sigma^+$ \cite{Amiot1999}, $B^1\Pi$ \cite{Kasahara1999}, $D^1\Pi$ \cite{Amiot2000}, $(3)^1\Pi$ \cite{Amiot1999}& 
			\cite{Marinescu1999a} & $(4,5)^1\Sigma^+$ & $(b/A)$ \cite{Docenko2007} & 
			\cite{Aymar2005}\linebreak  (tw) &
			$A^1\Sigma^+$, $b^3\Pi$ : spectroscopic data on a very limited range $\Rightarrow$ shift of \ab PECs to fit with data. 
			SOCME from NaRb is used.\linebreak
			$(1,2)^1\Pi$, $C^1\Sigma^+$  : RKR + \ab + long range\linebreak
			$(3)^1\Pi$ : \ab + experimental $T_e$
			\tabularnewline
		\hline
		RbCs & \cite{Docenko2011} & $A^1\Sigma^+$, $b^3\Pi$ \cite{Docenko2010} & \cite{Docenko2011,Marinescu1999a} & 
			$(3-7)^1\Sigma^+$, $(1-5)^1\Pi$ & $(b/A)$ \cite{Docenko2010} & \cite{Aymar2005}\linebreak  (tw) &  \tabularnewline
		\hline
		LiNa & \cite{Fellows1991} & $A^1\Sigma^+$ \cite{Fellows1989}, $C^1\Sigma^+$ \cite{Fellows1990}, 
			$E^1\Sigma^+$ \cite{Bang2005} & \cite{Fellows1991,Marinescu1999a} & $(5)^1\Sigma^+$, $(1,2,3)^1\Pi$ & -- & 
			\cite{Aymar2005}\linebreak  (tw) &
			Weak SO interaction $\Rightarrow$ not included in the calculations.\tabularnewline
		\hline
		LiK & \cite{Tiemann2009} & $A^1\Sigma^+$ \cite{Grochola2012}, $C^1\Sigma^+$ \cite{Jastrzebski2001}, 
			$B^1\Pi$ \cite{Jastrzebski2001} & \cite{Tiemann2009,Marinescu1999a} & 
			$(4,5)^1\Sigma^+$, $(2,3)^1\Pi$ & $(b/A)$ \cite{Docenko2007} &
			\cite{Aymar2005}\linebreak  (tw) &
			Rescaled SOCME from NaRb is used.\linebreak
			$A^1\Sigma^+$, $C^1\Sigma^+$, $B^1\Pi$ : RKR + \ab + long range\tabularnewline
		\hline
		LiRb & \cite{Ivanova2011} &  $B^1\Pi$ \cite{Ivanova2013}, $C^1\Sigma^+$ \cite{Ivanova2013}, $D^1\Pi$ \cite{Ivanova2013}  & \cite{Ivanova2011,Ivanova2013,Marinescu1999a} & 
			$(2,4,5)^1\Sigma^+$, $(3)^1\Pi$ & $(b/A)$ \cite{Docenko2007} & 
			\cite{Aymar2005}\linebreak  (tw) & SOCME from NaRb is used \tabularnewline
		\hline
		LiCs & \cite{Staanum2007} & $A^1\Sigma^+$, $b^3\Pi$ \cite{Kowalczyk2015}, $B^1\Pi$ \cite{Grochola2009} & \cite{Staanum2007,Grochola2009,Marinescu1999a} & 
			$(2-5)^1\Sigma^+$, $(2,3)^1\Pi$ & $(b/A)$ \cite{Kowalczyk2015} &
			\cite{Aymar2005}\linebreak  (tw) &  \tabularnewline
		\hline
		NaK & \cite{Gerdes2008} &  $A^1\Sigma^+$, $b^3\Pi$ \cite{Harker2015}, $B^1\Pi$ \cite{Kasahara1991}, $C^1\Sigma^+$ \cite{Ross2004}, 
			$D^1\Pi$, $d^3\Pi$ \cite{AdohiKrou2008} & \cite{Gerdes2008,AdohiKrou2008,Marinescu1999a} & 
			$(4,5)^1\Sigma^+$, $(2-4)^1\Pi$ & $(d/D)$ \cite{AdohiKrou2008}, $(b/A)$ \cite{Harker2015} & 
			\cite{Aymar2007} &
			\tabularnewline
		\hline
		NaRb & \cite{Docenko2004a} & $A^1\Sigma^+$, $b^3\Pi$ \cite{Docenko2007}, $B^1\Pi$ \cite{Pashov2006}, 
			$C^1\Sigma^+$ \cite{Jastrzebski2005}, $D^1\Pi$ \cite{Docenko2005},$(4)^1\Pi$\cite{Bang2009a} & 
			\cite{Docenko2004a,Docenko2007,Pashov2006,Jastrzebski2005,Docenko2005,Marinescu1999a} & 
			$(4-5)^1\Sigma^+$, $(3,5)^1\Pi$ & $(b/A)$ \cite{Docenko2007} & \cite{Aymar2007} & \tabularnewline
		\hline
		NaCs & \cite{Docenko2004} & $A^1\Sigma^+$, $b^3\Pi$ \cite{Zaharova2009}, $B^1\Pi$ \cite{Zaharova2007}, 
			$(3)^1\Pi$ \cite{Docenko2006a} & \cite{Docenko2004,Zaharova2007,Docenko2006a,Marinescu1999a} & 
			$(3,4,5)^1\Sigma^+$, $(2,4,5)^1\Pi$ & $(b/A)$ \cite{Zaharova2009} & \cite{Aymar2007} & \tabularnewline
		\hline	
	\end{tabularx}
\end{ruledtabular}
\end{table*}

We emphasize that table \ref{table:PEC} is by no mean an exhaustive listing of spectroscopic studies on bialkali molecules. For example recent studies have been made on LiNa \cite{Steinke2012, Bang2013, Bang2009a, Bang2007, Bang2006, Bang2005}, LiK \cite{Szczepkowski2010}, LiCs \cite{Stein2008}, NaK \cite{Burns2005}, NaRb \cite{Bang2009b}, NaCs \cite{Grochola2010}, KCs \cite{Birzniece2015b, Birzniece2015a, Szczepkowski2014, Kruzins2013, Szczepkowski2013, Klincare2012, Szczepkowski2012}, RbCs \cite{Yuan2015, Kruzins2014, Birzniece2013}. However we show in following sections that a more precise description of high excited electronic states does not modify significantly our results.

Table \ref{table:PEC} indicates that all the electronic PDMs and TDMs, as well as the experimentally unavailable PECs, are computed in our group using a quantum chemistry approach. These calculations, performed with the so-called CIPSI package, are explained in a previous paper \cite{Aymar2005,Aymar2006}. Briefly we model the alkali-atom diatomic molecule as two valence electrons moving in the field exerted by two polarizable ionic cores. The valence-core and core-core interactions are then described by a model potential. From there we first determine a basis of molecular orbitals with a Hartree-Fock calculation. Correlation between valence electrons is then taken into account by performing a full configuration-interaction (FCI) calculation. Electronic $R$-dependent PDMs and TDMs are determined by computing the expectation value of the dipole-moment operator (see Fig.~\ref{fig:etdm}). For each molecule, all the data (PECs, SO and $d_{\Lambda'-\Lambda}(e,e';R)$) used in our calculations are given in the supplementary material.

Our quantum-chemistry calculations are able to describe electronic states corresponding to the excitation of valence electrons, but not the states for which core electrons are excited. However the latter can account for up to a few percents of the DDP \cite{Derevianko2010}. Because the two alkali ionic cores $A^+$ and $B^+$ weakly perturb each other, we consider that the contribution $\alpha_c$ coming from core-electron excited states is the sum of the ionic $A^+$ and $B^+$ DDPs: $\alpha_c(\omega) \approx \alpha_{A^+}(\omega) + \alpha_{B^+}(\omega)$. A description of our method to obtain the ionic DDPs is given in \cite{Deiss2015}. Briefly we compare our atomic species (Cs, Rb, K) calculated DDPs to the very accurate atomic determination made by Derevianko \textit{et al.}~\cite{Derevianko2010}. From the differences, we extract effective parameters that allow us to reproduce their results (Table \ref{table:core}), assuming the expression (see also Section \ref{sec:effective})

\begin{equation}
  \label{eq:pola-eff-core}
\alpha_{A^+}(\omega)= \frac{2\omega_{1}d_{1}^2} {\omega_{1}^2 - \omega^2} 
   + \frac{2\omega_{2}d_{2}^2} {\omega_{2}^2 - \omega^2} ,
\end{equation}

(and a similar one for the $B^+$ core). The empirical ionic polarizabilities obtained in this way are described by one or two effective transitions. The polarizabilities of Na$^+$ and Li$^+$ are too small to be obtained by this procedure and are thus neglected in our polarizability calculations for molecules containing one of these two atoms.

 \begin{table}[htb]
 \caption{\label{table:core} \small Parameters for the dynamic dipole polarizabilities of the heaviest alkali-metal ions, written  as a sum of two effective transitions, see Eq.~(\eqref{eq:pola-eff-core}). The values for the static polarizability ($\omega=0$) are displayed for illustration. }
\begin{ruledtabular}
  \begin{tabular}{c|cccc|c} 
    Ionic core & \multicolumn{4}{c|}{effective parameters} & static polarizability \\
     &  $\omega_1$ & $d_1^2$   & $\omega_2$  & $d_2^2$& (a.u.)\\
     &   ($cm^{-1}$)&  (a.u.)  & ($cm^{-1}$) & (a.u.) & \\     
   \hline
   K$^+$ & 157105 & 0.304 & 199440 & 1.383 & 3.89\\
   Rb$^+$ & 165912 & 2.728 & 362918 & 1.401 & 8.91\\
   Cs$^+$ & 142044 & 4.488 & 553515 & 1.860 & 15.34\\   
 \end{tabular}
 \end{ruledtabular}
 \end{table}

Once we have the necessary information regarding electronic states, we compute the energy and the wave function of all their rovibrational levels and continuum-in-a-box levels, by means of a mapped Fourier grid method using a discrete variable representation \cite{Kokoouline1999}. Because resonant couplings with the continuum are not in the scope of this article, the discretization of the continuum is a good approximation. Note that these bound- and continuum-level calculations are not performed with core-electron excited states. For the lifetimes $\gamma_{e'v'}^{-1}$ of the excited levels, we take an arbitrary value of 10 ns, order of magnitude of the lifetime of a deeply bound vibrational level of the A$^1\Sigma^+$ state (see for example \cite{Debatin2011}). Accurate lifetimes are needed to obtain precise values of the imaginary part of the polarizability, especially in the case of a near-resonant light. However in the present study, we focus on non-resonant light for trapping purposes, and thus approximate lifetimes are sufficient, as further discussed in section \ref{sec:lifetime}.

Obviously there exists other approaches to compute DDPs of molecules, through a wealth of available quantum chemistry codes. Concerning the specific cases of alkali dimers, one can quote for instance the work by Byrd \textit{et al.} \cite{Byrd2011}, Buchachenko \textit{et al.} \cite{Buchachenko2012}, \.{Z}uchowski \textit{et al.} \cite{Zuchowski2013}. In contrast with these works, the present approach based on the sum-over-states formula of Eq.\eqref{eq:alpha-re} benefits from the ability of our quantum chemistry approach to compute highly-excited electronic states of the diatomic molecules up to a step which allows us to reach the convergence of our numerical results with respect to the number of included electronic states. Moreover, our approach is readily adapted to the calculations of long-range atom-atom $R^{-6}$ interactions based on DDPs expressed with imaginary frequencies \cite{Vexiau2015}.

\section{Polarizability in the low-frequency domain }
\label{sec:low-freq}

The sums in Eqs.~\eqref{eq:pola-JM}--\eqref{eq:pola-perp} can be separated into two parts related to two different kinds of contributions: the ones from the rovibrational transitions within the ground electronic state, gathered in the term $\alpha_\mathrm{gr}(\omega)$, and the ones from transitions towards excited electronic states, gathered in the term $\alpha_\mathrm{exc}(\omega)$. The term $\alpha_\mathrm{gr}(\omega)$, due to the non-zero PDM, appears therefore exclusively in heteronuclear molecules.

From the $(v, J=0)$ levels of the electronic ground state, we consider separately the transition towards the excited rotational level $J=1$ of the same level $v$, the so-called pure rotational transition, and transitions towards other ground-state vibrational levels $v'$, the so-called rovibrational ones. In most cases, the latter bring a small contribution, since the vibrational wave functions of a same electronic state are orthogonal, exhibiting then an overlap $\int \psi_v\psi_{v'}^*dR$ which is strictly equal to zero. Because the PDM is a slowly varying function of $R$ in comparison with the vibrational wave functions $\psi_v$ and $\psi_{v'}$, the integral $\int \psi_v d_0(\mathrm{XX};R) \psi_{v'}^*dR$, where $d_0(\mathrm{XX};R)$ is the $R$-dependent PDM of the X state, is close to zero. This is especially the case for deeply bound levels, for which the $R$-extension of the wave function is small. For more vibrationally excited levels, the above integral is non-zero. However our calculations show that rovibrational transitions always lead to contributions smaller than the purely rotational transition.

The TDM associated with the pure rotational transition ($\mathrm{X}^1\Sigma^+,v,J=0 \rightarrow \mathrm{X}^1\Sigma^+,v,J'=1$) is quite strong. It is for example comparable to the TDM of the most favorable transitions towards the $A^1\Sigma^+$ state. However the energy of a rotational transition is of the order of $0.01$~cm$^{-1}$, while the energy of an electronic transition is of the order of several thousands of cm$^{-1}$. Since the polarizability varies as the inverse of the difference between the squares of the transition energy and of the trapping laser energy (see Eqs.~\eqref{eq:pola-para} and \eqref{eq:pola-perp}), the contribution of the pure rotational transition is by far dominant in the microwave regime ($\omega < 1$~cm$^{-1}$), while the contribution from electronic transitions is dominant in the optical regime ($\omega \approx 10000$~cm$^{-1}$) relevant for trapping.

\begin{figure}[htbp]
 \begin{center}
 \includegraphics[width=0.45\textwidth]{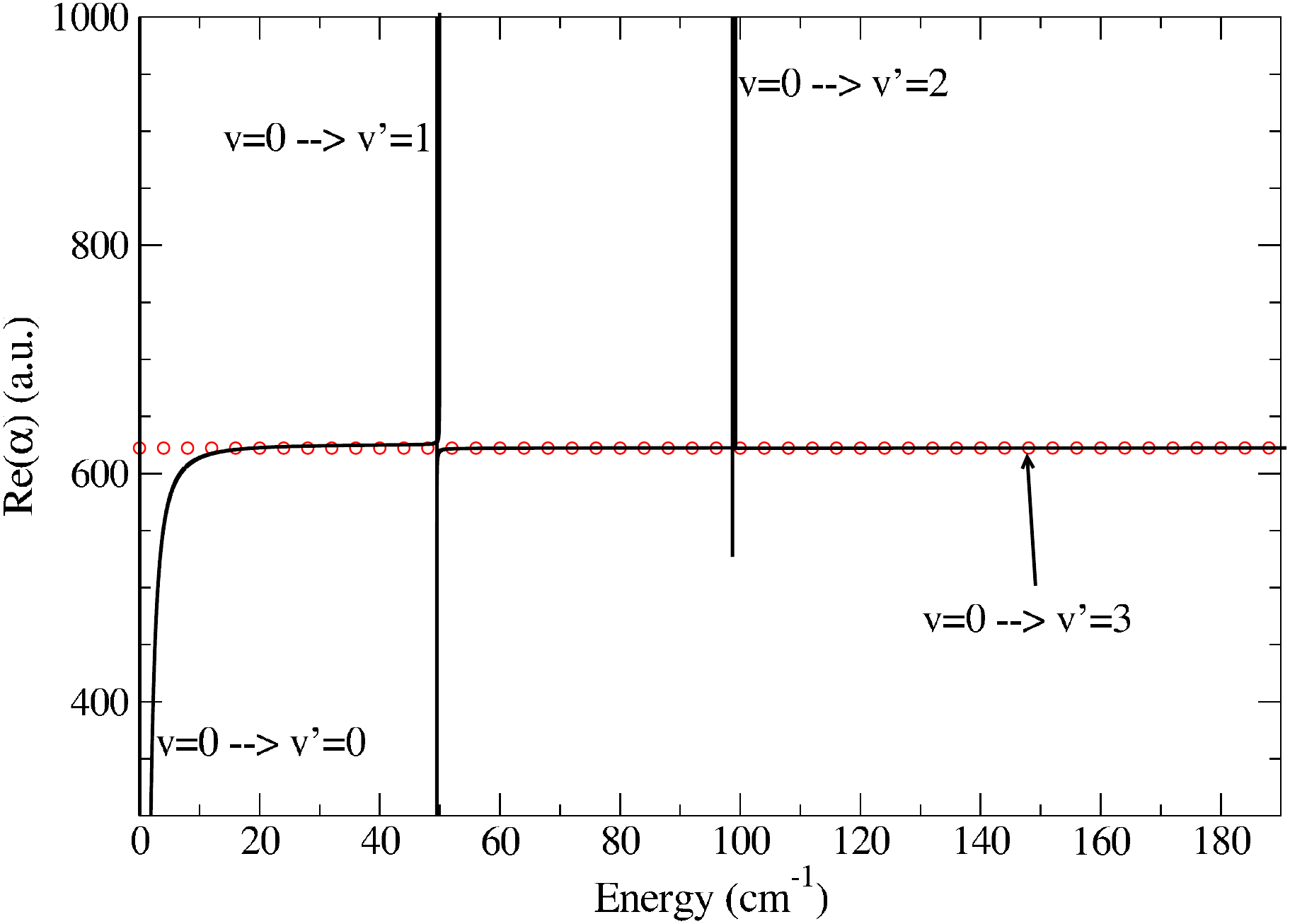}
 \caption{\label{fig:dynp_perm} \small Real part of the DDPs (black curve) and $\alpha_\mathrm{exc}$ contribution (open circles) of a ($X^1\Sigma^+,v=0,J=0$) RbCs molecule as a function of the energy of the laser. Resonant peaks are labeled by the relevant $v \rightarrow v'$ transition assuming that the rotational transition is $J=0 \rightarrow J'=1$.  }
 \end{center}
\end{figure}

Figure \ref{fig:dynp_perm} shows the real part of the DDP of a ($X^1\Sigma^+,v=0,J=0$) RbCs molecule for low frequencies, calculated by including all electronic states reported in Table \ref{table:PEC}. It shows that both $\alpha_\mathrm{gr}(\omega)$ and $\alpha_\mathrm{exc}(\omega)$ are identical and smoothly vary with the laser frequency, apart from a few resonant peaks in $\alpha_\mathrm{gr}(\omega)$ due to rovibrational transitions within the $X^1\Sigma^+$ state. Namely, a peak due to the pure rotational transition ($X^1\Sigma^+,v=0,J=0 \rightarrow X^1\Sigma^+,v'=0,J'=1$) is visible near zero energy, as well as two smaller peaks due to transitions towards vibrational levels $X^1\Sigma^+,v'=1,J'=1$ and $X^1\Sigma^+,v'=2,J'=1$. At the scale of the graph the transition towards the $X^1\Sigma^+,v'=3,J'=1$ level is not visible. For trapping frequencies larger than $150$~cm$^{-1}$ the contribution from all the rovibrational transitions and from the pure rotational transition is negligible.

The controversy regarding the definition of the imaginary part of the DDP is illustrated in Fig.~\ref{fig:Im-comparaison} for RbCs, where we show $\mathrm{Im}(\alpha_n^-(\omega))$ and $\mathrm{Im}(\alpha_n^+(\omega))$, given by Eqs.~\eqref{eq:alpha-im-herm} and \eqref{eq:alpha-im-non-herm} respectively. A significant difference is visible at low frequencies, as the opposite-sign formula tends to zero, while the constant-sign formula tends to a constant (~10$^{-5}$ a.u.). However, it is important to note that the frequencies for which that difference is visible are not used for optical trapping. In the rest of the article, we will use the constant-sign formula, Eq.~\eqref{eq:alpha-im-non-herm}, to compute the imaginary part of the DDP.

\begin{figure}[htbp]
\begin{center}
\includegraphics[width=0.45\textwidth]{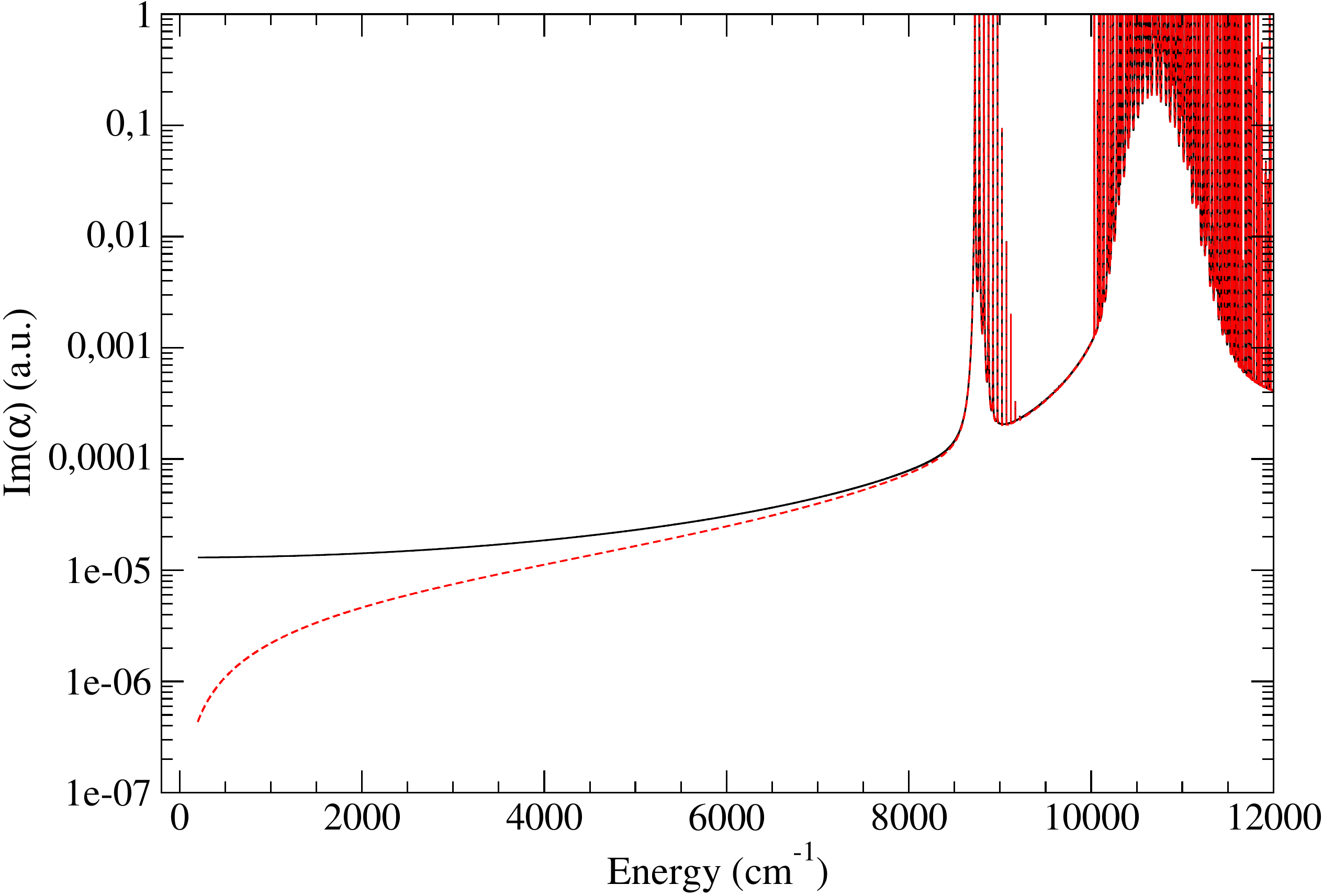}
\caption{\label{fig:Im-comparaison} \small Imaginary part of the DDP of RbCs computed with the constant-sign formula of Eq.~\eqref{eq:alpha-im-herm} (solid black curve), or with the opposite-sign formula of Eq.~\eqref{eq:alpha-im-non-herm} (dashed red curve) as a function of the energy of the laser. }
\end{center}
\end{figure}

\section{Polarizability in the optical-frequency domain}
\label{sec:opt-freq}

In this section we discuss the DDP of heteronuclear molecules in a ($X^1\Sigma^+, v=0, J=0$) level focusing more on optical, \textit{i.e.}~near-infrared and visible, trapping frequencies. The case of RbCs is first investigated in detail for illustration, and the resulting conclusions can be readily extended for all the heteronuclear dimers, unless otherwise stated in the rest of the paper.

We start with analyzing the convergence of our calculations with respect to the transitions involving different electrically-excited states and core electrons in RbCs. On Fig.~\ref{fig:rbcs-converg}, we show, as functions of the trapping frequency, their relative contributions with respect to the total real part of the DDP, calculated with all the excited states given in Table \ref{table:PEC}, and shown in Fig.~\ref{fig:dynpol_rbcs}. Apart from the peaks, where the total DDP vanishes, and which are therefore irrelevant, the first two excited states, $A^1\Sigma^+$ and $B^1\Pi$, represent at least $91~\%$ of the total DDP. Their contribution increases as the trapping frequency gets closer to the resonances toward vibrational levels of the $A$ state, because then the denominator of Eqs.~\eqref{eq:pola-para} and \eqref{eq:pola-perp} decreases. The second largest contribution, which is at most $4~\%$ (see the difference between the red and black curves), is due to transitions involving core electrons. Then, higher excited states bring smaller contributions, because their energy difference with the ground state increases, and their transition dipole moment with this state decreases.

\begin{figure}[htbp]
  \begin{center}  
  \includegraphics[width=0.45\textwidth]{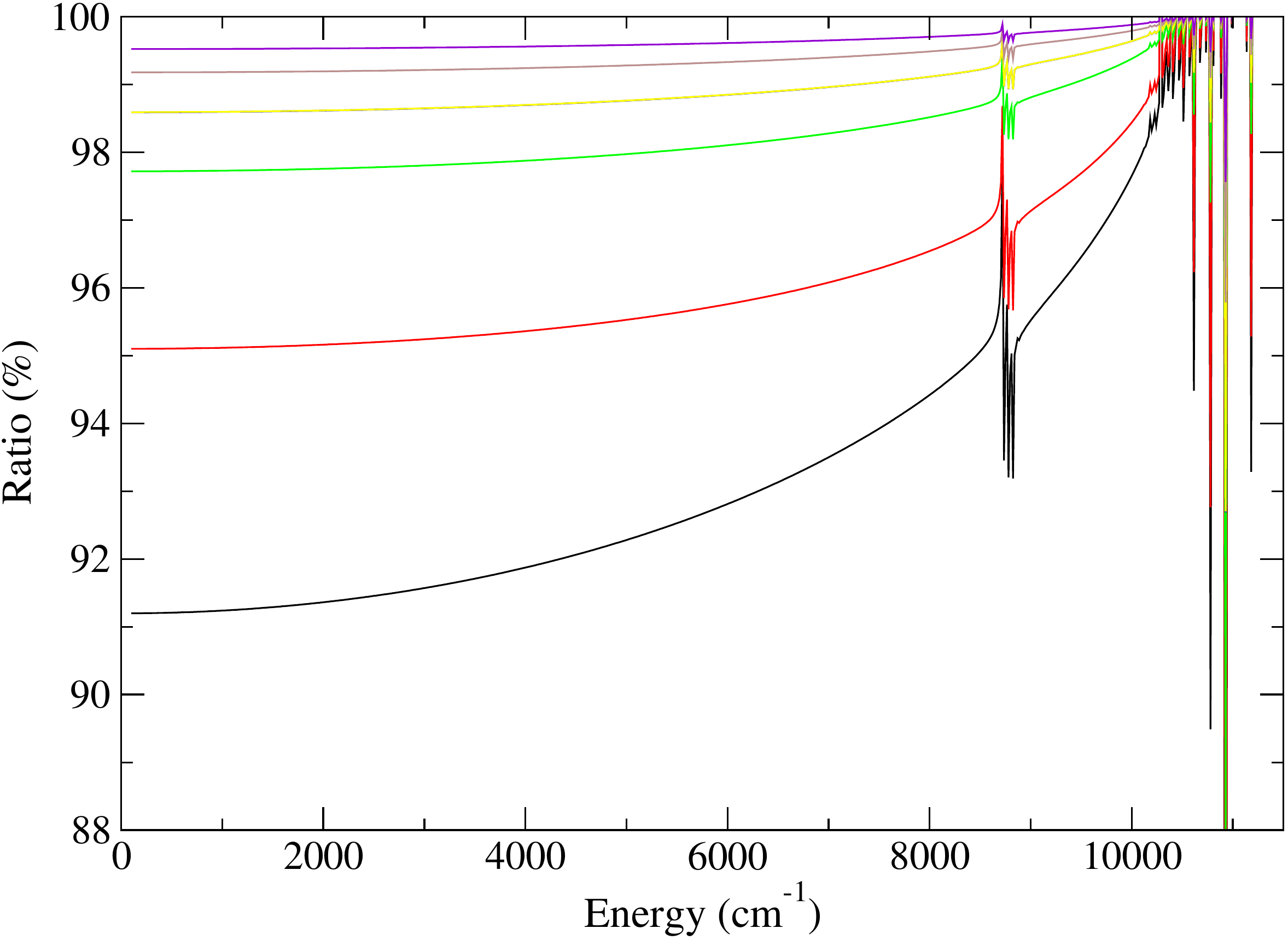}
  \caption{\label{fig:rbcs-converg} Relative contributions to the real part of the DDP (see Eqs.~\eqref{eq:pola-para}) and \eqref{eq:pola-perp}), of transitions involving different electronically-excited states and core electrons, as functions of the trapping frequency. The total DDP, correspoding to $100~\%$, is calculated with all the excited states given in Table \ref{table:PEC}. Each curve displays the cumulative contribution of a restricted number of states, namely: 2 states ($A$ and $B$) for the black curve; 2 states + core transitions for the red curve; 4, 6, 7, 9 and 10 states + core transitions for the green, blue, yellow, brown and purple curve respectively.}
  \end{center}
\end{figure}

\begin{figure*}[htbp]
  \begin{center}  
  \includegraphics[width=0.9\textwidth]{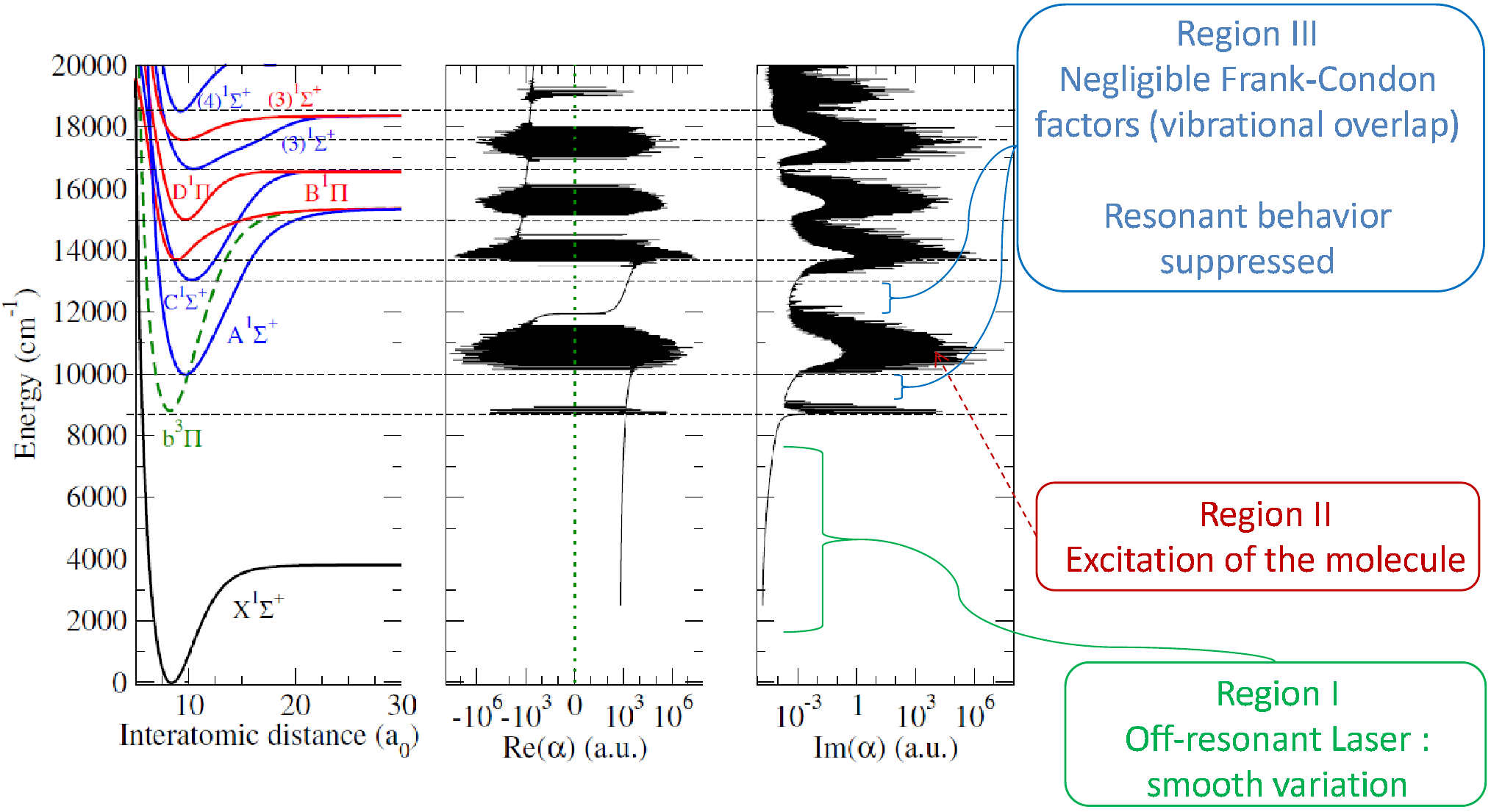}
  \includegraphics[width=0.9\textwidth]{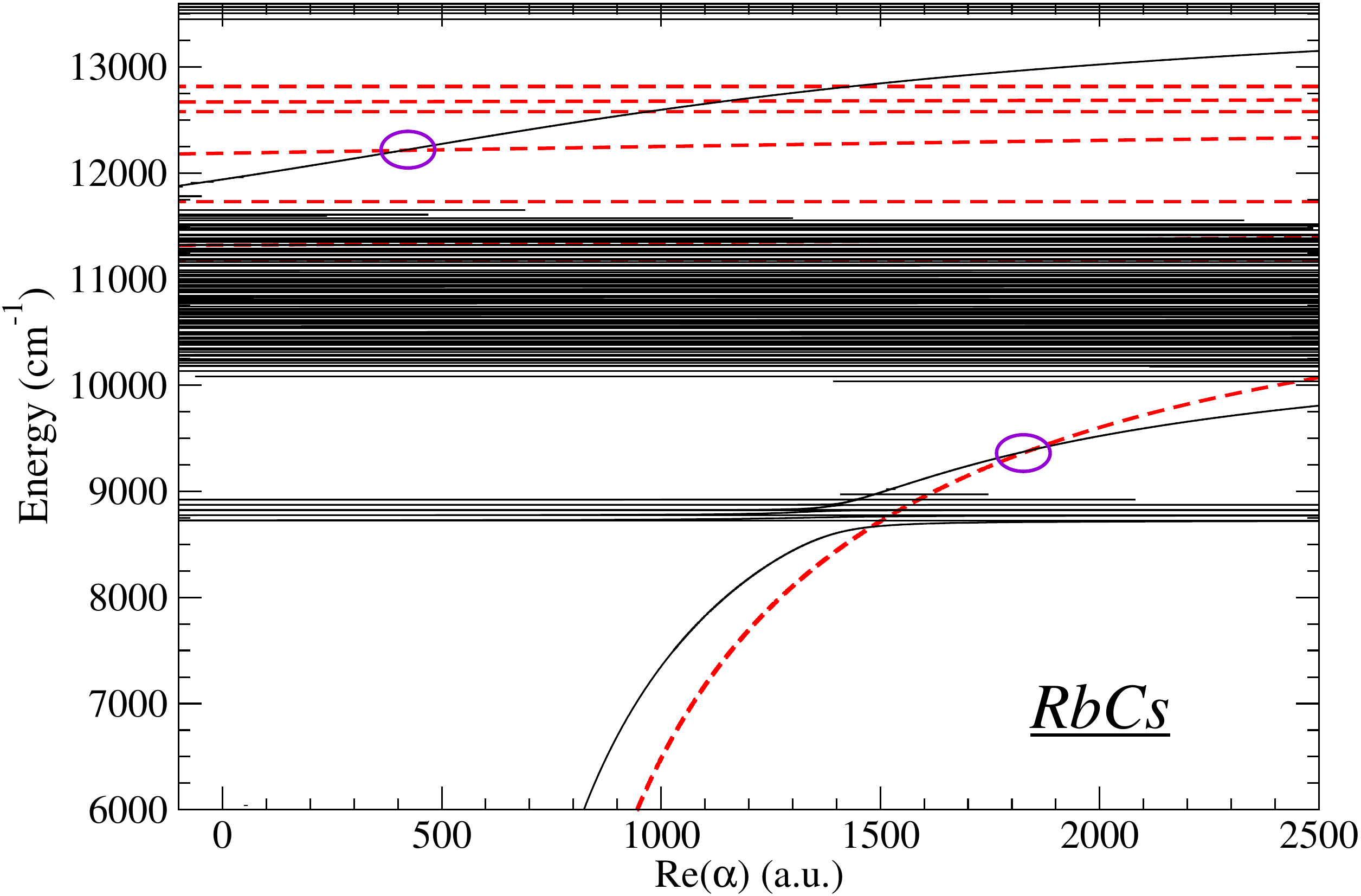}
  \caption{\label{fig:dynpol_rbcs} \small \underline{RbCs :} Upper panel: Potential energy curves of RbCs as a function of the internuclear distance (left graph). Real part (middle graph) and imaginary part (right graph) of the isotropic dynamic dipole polarizability (DDP) of a $X^1\Sigma^+,v=0, J=0$ RbCs molecule as a function of the energy of the laser. Specific regions discussed in the text are highlighted in boxes.  Lower panel: Zoom on the real part of the isotropic DDP (black solid curve) and comparison with the sum of the atomic DDP of Rb and Cs (red dashed curve). The circle points out the magic frequencies where the two quantities are equal.}
  \end{center}
\end{figure*}

Our results are displayed on similar figures for all the molecules, namely on the panel (a) of Fig.~\ref{fig:dynpol_rbcs} for RbCs, Fig.~\ref{fig:dynpol_lina} for LiNa, Fig.~\ref{fig:dynpol_lik} for LiK, Fig.~\ref{fig:dynpol_lirb} for LiRb, Fig.~\ref{fig:dynpol_lics} for LiCs, Fig.~\ref{fig:dynpol_nak} for NaK, Fig.~\ref{fig:dynpol_narb} for NaRb, Fig.~\ref{fig:dynpol_nacs} for NaCs, Fig.~\ref{fig:dynpol_krb} for KRb and Fig.~\ref{fig:dynpol_kcs} for KCs. 

The case of RbCs are shown in Fig.~\ref{fig:dynpol_rbcs}, for the real and imaginary parts of the DDP computed every 0.02~cm$^{-1}$ (or 600 MHz) for a range of frequencies between 0 and 20000~cm$^{-1}$. Note that as the real part can be positive or negative, the middle graph displayed on a logarithmic scale actually shows $sgn(\Re(\alpha(\omega))) \times \log(|\Re(\alpha(\omega))|)$, so that values of the real part with an absolute value lower than 1 a.u. are disregarded.

We can observe three kinds of behavior labeled I, II or III on figure \ref{fig:dynpol_rbcs}, and depending on the frequency of the trapping laser:
\begin{itemize}
 \item region I: the laser frequency is smaller than the lowest electronic transition frequencies, below the vibrational levels located in the bottom of the $b^3\Pi$ state potential well. The real and imaginary parts evolve smoothly with the laser frequency. The small magnitude of the imaginary part indicates a low photon scattering rate.
 \item region II: the laser frequency leads to strong absorption, in particular at the peaks observed both in the real and imaginary parts. The amplitude and width of these peaks depend on the TDM and on the lifetime of the reached excited level. As we took an arbitrary lifetime of 10 ns for all the excited levels, the value of the imaginary part is not relevant in itself; but the relative height of the peaks gives an indication of the relative strength of the different transitions.
 \item region III: in principle, such laser frequencies could induce electronic transitions. But it is actually not the case, because of unfavorable Franck-Condon factors (FCFs). This results in a smooth variation of the DDP, and in a relatively low imaginary part.
\end{itemize}

An important criterion for the choice of a lattice frequency to trap molecules is the sign of the corresponding real part of the DDP. The relation between the DDP and the potential seen by the molecules at a position $\vec{r}$ of the optical lattice being $U(\vec{r}) \propto -\Re(\alpha(\omega)) \times I(\vec{r})$, with $I(\vec{r})$ the $\vec{r}$-dependent intensity of the standing wave creating the lattice. For positive values of $\Re(\alpha)$, the molecules are trapped in the regions of maximum laser intensity, while for negative values of $\Re(\alpha)$, they are trapped in the regions of minimum laser intensity. For a two-level system, we have the well-known result that the DDP is positive for a red-detuned laser with respect to the transition, and negative for a blue-detuned one. For a multilevel system the problem is more complex because different transitions can compensate each other in the sum of Eqs.~\eqref{eq:pola-para} and \eqref{eq:pola-perp}, making no obvious the choice of the appropriate trapping frequency.

Nevertheless, we can extract general conclusions from Fig.~\ref{fig:dynpol_rbcs}. The real part of the DDP is positive if the laser is red-detuned with respect to the major transitions and negative if it is blue-detuned. Our calculations show that these major transitions are towards levels of the $A^1\Sigma^+$ and $B^1\Pi$ states, showing the best Franck-Condon factors with the ground state level $v$. In particular, the real part of the DDP is negative for frequencies slightly larger than the transition frequencies towards the levels of A$^1\Sigma^+$, but is positive for frequencies slightly smaller than the transition frequencies towards the levels of $B^1\Pi$. In between, because the DDP varies continuously, there exists one frequency, the so-called tune-out frequency, for which the real part of the DDP is exactly zero.

It is worthwhile to notice that for RbCs the deepest levels of the $b^3\Pi$ state have enough singlet character, due to spin-orbit coupling with A$^1\Sigma^+$, and sufficiently strong FCFs with the ground level, to induce a resonant behavior. However there is an optical window around 9000 cm$^{-1}$, corresponding to frequencies reaching the levels of the $b^3\Pi$ state, but below the bottom of the A$^1\Sigma^+$ state, where the contributions of individual resonances cancel out. Within that window, which includes the laser frequency corresponding to the widely used 1064-nm wavelength, the trapping of the molecules can be achieved, as shown in our previous paper on Cs$_2$ \cite{Vexiau2011}. We predict that it will also be the case for KRb and KCs.

\section{Influence of the excited levels lifetime}
\label{sec:lifetime}

\begin{figure}[htbp]
\begin{center}
\includegraphics[width=0.45\textwidth]{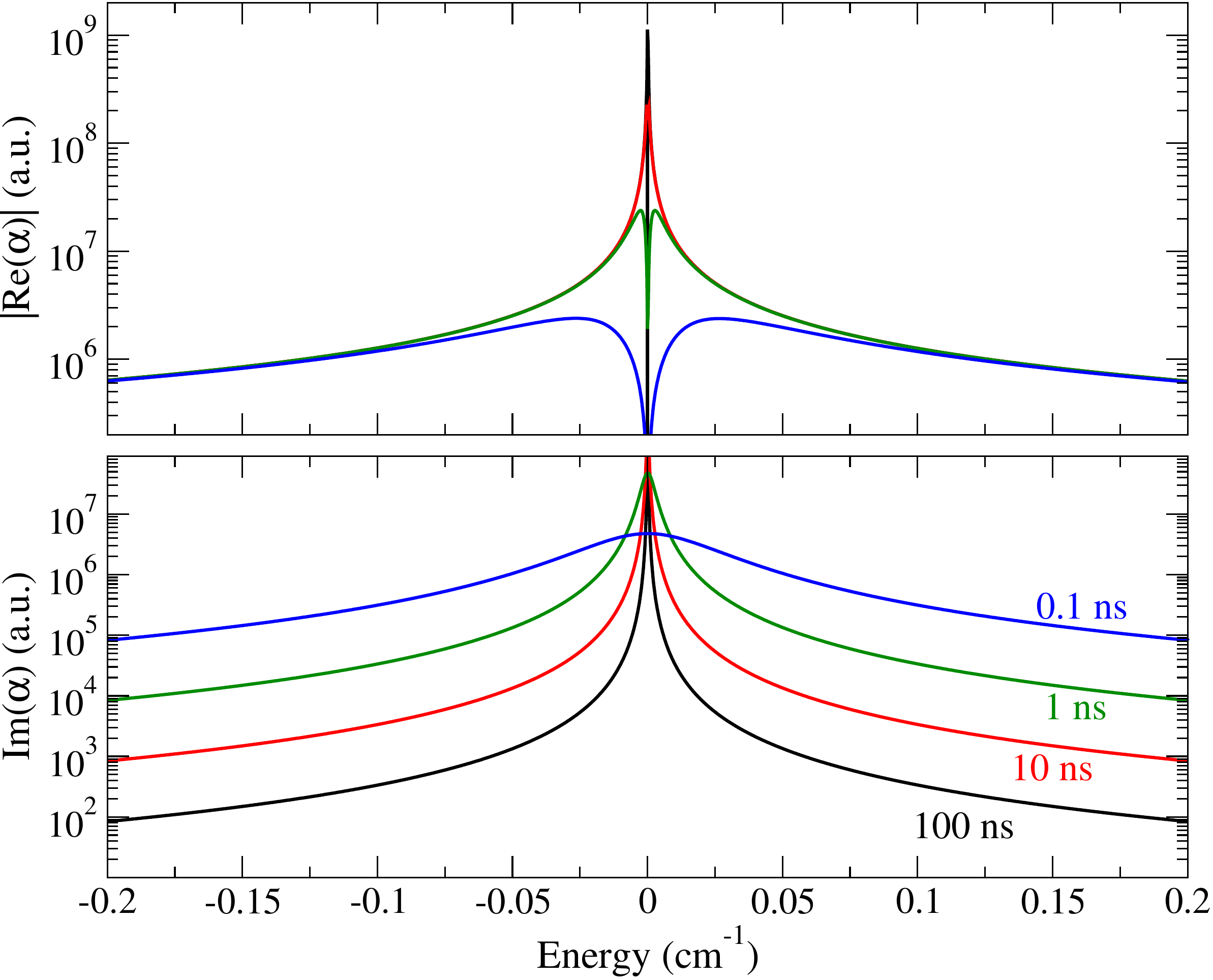}
\caption{\label{fig:lifetime} \small Real part (top panel) and imaginary part (bottom panel) of the DDP of a $X^1\Sigma^+,v=0,J=0$ RbCs molecule as a function of the absolute difference between the energy of the laser and an electronic transition. Calculations were made with a fixed excited level lifetime of $0.1~ns$ (blue curve), $1~ns$ (green curve), $10~ns$ (red curve), $100~ns$ (black curve). }
\end{center}
\end{figure}

Up to now, we performed our calculation using an arbitrary radiative lifetime of 10 ns for all excited levels. To understand the influence of this choice, we run several calculations using lifetimes equal to 0.1, 1, 10 and 100 ns. Figure \ref{fig:lifetime} shows the result of these calculations centered around the energy of one allowed electronic transition chosen arbitrarily. As we can see on the real part of the DDP, the height of the resonant peak is proportional to the lifetime, while the difference between the curves quickly vanishes away from the resonance. For more than 0.1 cm$^{-1}$ away from the resonance, the real part of the DDP is, to a good approximation, independent from the lifetime, which justifies our choice of an arbitrary value.

The imaginary part of the DDP is however inversely proportional to the lifetime outside of resonances, which supports our statement that our results should not be understood as giving quantitative information about the photon scattering rate. Instead, it indicates the frequencies inducing excitation, the frequency ranges of low absorption and the frequency ranges of high absorption.

\section{Comparison with other theoretical and experimental results}
\label{sec:compar}

\begin{table}[htbp]
 \caption{\label{pola:hetero:static} \small Comparison of permanent dipole moment and static dipole polarizabilities  $\alpha_\mathrm{exc}(0)$ (\textit{i.e.}~without taking into account the pure rotational transition) and $\alpha(0)$, of heteronuclear bialkali molecules in the $(X^1\Sigma^+ , v=0 , J=0$) level. Note that the core contribution $\alpha_c$ to the static polarizability is missing in the work of ref. \cite{Deiglmayr2008a}. In order to compare similar quantities, we added this contribution, computed using the values of Table \ref{table:core}, to their polarizabilities.}
\begin{ruledtabular}
\begin{tabular}{|c|c|c|c|c|}
   Molecule & Reference & PDM (Debye) & $\alpha_\mathrm{exc}(0)$ (a.u.)  & $\alpha(0)$ (a.u.)  \\
  \hline
  LiCs &
  This work & $5.59$ & $377.1$ & $1.890~\times10^6$ \\
  &ref. \cite{Deiglmayr2008a} & $5.523$ & $384.1$ & \\
  &ref. \cite{Quemener2011} & $5.355$ & $389.7$ & \\
  &ref. \cite{Byrd2012} &  & $367.8$ & \\
  \hline
  NaCs &
  This work & $4.69$ & $423.2$ & $4.298~\times10^6$ \\
  &ref. \cite{Deiglmayr2008a} & $4.607$ & $437.2$ & \\
  &ref. \cite{Byrd2012} &  & $411.2$ & \\ 
  \hline
  LiRb &
  This work & $4.18$ &  $346.1$ & $9.198~\times10^5$\\
  &ref. \cite{Deiglmayr2008a} & $4.165$ & $349.3$ & \\
  &ref. \cite{Quemener2011} & $4.046$ & $346.2$ & \\
  &ref. \cite{Byrd2012} &  & $319.2$ & \\
  \hline
  LiK &
  This work & $3.58$ & $318.8$ & $5.675~\times10^5$ \\
  &ref. \cite{Deiglmayr2008a} & $3.565$ & $322.6$ & \\
  &ref. \cite{Quemener2011} & $3.513$ & $324.9$ & \\
  &ref. \cite{Byrd2012} &  & $326.3$ & \\
  \hline
  NaRb &
  This work & $3.31$ &  $378.6$ & $1.784~\times10^6$ \\
  &ref. \cite{Deiglmayr2008a} & $3.306$ & $384.5$ & \\
  &ref. \cite{Byrd2012} & & $358.4$ & \\
  \hline
  NaK &
  This work & $2.78$ & $350.4$ & $9.237~\times10^5$ \\
  &ref. \cite{Deiglmayr2008a} & $2.759$ & $356.2$ & \\
  &ref. \cite{Byrd2012} &  & $344.6$ & \\
  \hline
  KCs&
  This work & $1.84$ & $572.5$ & $1.259~\times10^6$  \\
  &ref. \cite{Deiglmayr2008a} & $1.906$ & $585.2$ & \\
  &ref. \cite{Byrd2012} &  & $591.3$ & \\
  \hline
  RbCs&
  This work & $1.25$ & $621.5$ & $1.076~\times10^6$ \\
  &ref. \cite{Deiglmayr2008a} & $1.237$ & $621.9$ & \\
  &ref. \cite{Byrd2012} & & $635.0$ & \\
  \hline
  KRb&
  This work & $0.62$ & $513.1$ & $1.141~\times10^5$ \\
  &ref. \cite{Deiglmayr2008a} & $0.615$ & $514.8$ & \\
  &ref. \cite{Byrd2012} & & $523.8$ & \\
  \hline
  LiNa&
  This work & $0.57$ & $233.5$ & $9.950~\times10^3$\\
  &ref. \cite{Deiglmayr2008a} & $0.566$ & $236.5$ & \\
  &ref. \cite{Quemener2011} & $0.531$ & $237.8$ & \\
  &ref. \cite{Byrd2012} & & $223.7$ & \\ 
 \end{tabular}
 \end{ruledtabular}
 \end{table}

To test the validity and the precision of our calculations we compare the real part of our DDP values with the few data  published in the literature. Firstly, Table \ref{pola:hetero:static} shows such a comparison in the static limit $\omega=0$. The agreement with other (theoretical) values of the ground-state PDM and static polarizability is good. One can also quote the work on KRb of Ref.\cite{Buchachenko2012} which report results at equilibrium distance using several levels of approximation, ranging between 524 and 539 a.u. . Note that the polarizability values from our previous work \cite{Deiglmayr2008a} was obtained with the finite field method applied on a theoretical ground state PEC, while the sum-over-states formula employed here involves an experimentally-derived ground state PEC. Similarly, the ground-state PDM is obtained after averaging the $R$-dependent PDM function of Ref.\cite{Aymar2005} onto the $v=0$ level of the experimentally-derived ground state PEC. These statements explain the small differences observed among all values for a given species.

\begin{table}[htbp]
 \caption{\label{pola:hetero:exp} \small Comparison with available experimental values of various components of the computed DDP (either isotropic, parallel, or perpendicular) of bialkali molecules in the ($v=0 ; J=0$) level of the electronic ground state of KRb and Cs$_2$, and of the lowest  triplet electronic state for Rb$_2$. The results are given in $10^{-5}$~MHz/(W/cm$^2$), or in units of separated atoms polarizability for Cs$_2$ $\alpha_\textrm{Cs}$ (see text). Note that the two displayed values of Ref.\cite{Neyenhuis2012} correspond to an experimental and a theoretical determination.} 
\begin{ruledtabular}
  \begin{tabular}{cccc} 
    Molecule & Wavelength & {}DDP component & polarizability  \\
     & (nm) & & ($10^{-5}$~MHz/(W/cm$^2$))\\
   \hline
  KRb & 1090 & isotropic & $4.2$ \cite{Kotochigova2009}   \\
  & & & $5.1$ (this work)\\
  & & & $4.8$ \cite{Ospelkaus2009} \\
  & 1064 & parallel & $10.0(3)$ \cite{Neyenhuis2012}\\
  & & &$12$ \cite{Neyenhuis2012} \\
  & & & $10.0$ (this work) \\
  & & perpendicular & $3.3(1)$ \cite{Neyenhuis2012}\\
  & & & $2.0$ \cite{Neyenhuis2012} \\
  & & & $2.8$ (this work) \\
  Cs$_2$ & 1064.5 & isotropic & $2.42~\alpha_\textrm{Cs}$ (priv. comm.)\\
   & & & $2.49~\alpha_\mathrm{Cs}$ (this work)\\
  Rb$_2$ & $830.4$ & isotropic & $2.809$ \cite{Deiss2015} \\
  & & & $4.106$ (this work)\\
  & $1064.5$ & isotropic & $16.08 \pm 0.67$ \cite{Deiss2015} \\
  & & & $14.75$ (this work) \\
  & & parallel & $41.7 \pm 4.2$ \cite{Deiss2014} \\
  & & & $35.2 \pm 5.6$ \cite{Deiss2014} \\
  & & & $34.97$ (this work) \\
  & & perpendicular & $4.2 \pm 1.9$ \cite{Deiss2014} \\
  & & & $4.7 \pm 0.5$ \cite{Deiss2014} \\
  & & & $4.75$ (this work) \\
 \end{tabular}
 \end{ruledtabular}
 \end{table}

In table \ref{pola:hetero:exp} we present all the  experimental and theoretical DDPs that we have found in the literature and we compare them to our calculated values. For experimentalist convenience, the results are given in MHz/(W/cm$^2$ ($1 \mathrm{a.u.} = 4.6883572 \times 10^{-8}$~MHz/(W/cm$^2$)), or sometimes in units of separated atoms polarizability. Again the overall agreement is good, except for Rb$_2$ at 830.4 nm (which was discussed in Ref.\cite{Deiss2014}.

Among all species, KRb is the most studied one in the context of ultracold molecules. Extensive experiments and calculations have been made \cite{Kotochigova2006, Ospelkaus2009}, to extract the DDP of the molecule in its absolute ground state for an optical lattice of wavelength 1090~nm. Later, the same experimental group made another measurement using a 1064-nm laser \cite{Neyenhuis2012}. In this case the lattice was not an isotropic 3D-lattice but a 1D-lattice, which allowed to tilt the axis of the laser from the quantization axis defined by the magnetic field. To explore the anisotropic behavior of the DDP, measurements were made for various tilting angles, and so the parallel and perpendicular polarizabilities were extracted.
Note that the experimental values were given in units of DDP of Feshbach molecules $\alpha_\mathrm{Fesh}$ (see Sec.~\ref{sec:Feshbach}). For example, the measured isotropic DDP of KRb, is $0.85~\alpha_\mathrm{Fesh}$, which gives $4.8\times10^{-5}$ MHz/(W/cm$^2$), by taking the sum of the atomic DDP for $\alpha_\mathrm{Fesh}$. It is in good agreement with our value of $5.1\times10^{-5}$~MHz/(W/cm$^2$).

On the theoretical side, Kotochigova and coworkers \cite{Kotochigova2006, Kotochigova2010} have computed DDPs for a broad range of frequencies for both KRb and RbCs molecules. Nevertheless, the values reported in Table \ref{pola:hetero:exp} are the only numerical ones with which we can compare our results. Their calculations were done using the same formalism, but relying completely on PECs calculated with quantum chemistry. This probably explains the slight differences between their results and ours.

Even if the present study focuses on heteronuclear molecules, we report in Table \ref{pola:hetero:exp} on two other relevant experiments, on Cs$_2$ and Rb$_2$ molecules. In Ref.~\cite{Danzl2010} we report a difference of $18~\%$ between the experimental and the theoretical values for Cs$_2$. However it has been shown later by the same group \cite{Markp} that the difference was not due to a lack of accuracy in the experiment or in our calculation. Indeed our calculation was made in the lowest rotational level $J=0$, while the measurement was made in the excited rotational level $J=2$. At the time, it was believed that due to the geometry of the 3D lattice, only the isotropic part of the DDP (equal to the $J=0$ value) was accessible experimentally, justifying a direct comparison of the $J=2$ experimental value and the $J=0$ theoretical one. A new measurement for molecules in $J=0$ invalidated this assumption. Indeed, the new measured value of $2.42~\alpha_\mathrm{Cs}$, $\alpha_\mathrm{Cs}$ being the DDP of the Cesium atom, is very close to our theoretical value of $2.49~\alpha_\mathrm{Cs}$. 
	
In Ref.~\cite{Deiss2015} measurements of the DDP for Rb$_2$ molecules in the (a$^3\Sigma^+_u,v=0$) level are compared to our theoretical predictions using the same formalism. For a trapping wavelength of 1064.5~nm, our calculations are in very good agreement with two of the three measured values, \textit{i.e.}~the isotropic  and the perpendicular DDPs. The discrepancy on the parallel part can be easily explained: as the trapping frequency is close to resonances due to parallel transitions ($\Sigma-\Sigma$) towards the $(1)^3\Sigma^+_g$ excited state, the parallel part of the DDP is very sensitive to the description of the levels of that state. However in our calculation we used a PEC calculated with quantum chemistry for the $(1)^3\Sigma^+_g$ state, which altered the accuracy of our calculations. In Ref.~\cite{Deiss2015} we also report the measured value at a wavelength of 830.4~nm issued from an earlier, less accurate experiment, which only shows a rough agreement with our calculated value.

\section{Polarizability of Feshbach molecules}
\label{sec:Feshbach}

The efficiency of the transfer of Feshbach molecules (FMs) to the absolute ground state is enhanced by reducing the difference between the lattice depth seen by the FM and the one seen by the molecule in the absolute ground state ($X^1\Sigma^+,v=0,J=0$).
Because the trap depth is proportional to the DDP in each state, an optimal transfer is achieved if the two DDPs are equal, or at least, close to each other. Before searching for the trapping frequencies at which this equality occurs (see Sec.~\ref{sec:magic}), we calculate, in the present section, the DDP of Feshbach molecules.

Because FMs are characterized by vibrational levels just below the ground state dissociation limit $n_As + n_Bs$, thus with large elongation, it is natural to express their DDP as the sum of the atomic DDPs, $\alpha_\mathrm{Fesh}(\omega) = \alpha_A(\omega) + \alpha_B(\omega)$. To check this {}``atom-pair'' approximation, we compare its results with a {}``molecular'' calculation, in which the FM is assumed to be in the last vibrational level of the ground electronic state $X^1\Sigma^+$.

We first need to calculate the DDP of each ground-state ($ns$) alkaline atom, which is obtained using the sum-over-state formula (see Eq.~\eqref{eq:alpha-re}), taken over excited atomic levels. As input data, we take the TDMs and transition energies from the NIST database \cite{NIST2013}. At infrared frequencies the main contributions come from the $np_{1/2}$ and $np_{3/2}$ levels. Transitions towards $(n+1)p_{1/2,3/2}$ levels are also taken into account for Li, Na, Rb and Cs. Regarding the molecular calculation, the PECs and TDMs included in the sum are the same as those used for the ($X^1\Sigma^+ , v=0 , J=0$) level (see Sec.~\ref{sec:PEC}).

\begin{figure*}[htbp]
\begin{center}
\includegraphics[width=0.9\textwidth]{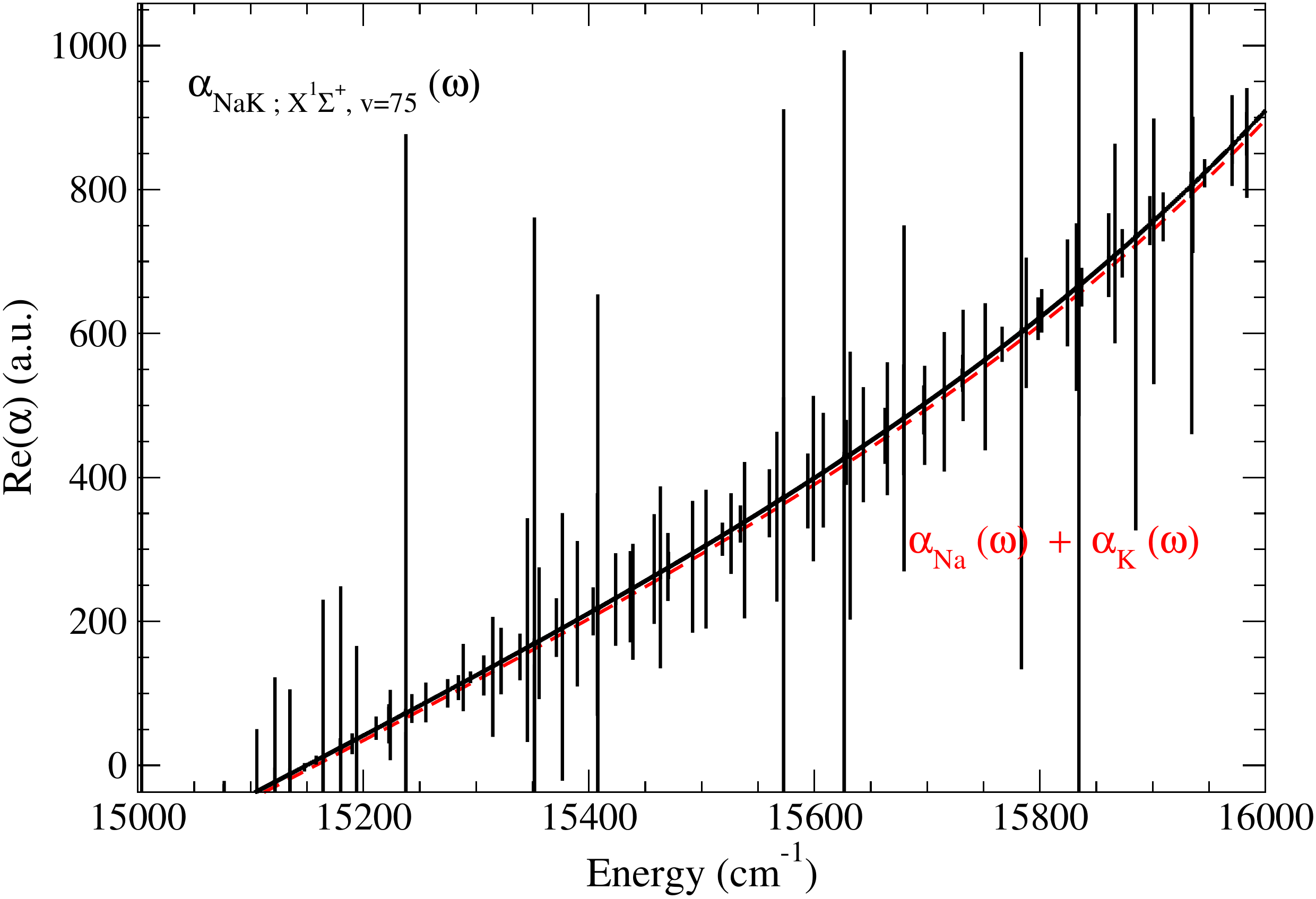}
\caption{\label{fig:magic_nak_fesh} \small Real part of the polarizability of the last bound vibrational level $v=75$ of the $X^1\Sigma^+$ state in NaK (black curve) and sum of the real part of the atomic polarizability of Na and K (red curve) as a function of the energy of the laser.}
\end{center}
\end{figure*}

The result of our analysis is illustrated in the case of NaK ($n_A=3$, $n_B=4$). We show on Fig.~\ref{fig:magic_nak_fesh} the real part of the molecular DDP (black curve) and compare it to the sum of the Na and K atomic DDPs (red curve). The overall variation of the two curves are very similar, while narrow peaks in the molecular DDP denote transitions towards vibrational levels of excited electronic states. Away from these resonances, the difference between the two DDPs is roughly of $3~\%$, which probably originates from the limited precision of the molecular input data. Indeed we used for the nearby $B^1\Pi$ state dissociating to Na($3s$)-K($4p$) a Hund's case (a) PEC without SO coupling. However we know that for weakly-bound levels of this state, which have large TDMs with the Feschbach molecular level, the SO coupling cannot be neglected and a Hund's case (c) representation would be more appropriate.

The resonances of the molecular DDP illustrate the limitation of the atom-pair approximation. For an optimal trapping experiment, one has to make sure that the frequency of the lattice laser will not induce resonant absorption for the FMs. The position of those resonances strongly depends on the details of the vibrational wave function of the FM, and in particular on its singlet or triplet character. Outside the resonances, the DDPs of a singlet and of a triplet FM coincide and are well described in the atom-pair approximation.

It is worthwile mentioning that the dependence of the parallel and perpendicular components of the static ($\omega=0$) dipole polarizability at large interatomic distances has been discussed in several papers \cite{crubellier2015,heijmen1996,jensen2002}

\begin{align}
\alpha_{\parallel} & =\alpha_1+\alpha_2+\frac{4\alpha_1 \alpha_2}{R^3}+\frac{4\alpha_1 \alpha_2(\alpha_1+\alpha_2)}{R^6} \nonumber \\
\alpha_{\perp} & =\alpha_1+\alpha_2-\frac{2\alpha_1 \alpha_2}{R^3}+\frac{\alpha_1 \alpha_2(\alpha_1+\alpha_2)}{R^6}
\label{eq:alpha_LR}
\end{align}

where $\alpha_1$ and $\alpha_2$ are the atomic static dipole polarizabilities of the individual atoms. We applied this formula to the uppermost vibrational levels of the NaRb ground state by averaging Eq.~\eqref{eq:alpha_LR} on the related vibrational wavefunctions, and we compared the results to the ones of the sum-over-states approach (Fig. \ref{fig:pola-NaRb-asympt}). We see that for the uppermost level, this formula seems to agree with the numerical result from the present study. Note however that the correction to the isotropic polarizability $\alpha=(\alpha_{\parallel}+2*\alpha_{\perp})/3$ induced by the $R^{-6}$ term in Eq.~\eqref{eq:alpha_LR} accounts for about $0.1~\%$ compared to the sum of atomic polarizabilities. Given that our numerical calculation relies on a drastically different approach based on theoretical transition dipole moments, it is tedious to argue about a good agreement between the two approaches.
In contrast these two models yield different results already from the last-but-one level, and increase for even deeper levels. This is due to the short range of the interactomic potential (varying as $R^{-6}$ for different atoms), so that already for the last-but-one vibrational levels, one needs to consider its radial wavefunction over the entire range of distances, \textit{i.e.} where the interaction potential is governed by exchange interaction. This specificity of $R^{-6}$ asymptotic potentials was already pointed out for instance for the calculation of photoassociation rates for heteronuclear alkali-atom pairs \cite{azizi2004}.

\begin{figure}[htbp]
\begin{center}
\includegraphics[width=0.45\textwidth]{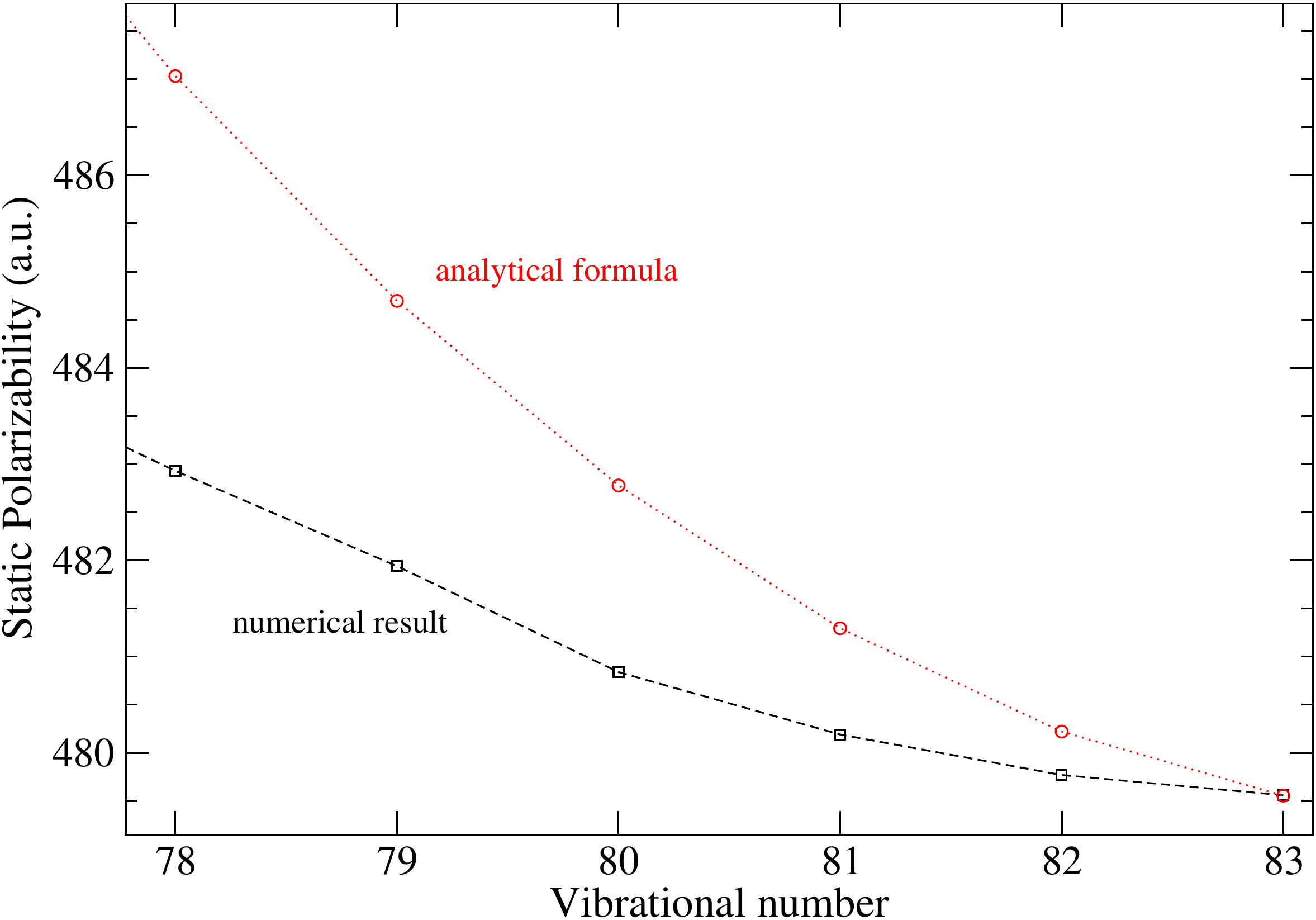}
\caption{\label{fig:pola-NaRb-asympt} Comparison of the DDP of the uppermost vibrational levels of the NaRb ground state, as obtained from the present numerical calculations, and as resulting from the application of the asymptotic expansion of Eq.~\eqref{eq:alpha_LR}. }
\end{center}
\end{figure}

\section{Magic frequencies}
\label{sec:magic}

Now we compare the DDP of the ($X^1\Sigma^+,v=0,J=0$) molecules and the one of the FMs, expressed in the atom-pair approximation. These two quantities are displayed over a broad range of frequencies on panels b of Fig.~\ref{fig:dynpol_rbcs} for RbCs, Fig.~\ref{fig:dynpol_lina} for LiNa, Fig.~\ref{fig:dynpol_lik} for LiK, Fig.~\ref{fig:dynpol_lirb} for LiRb, Fig.~\ref{fig:dynpol_lics} for LiCs, Fig.~\ref{fig:dynpol_nak} for NaK, Fig.~\ref{fig:dynpol_narb} for NaRb, Fig.~\ref{fig:dynpol_nacs} for NaCs, Fig.~\ref{fig:dynpol_krb} for KRb, and Fig.~\ref{fig:dynpol_kcs} for KCs. Focusing mainly on frequencies below the one reaching the bottom of the $B^1\Pi$ state, we see that, outside the regions of resonances towards the A$^1\Sigma^+$ and $b^1\Pi$ states, the DDPs are both positive and have the same order of magnitude. For several molecules (RbCs, KCs, KRb, NaK), we note the existence of particular frequencies at which the DDPs are equal: these are the so-called ``{}magic frequencies''. When they exist, the magic frequencies (\textit{i.e.} intersections located away from resonances) are given for each molecule in table \ref{pola:hetero:table}.

 \begin{table*}[htb]
 \caption{\label{pola:hetero:table} \small Magic frequencies (when present), DDP at the magic frequency $\omega_0$, and DDP at two standard frequencies $\omega_{L1}$ and $\omega_{L2}$ corresponding to the wavelengths 1064 and 1550 nm respectively. All the DDPs are given in units of 10$^{-5}$~MHz/(W/cm$^2$). Only frequencies outside the resonant regions are given. The column {}``sensitivity'' contains the relative difference between the DDPs of ground-state and Feshbach molecules at an arbitrary value of $5~cm^{-1}$ above the magic frequencies. The last column gives the ratio at $\omega_{L2}$ between the DDP of ground-state molecules and the DDP of FMs. the star symbol for LiCs indicates that the frequency $\omega_{L1}$ is close to strong transitions towards levels of the $b^3\Pi$ state.}
\begin{ruledtabular}
  \begin{tabular}{c|ccc|ccc} 
    Molecule & magic frequencies & $\alpha(\omega_0)$ & sensitivity & $\alpha(\omega_{L1})$ & $\alpha(\omega_{L2})$ & $\alpha / \alpha_\mathrm{Fesh}(\omega_{L2})$ \\
     & $\omega_0$ ($cm^{-1}$) & & ($\%$) & & & \\
   \hline
   RbCs & $9390$ and $12214$ & $8.645$ and $1.923$ & $0.11$ and $-14.4$ & $8.667$ & $4.082$ & $0.874$\\
   KCs & $9639$ and $12394$ &$9.047$ and $1.421$ &$0.14$ and $-10.7$ & $7.783$ & $3.760$ & $0.842$\\
   KRb & $10287$  & $7.760$ &$0.12$& $5.385$ & $3.188$ & $0.846$ \\
   NaCs & - &- &- & $4.379$ & $2.586$ & $0.725$\\
   NaRb & - &- &-& $3.158$ & $2.207$ & $0.768$ \\
   NaK & $15718$ & $2.414$ & $-0.46$& $2.790$ & $2.017$ & $0.757$ \\
   LiCs & - & - & - & $3.695$* & $2.299$ &$0.635$\\ 
   LiRb & - & - & - & $3.044$ & $2.045$ &$0.700$\\
   LiK & - & - & - & $2.584$ & $1.844$ &$0.680$\\
  LiNa & - & - & - & $1.582$ & $1.275$ &$0.702$\\
 \end{tabular}
 \end{ruledtabular}
 \end{table*}

Around magic frequencies the DDP of the molecule varies smoothly. As an indication we compute in table \ref{pola:hetero:table} the difference between the DDPs of the absolute ground state molecule and the FM one, when the laser is detuned by 5~cm$^{-1}$ from the magic frequency. When that difference is smaller than 0.5 \% in absolute value,  the differences in the trapping efficiency due to this 5-cm$^{-1}$ detuning should be negligible compared to other limitations of the experimental setup, allowing experimentalists to choose an available laser wavelength near the magic frequency without a significant loss in the transfer efficiency. 

During the experiments it is easier to use the same laser to trap the separated atoms, the FMs and the ground-state molecules. This implementation is problematic for  NaK and LiCs molecules, because at the determined magic frequency, the DDP of the K atom in the case of NaK (resp.~the Cs atom in the case of LiCs) is negative, while all the other DDPs are positive. In an optical lattice, the K and Cs atoms would be trapped in the region of minimum intensity, whereas all the other species would be trapped in the region of maximum intensity, which would prevent Na-K and Li-Cs magneto-association. Therefore, NaK and LiCs can be considered as having a workable magic frequency, provided that two different lattices can be implemented subsequently: one lattice for the magneto-association step, running at a frequency such that the atomic DDPs are positive, and another lattice for the STIRAP step, running at the magic frequency.

For the molecules without magic frequency located in smoothly-varying regions of the DDP (NaCs, NaRb, LiRb, LiK, LiCs, LiNa), we propose two alternatives in order to successfully trap the molecules during the STIRAP process, implying two different kinds of trapping wavelengths:
\begin{itemize}
 \item Infrared wavelengths corresponding to frequencies far away from any resonance which ensure a very low absorption rate and a high lifetime for the molecules. The FM and the ground-state molecules DDPs are not equal, leading to a limitation in the efficiency of the transfer process. However as shown in our recent paper \cite{Deiss2015}, a ground-state molecule DDP  $2.5$ greater than the FM one still allows for an acceptable transfer efficiency.
 \item Visible or near infrared wavelengths satisfying the {}``magic conditions'', but firstly not retained and not included in Table \ref{pola:hetero:table}. Indeed their vicinity to resonances induces non-negligible absorption rate, and limits the trap lifetime of the molecules down to a few milliseconds. This alternative was tested with Cs$_2$ \cite{Vexiau2011} where the FMs were first transferred  towards the $X,v=73$ level with a high efficiency, even if the lifetime of the sample was only of 19~ms \cite{Vexiau2011}. The determination of a magic frequency so close to resonances requires an accurate knowledge of their position and width, which must thus be determined with spectroscopic precision.
\end{itemize}

An alternative solution to circumvent the lack of magic frequency would be to add an additional control mechanism during the STIRAP process, which would consist in keeping the lattice depth identical for molecules in the (initial) Feshbach state, in the (final) ground state, or in any (intermediate) superposition of the two states occurring during the transfer. This requires to adjust the laser intensity in order to compensate the variation of DDP of a molecule during the transfer, imposing that the product (DDP $\times$ intensity) is constant.
To illustrate this proposal we consider LiK molecules trapped by a telecom laser at 1550~nm. Since the DDP of ground-state molecules represents $68~\%$ of the DDP of FMs (see Table \ref{pola:hetero:table}), the intensity should be increased by 47 \% within the transfer duration of a few microseconds \cite{Takekoshi2014}. It would be particularly interesting to study the experimental feasibility of this proposal.

\section{Effective polarizability for all vibrational levels of the electronic ground state}
\label{sec:effective}

The above results consist in thousands of computed DDP values for the ten heteronuclear molecules in their absolute ground state, and for frequencies from 0 to 25000~cm$^{-1}$ with a step of 0.02~cm$^{-1}$. Even though we have all the necessary data to obtain the DDPs for any vibrational $v$ level of the ground state, the number of computed DDP values) (and subsequently, of anistropy values) would be too big to build an exhaustive table, even as supplementary material. In this section, we rather give an analytical expression of the isotropic DDPdepending on a few parameters, which is a reliable estimate of the DDP outside the regions of resonances, as already suggested in Ref.\cite{Deiss2015} for Rb$_2$. Such an expression also gives further insight into the limited number of transitions that mainly contribute to Eqs.~\eqref{eq:pola-JM}--\eqref{eq:pola-perp}.

The effective isotropic DDP is then given by: 
\begin{equation}
  \label{eq:pola-eff}
  \alpha_\mathrm{eff}(\omega) = \frac{2\omega_{\Sigma}
   d_{\Sigma}^2} {\omega_{\Sigma}^2 - \omega^2} 
   + \frac{2\omega_{\Pi}d_{\Pi}^2} {\omega_{\Pi}^2 - \omega^2} 
   + \alpha_c(\omega)
\end{equation}
where $\omega_{\Sigma}$ and $d_{\Sigma}$ (resp.~$\omega_{\Pi}$ and $d_{\Pi}$) represent an effective transition energy and an effective TDM towards the first excited $\Sigma$ (resp.~$\Pi$) states, which bring the main contribution to the DDP. The parameters $\omega_{\Sigma}$, $\omega_{\Pi}$, $d_{\Sigma}$ and $d_{\Pi}$ represent TDM averages on the transitions towards levels sharing a good FCF with the ground state level ($X^1\Sigma^+, v$). As the vibrational wave function of the ground state level greatly varies with $v$, we can thus expect the parameters of Eq.~\eqref{eq:pola-eff} to be $v$-dependent. The only exception is $\alpha_c(\omega)$, which reflects the contribution of transitions involving the core electrons (see Sec.~\ref{sec:PEC}), and which are also described in terms of effective transition parameters .
This core contribution is isolated in Eq.~\eqref{eq:pola-eff} as its energy dependence is significantly different from the other contributions.

\begin{table*}[htbp]
 \caption{\label{tab:effective} \small Effective transition energy and dipole moment for heteronuclear bialkali molecules in their absolute ground state $(X^1\Sigma^+, v=0, J=0)$, obtained by fitting the computed polarizabilities with Eq.~\eqref{eq:pola-eff}. For each molecule the range of frequencies excluded from the fit are given. In the case of RbCs, $\alpha(\omega)$ vanishes at $\omega=\numprint{11956}$ cm$^{-1}$ giving an abnormally high rms. Excluding the near zero values from the fit gives similar effective parameter with an rms of 0.67 \%. }
\begin{ruledtabular}
  \begin{tabular}{|c|c|c|c|c|c|c|c|c|} 
    Molecule & \multicolumn{2}{c|}{$\Sigma$ state} & \multicolumn{2}{c|}{$\Pi$ state} & \multicolumn{3}{c|}{Frequencies excluded (cm$^{-1}$)} & rms ($\%$)\\
   \hline
     & $\omega_{\Sigma}$  & $d_{\Sigma}$ & $\omega_{\Pi}$  & $d_{\Pi}$ & $b^3\Pi$  & A$^1\Sigma$ & $B^1\Pi$& \\
     & (cm$^{-1}$) & (a.u) & (cm$^{-1}$) & (a.u) & resonances  & resonances & resonances &\\
   \hline
   RbCs & $10694.77$ & $2.68$ &$13854.92$ & $3.08$ & $8574.19$-$9123.21$& $9886.05$-$12030.45$& $> 13215.42$& $0.466$\\
   KCs & $10758.38$ & $2.62$ &$14225.92$ & $2.99$ & $8675.89$-$9314.17$& $9940.45$-$12326.32$& $> 13635.89$& $0.404$\\
   KRb & $11583.60$ & $2.58$ &$15069.77$ & $2.90$ & $9584.94$-$10115.96$& $10767.00$-$13325.37$& $> 14318.42$& $0.428$\\
   NaCs & $11450.68$ & $2.32$ &$15681.56$ & $2.63$ & \multicolumn{2}{c|}{ $10069.66$-$13419.82$}& $> 15013.11$& $1.755$\\
   NaRb & $12725.97$ & $2.33$ &$16891.66$ & $2.61$ & \multicolumn{2}{c|}{ $11156.77$-$14890.54$}& $> 16276.57$& $1.083$\\
   NaK & $13164.20$ & $2.30$ &$17399.39$ & $2.59$ & \multicolumn{2}{c|}{ $11396.48$-$15463.52$}& $> 16798.45$& $1.127$\\
   LiCs & $11683.62$ & $2.31$ &$16088.54$ & $2.50$ & \multicolumn{2}{c|}{ $9018.60$-$14510.36$}& $> 15637.97$& $1.338$\\
   LiRb & $12326.08$ & $2.26$ &$17237.94$ & $2.42$ &  \multicolumn{2}{c|}{ $9938.34$-$15210.24$}& $> 16837.77$& $1.316$\\
   LiK & $12918.45$ & $2.22$ &$17783.19$ & $2.41$ & $10281.32$-$11397.07$& $11880.73$-$16416.37$& $> 17305.20$& $1.019$\\
   LiNa & $14862.75$ & $2.00$ &$20475.26$ & $2.25$ & (see text) & $14033.41$-$18589.43$& $> 19929.10$& $1.979$\\
 \end{tabular}
 \end{ruledtabular}
 \end{table*}

In order to determine the parameters of Eq.~\eqref{eq:pola-eff}, we first achieved the full numerical calculation of the isotropic DDP for each vibrational level $v$, and then  we fitted the results with Eq.~\eqref{eq:pola-eff}, for frequencies from 1000 to 20000~cm$^{-1}$, while excluding the regions of resonances. The corresponding parameters are given in Table \ref{tab:effective} for the lowest vibrational level $v=0$ and in the supplementary material for all vibrational levels, along with the energy domains for which Equation \eqref{eq:pola-eff} cannot be applied. For five molecules (NaCs, NaRb, NaK, LiRb, and LiCs), the excluded frequencies due to the $b^3\Pi$ and $A^1\Sigma^+$ are not separated, because the vibrational levels of $b$ and of $A$ presenting favorable FCFs towards the ground state do not belong to distinct energy ranges. For LiNa, as the spin-orbit interaction is very weak, and the $b^3\Pi$ state was not included in the calculation.

The choice of the lower bound 1000~cm$^{-1}$ of the frequency range arises from the specificity of heteronuclear bialkali molecules, which present low-frequency transitions within the electronic ground state $X^1\Sigma^+$ due to their permanent dipole moment. As mentioned in section \ref{sec:low-freq}, these transitions have negligible contributions at infrared frequencies, but are dominant at microwave frequencies. The choice of $\numprint{1000}$~cm$^{-1}$ allows to safely exclude the pure rotational and rovibrational transitions from the fitting procedure.

The $C^1\Sigma^+$ and $D^1\Pi$ states, correlated to a $n's + np$ asymptote, have strong TDMs with the ground state. One could then think about taking these states into account in the effective DDP, \textit{i.e.}~including four effective transitions in Eq.~\eqref{eq:pola-eff}. However the $C^1\Sigma^+$ and $D^1\Pi$ states are slightly higher in energy than the $A^1\Sigma^+$ and $B^1\Pi$ states, resulting in a smaller contribution at optical lattice frequencies. We have checked that, for infrared frequencies below the first excited asymptote, adding further effective transitions in Eq.~\eqref{eq:pola-eff} does not improve the quality of the fit. For larger frequencies, it did improve the fit results; but the obtained effective transition energy, above $\numprint{60000}$~cm$^{-1}$, had no physical meaning, and were disregarded. In these cases, the effective polarizabilities can still be useful, as its accuracy compared to the pure numerical results is around $2~\%$ in the worst case, whereas it is around $1~\%$ otherwise.

 \begin{table*}[htbp]
 \caption{\label{tab:dip-effective} \small Comparison between the effective TDMs $d_\Sigma$ (respectively $d_\Pi$) of Table \ref{tab:effective}, and the electronic TDMs between the electronic ground state $X^1\Sigma^+$ and the excited state A$^1\Sigma^+$ (respectively $B^1\Pi$) at the $X$ equilibrium distance $R_e$. The electronic TDM are multiplied by a rotational factor $\tilde{d_z}=d_z\sqrt{\frac{1}{3}}$ (resp. $\tilde{d_x}=d_x\sqrt{\frac{2}{3}}$) to assure compatibility between Eq.~\eqref{eq:pola-iso} and Eq.~\eqref{eq:pola-eff}.}
\begin{ruledtabular}
  \begin{tabular}{|c|c|c|c|c|c|c|} 
    Molecule & \multicolumn{3}{c|}{$\Sigma$ state} & \multicolumn{3}{c|}{$\Pi$ state} \\
   \hline
     & $d_\Sigma$ & $\tilde{d_z}(A,X;R_e)$ & $d_\Sigma/\tilde{d_z}$ & $d_\Pi$ & $\tilde{d_x}(B,X;R_e)$ & $d_\Pi/\tilde{d_x}$ \\
     & (a.u.) & (a.u.) & ($\%$) & (a.u.) & (a.u.) & (\%) \\
   \hline
   RbCs & $2.68$ & $2.64$ & $1.5$ & $3.08$ & $2.97$ &  $3.7$\\
   KCs  & $2.62$ & $2.58$ & $1.6$ & $2.99$ & $2.88$ &  $3.8$\\
   KRb  & $2.58$ & $2.56$ & $0.8$ & $2.90$ & $2.92$ & -$0.7$\\
   NaCs & $2.32$ & $2.22$ & $4.5$ & $2.63$ & $2.39$ &  $10.0$\\
   NaRb & $2.33$ & $2.24$ & $4.0$ & $2.61$ & $2.45$ &  $6.5$\\
   NaK  & $2.30$ & $2.20$ & $4.5$ & $2.59$ & $2.41$ &  $7.5$\\
   LiCs & $2.31$ & $2.16$ & $6.9$ & $2.50$ & $2.37$ &  $5.8$\\
   LiRb & $2.26$ & $2.20$ & $2.7$ & $2.42$ & $2.44$ & -$0.8$\\
   LiK  & $2.22$ & $2.15$ & $3.3$ & $2.41$ & $2.43$ & -$0.8$\\
   LiNa & $2.00$ & $1.96$ & $2.0$ & $2.25$ & $2.27$ & -$0.9$\\
 \end{tabular}
 \end{ruledtabular}
 \end{table*}

Since the wave function of the ($X^1\Sigma^+, v=0$) level is strongly localized around the equilibrium distance $R_e$ of the electronic ground state $X^1\Sigma^+$, in Table \ref{tab:dip-effective} we compare the electronic TDM from $X$ to $A$ (resp.~from $X$ to $B$) at $R_e$, with the effective TDM $d_\Sigma$ (resp.~$d_\Pi$) given in Table \ref{tab:dip-effective}. The close values taken by these quantities further illustrate the predominance of the levels belonging to the $A$ and $B$ states in the sums of Eqs.~\eqref{eq:pola-JM}--\eqref{eq:pola-perp}.

Finally, the present effective model designed for reproducing the isotropic DDP, also provides a good estimate of the DDP anisotropy over the same range of frequencies. Indeed one can assume from above that the effective $\Sigma$ (resp. $\Pi$) transition is mostly related to the parallel (resp. perpendicular) component of the DDP. With this assumption we can write $d_{\Sigma}^2=\frac{1}{3}d_{z}^2$ (resp. $d_{\Pi}^2=\frac{2}{3} d_{x}^2$) and thus infer the effective anisotropy of the DDP : 
\begin{equation}
  \label{eq:aniso-eff}
  \gamma_\mathrm{eff}(\omega) = \frac{6\omega_{\Sigma}
   d_{\Sigma}^2} {\omega_{\Sigma}^2 - \omega^2} 
   - \frac{3\omega_{\Pi}d_{\Pi}^2} {\omega_{\Pi}^2 - \omega^2} 
\end{equation}

This is illustrated in Figure \ref{fig:anis}: the anisotropy deduced from the numerically computed $\alpha_{\parallel}$ and $\alpha_{\perp}$ is compared to the effective one, through their ratio for the RbCs and LiNa molecules. We immediately see that despite that the model is optimized on the isotropic DDP, it delivers an effective anisotropy in agreement with the numerical one within 2\% to 15\%. The present model is thus undoubtedly useful for upcoming experiment aiming at exploring anisotropic trapping of heteronuclear alkali-metal diatomics.
 
\section{Conclusion}
\label{sec:conclu}

In this paper we have reviewed the experimental and theoretical results for the dynamic dipole polarizabilities (DDPs) of the ten alkali-metal heteronuclear diatomics composed of two of ($^7$Li,$^{23}$Na,$^{39}$K,$^{87}$Rb,$^{133}$Cs) alkali atoms in their electronic ground state. We recalled the general approach for such calculations, involving a detailed knowledge of large number of excited electronic states of these molecules. The set of relevant potential energy curves is built on a combination of up-to-date experimental data with theoretical data deduced from high-level quantum chemistry computations performed in our group. The corresponding set of transition dipole moments are exclusively obtained from the same calculations.  We also proposed an effective model which allows for a compact modeling of the DDPs involving a limited number of parameters (five) describing a small number (two in the present case) of effective electronic transitions, which reproduces with a good accuracy the DDP for all vibrational levels of the ground state ($X^1\Sigma^+ , v , J=0$) of the ten studied molecules. The comparison of our calculated DDPs with the existing experimental values for some molecules like Cs$_2$, Rb$_2$ and KRb highlights the quality and the precision of our work and allows to be confident in our predicted values for the ground state DDPs not yet determined experimentally.

The precise knowledge of the DDP enables to determine the response of the molecules when interacting with an oscillating electric field. In particular, we invoked the current developments in optical trapping of ultracold molecules using for instance optical lattices. This study leads to the determination of the optimal frequencies to be chosen for an optical lattice in order to optimize the transfer efficiency of trapped  molecules from an initial weakly-bound state down to the lowest rovibrational level of the ground state.

For some molecules (RbCs, KCs, KRb and NaK) there exists so-called {}``magic frequencies'' where the related DDPs are close enough together that the molecules are not excited into high motional modes of the lattice during the transfer. For NaK this magic frequency is in the vicinity of a resonance with the possibility that the  trapping laser excites the molecule which may thus escape from the trap. The success of the trapping relies on the precision of the knowledge of the resonance energies and in the ability of the experiments to keep the magic frequency of the laser used for the transfer far enough from this resonance. For some molecules (NaCs, NaRb, LiRb, LiK, LiCs and LiNa) such well-defined magic frequencies do not exist and we propose several alternatives to circumvent this drawback. Therefore this work also gives confident guidelines to the choice of the frequencies of the optical lattices used to control the ultracold bialkali-molecules during their transfer to the absolute ground state.

\section*{Acknowledgments}

The authors acknowledge the support of ''Agence Nationale de la Recherche'' under the COPOMOL project (Grant No. ANR-13-IS04-0004-01), and the BLUESHIELD project (contract 
ANR-14-CE34-0006).


	\begin{figure*}[htbp]
  \begin{center}

        \includegraphics[width=0.9\textwidth]{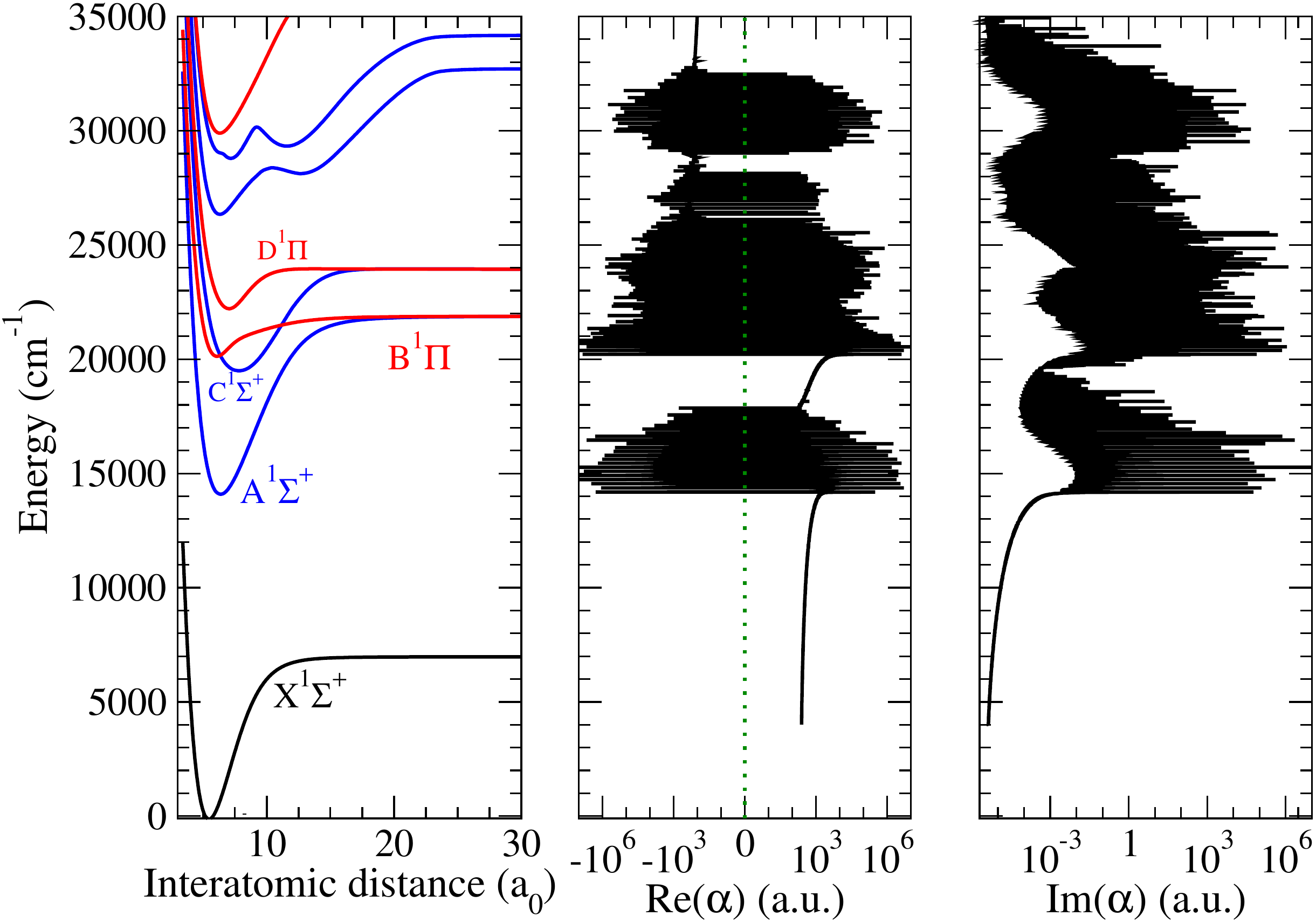}

        \includegraphics[width=0.9\textwidth]{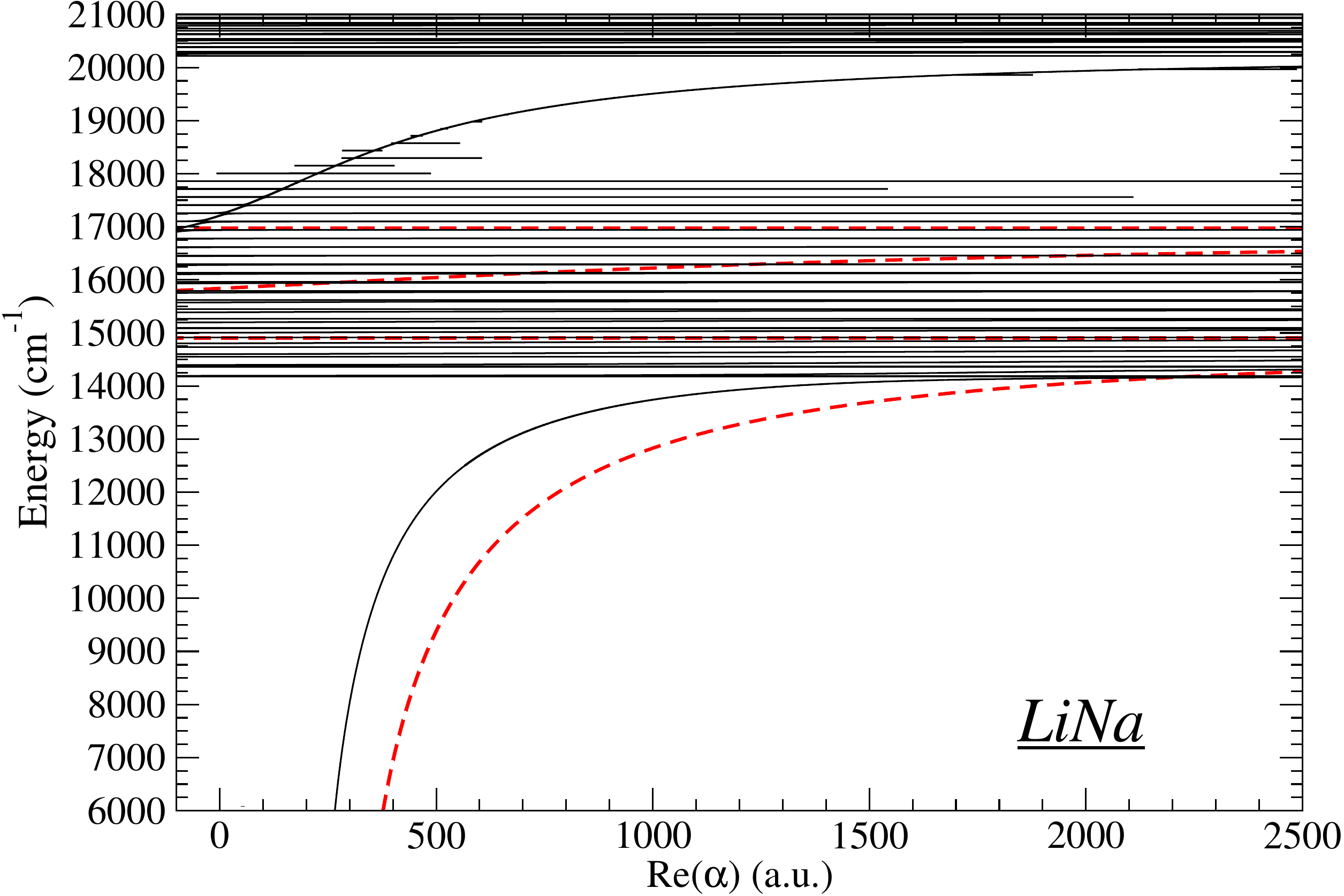}

    \caption{\label{fig:dynpol_lina} \small \underline{LiNa :} (left Fig. ) Potential energy curves of LiNa as a function of the internuclear distance (left panel). Real part of the polarizability of a $X^1\Sigma^+,v=0$ LiNa molecule (middle panel) and imaginary part of the polarizability of a $X^1\Sigma^+,v=0$ LiNa molecule (right panel) as a function of the energy of the laser. (Right Fig. ) zoom on the real part of the polarizability and comparaison with the sum of the atomic polarizability Li and Na (red dashed curve). Purple circle point out the magic frequencies.}
  \end{center}
\end{figure*}

 \begin{figure*}[htbp]
  \begin{center}  

        \includegraphics[width=0.9\textwidth]{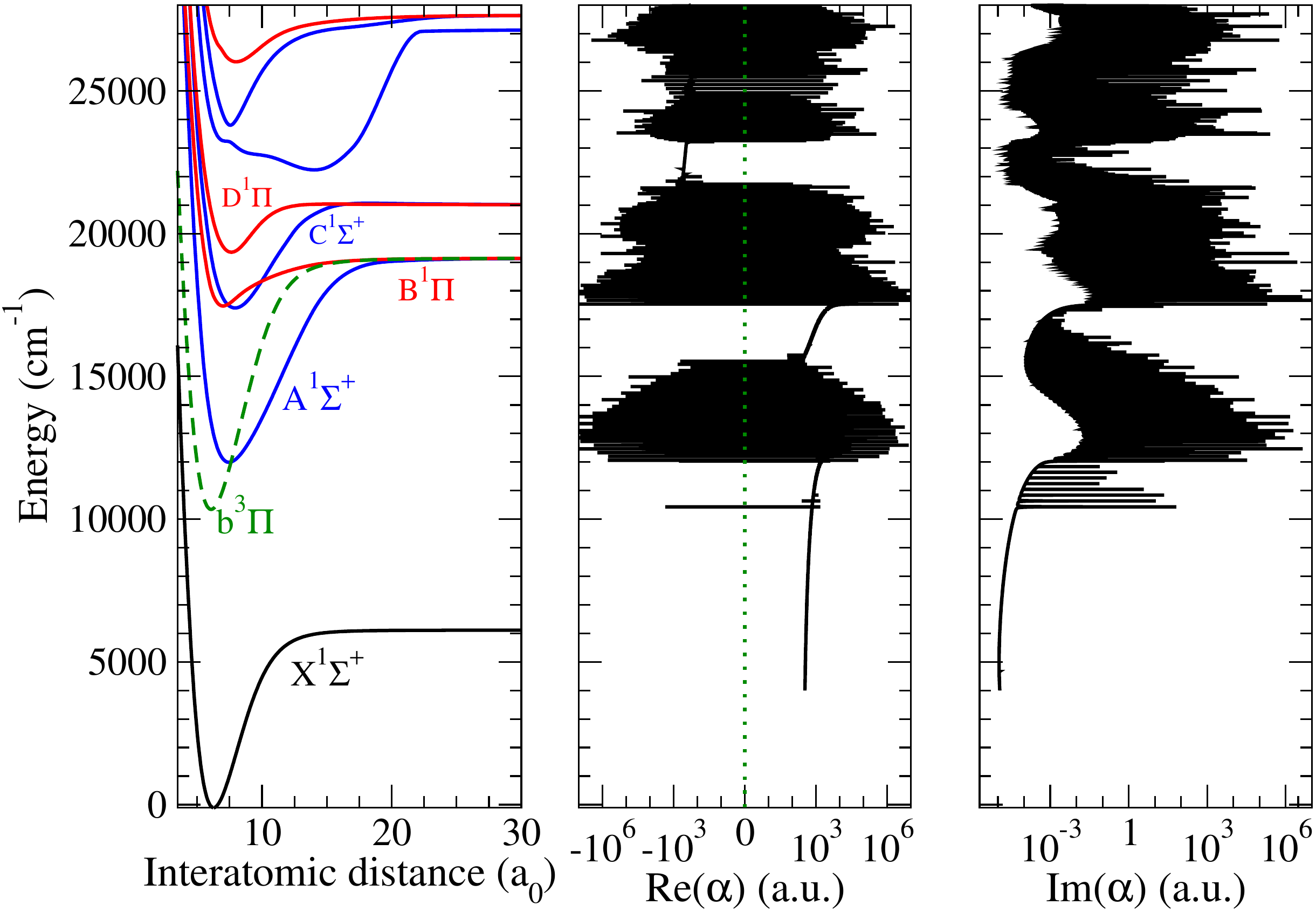}

        \includegraphics[width=0.9\textwidth]{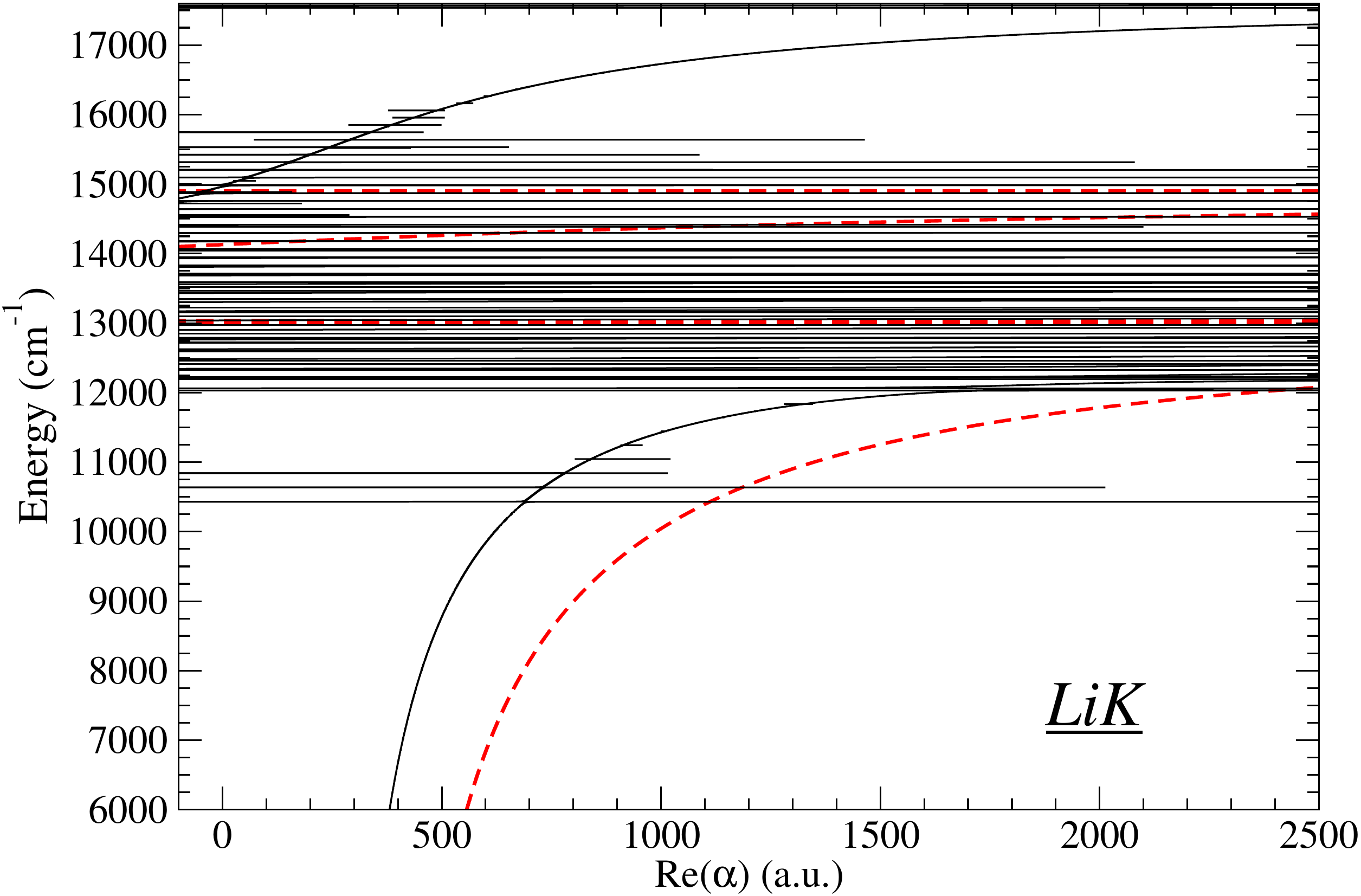}

    \caption{\label{fig:dynpol_lik} \small \underline{LiK :} (left Fig. ) Potential energy curves of LiK as a function of the internuclear distance (left panel). Real part of the polarizability of a $X^1\Sigma^+,v=0$ LiK molecule (middle panel) and imaginary part of the polarizability of a $X^1\Sigma^+,v=0$ LiK molecule (right panel) as a function of the energy of the laser. (Right Fig. ) zoom on the real part of the polarizability and comparaison with the sum of the atomic polarizability Li and K (red dashed curve). Purple circle point out the magic frequencies.}
  \end{center}
\end{figure*}

 
 \begin{figure*}[htbp]
  \begin{center}

        \includegraphics[width=0.9\textwidth]{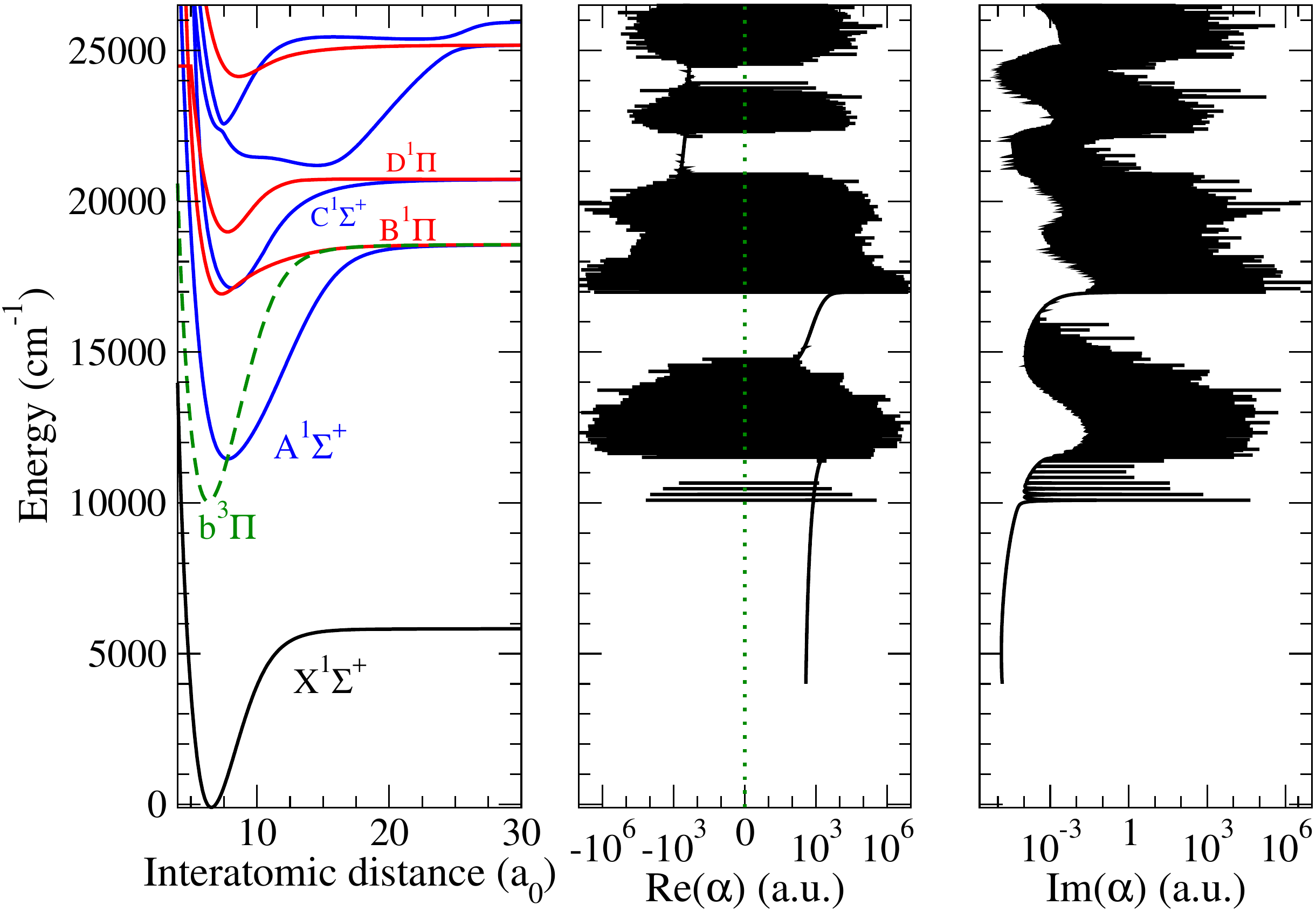}

        \includegraphics[width=0.9\textwidth]{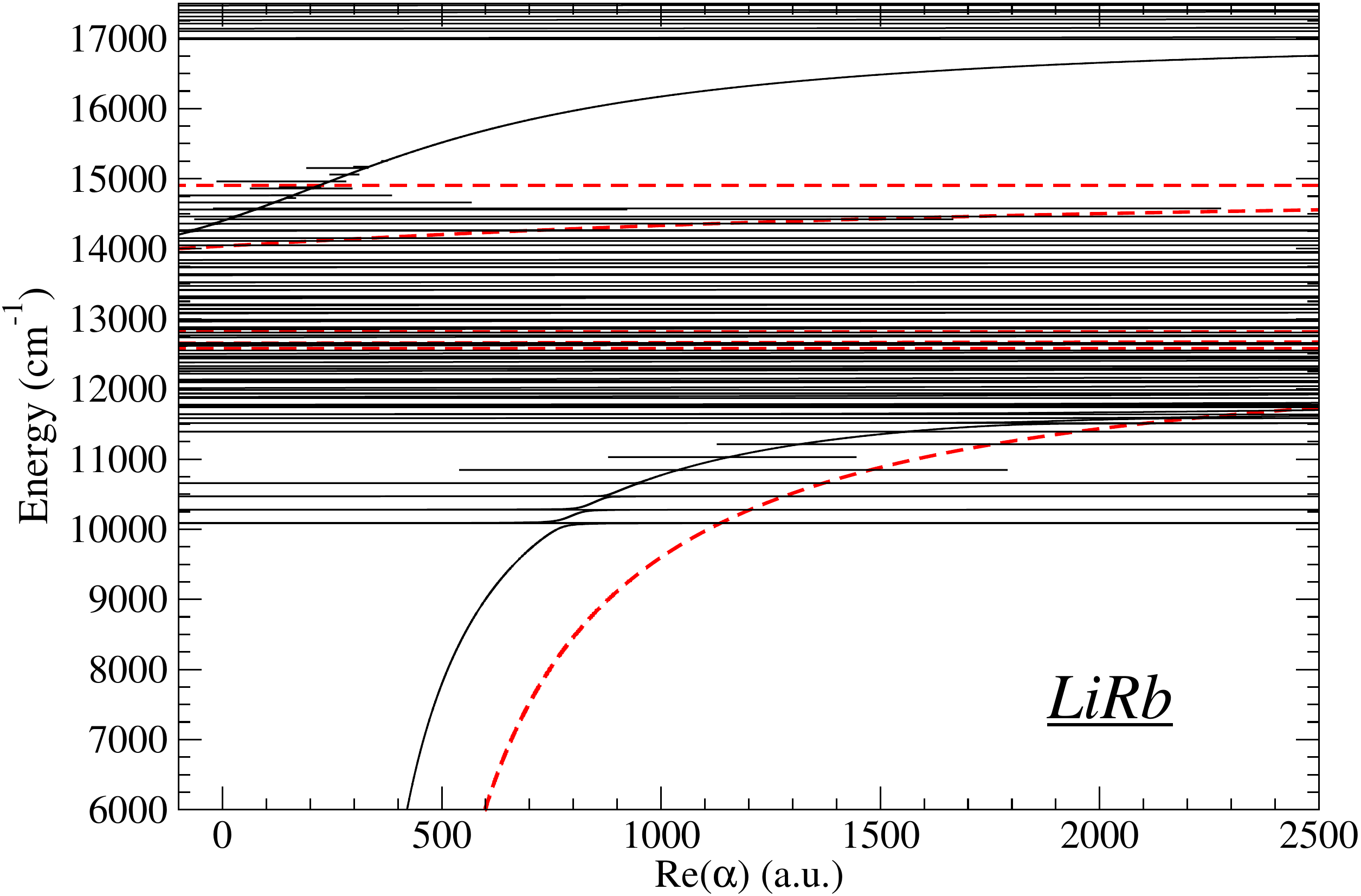}

    \caption{\label{fig:dynpol_lirb} \small \underline{LiRb :} (left Fig. ) Potential energy curves of LiRb as a function of the internuclear distance (left panel). Real part of the polarizability of a $X^1\Sigma^+,v=0$ LiRb molecule (middle panel) and imaginary part of the polarizability of a $X^1\Sigma^+,v=0$ LiRb molecule (right panel) as a function of the energy of the laser. (Right Fig. ) zoom on the real part of the polarizability and comparaison with the sum of the atomic polarizability Li and Rb (red dashed curve). Purple circle point out the magic frequencies.}
  \end{center}
\end{figure*}

 \begin{figure*}[htbp]
  \begin{center}   

        \includegraphics[width=0.9\textwidth]{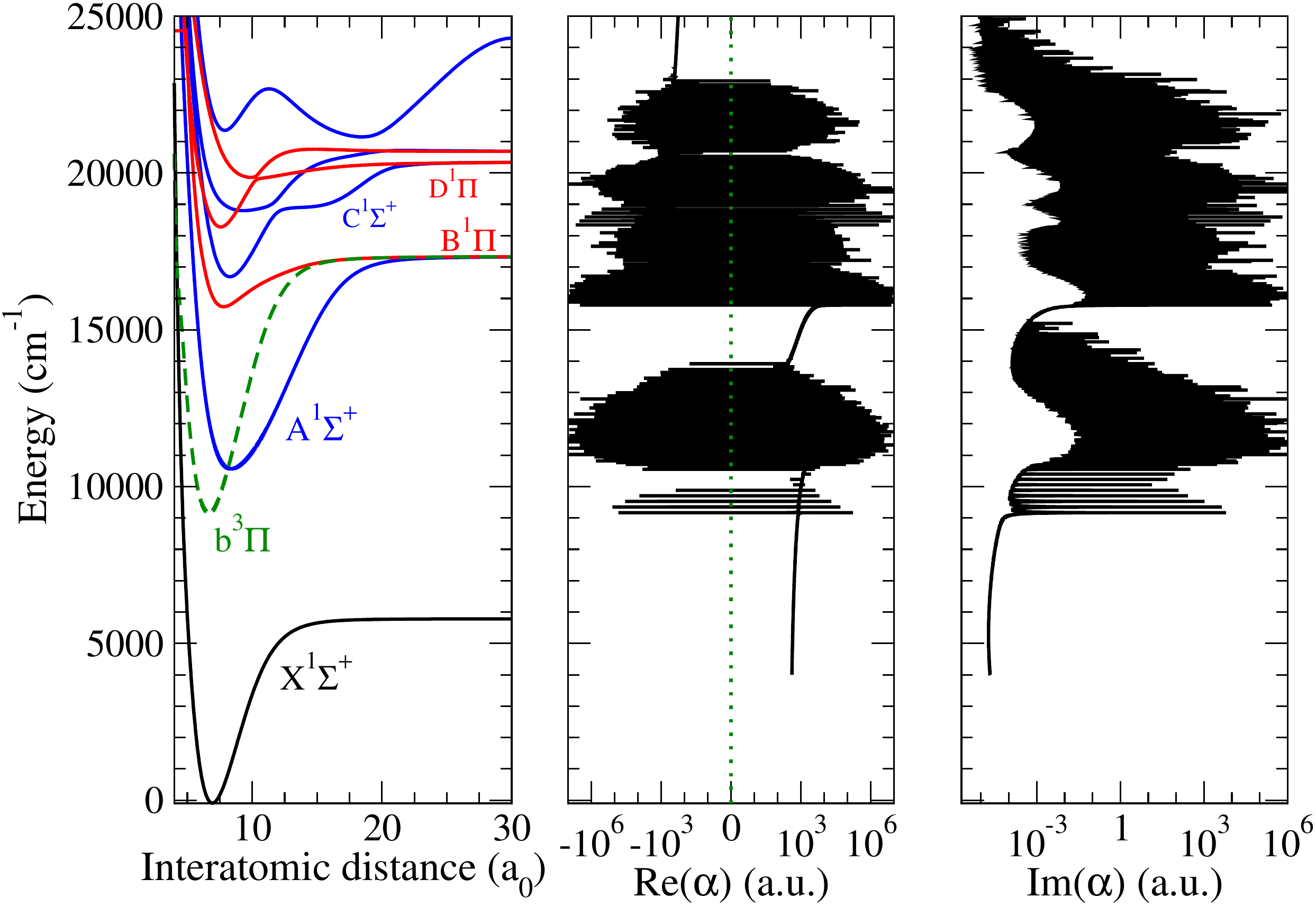}

        \includegraphics[width=0.9\textwidth]{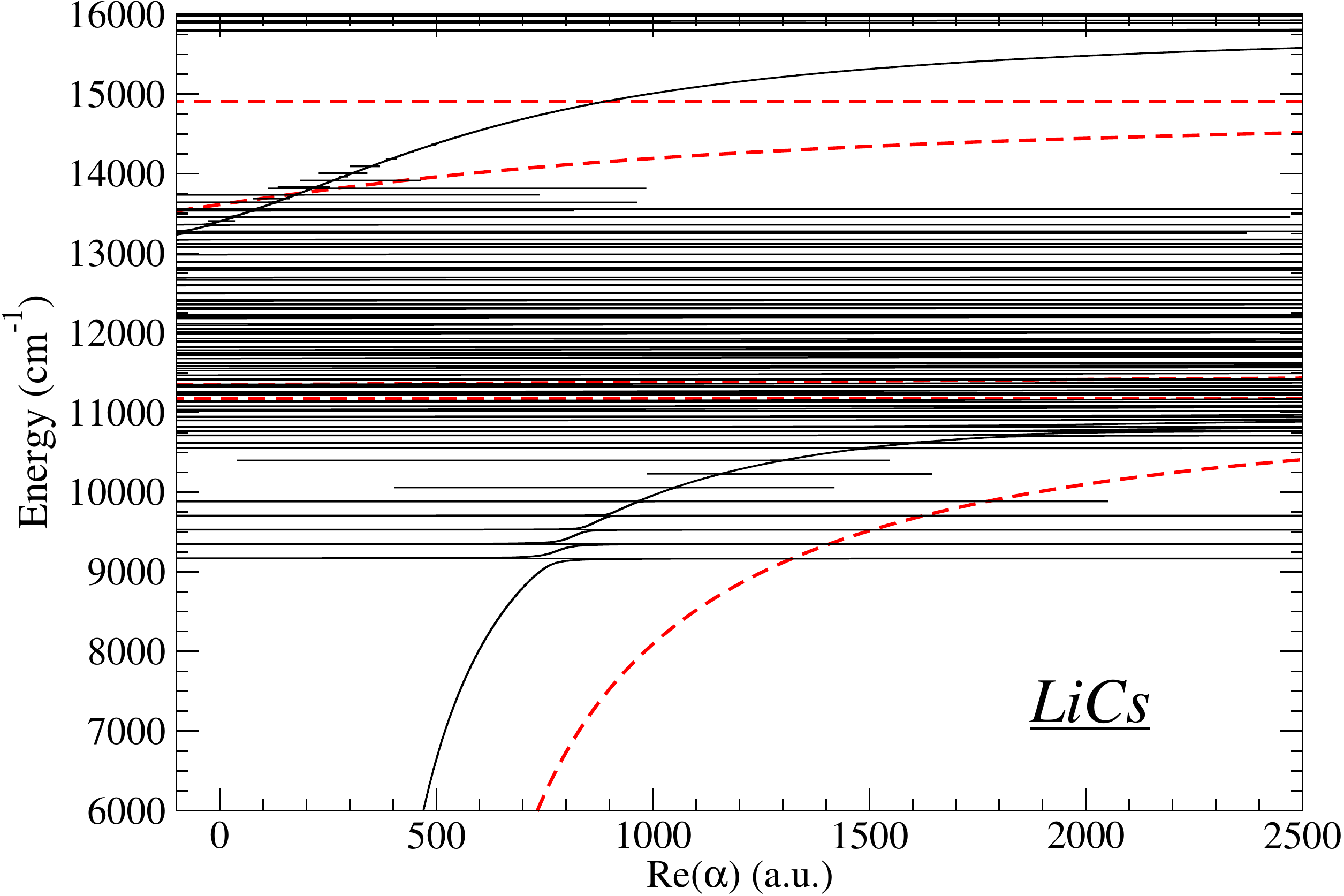}

    \caption{\label{fig:dynpol_lics} \small \underline{LiCs :} (left Fig. ) Potential energy curves of LiCs as a function of the internuclear distance (left panel). Real part of the polarizability of a $X^1\Sigma^+,v=0$ LiCs molecule (middle panel) and imaginary part of the polarizability of a $X^1\Sigma^+,v=0$ LiCs molecule (right panel) as a function of the energy of the laser. (Right Fig. ) zoom on the real part of the polarizability and comparaison with the sum of the atomic polarizability Li and Cs (red dashed curve). Purple circle point out the magic frequencies.}
  \end{center}
\end{figure*}

 \begin{figure*}[htbp]
  \begin{center}

        \includegraphics[width=0.9\textwidth]{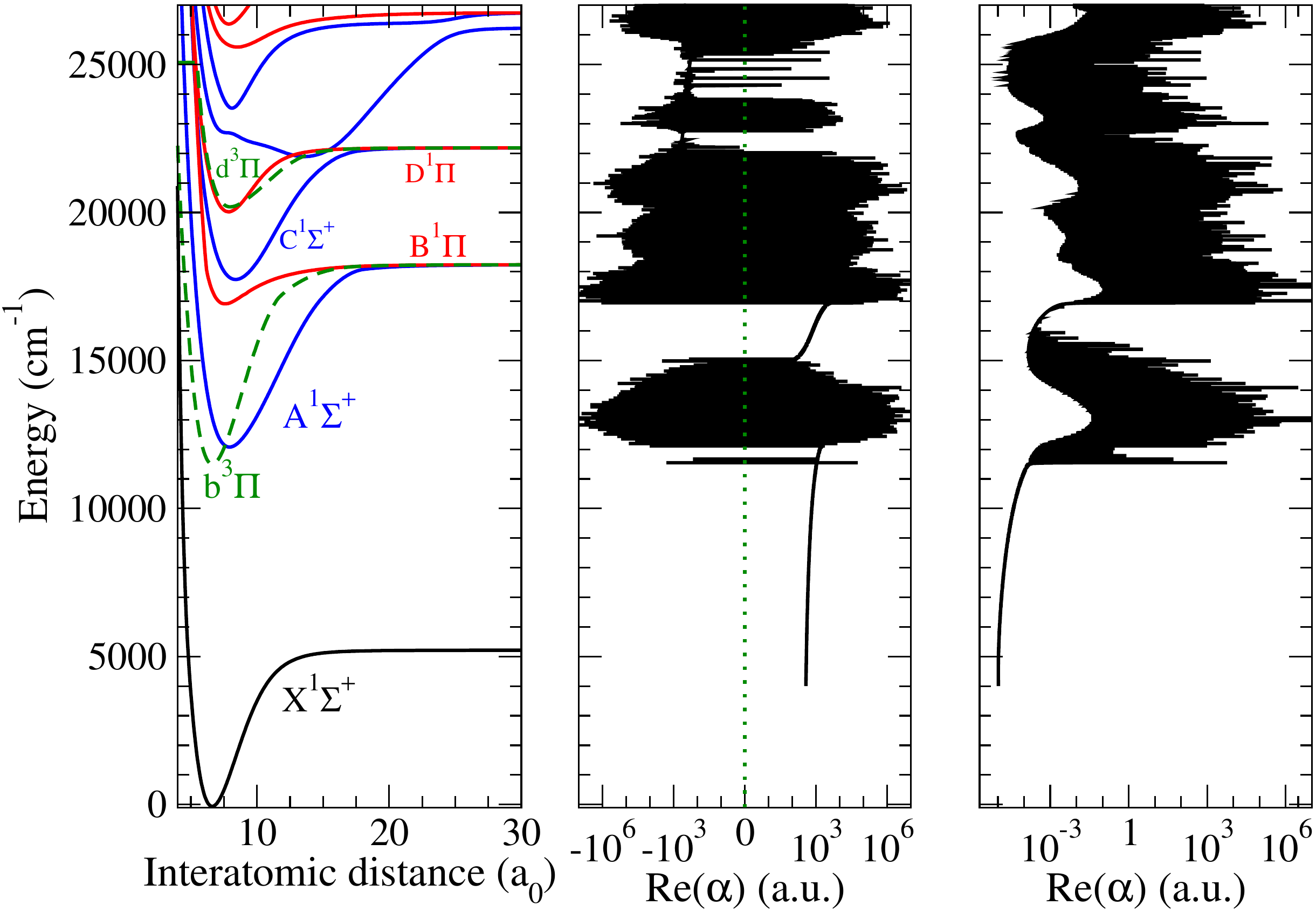}

        \includegraphics[width=0.9\textwidth]{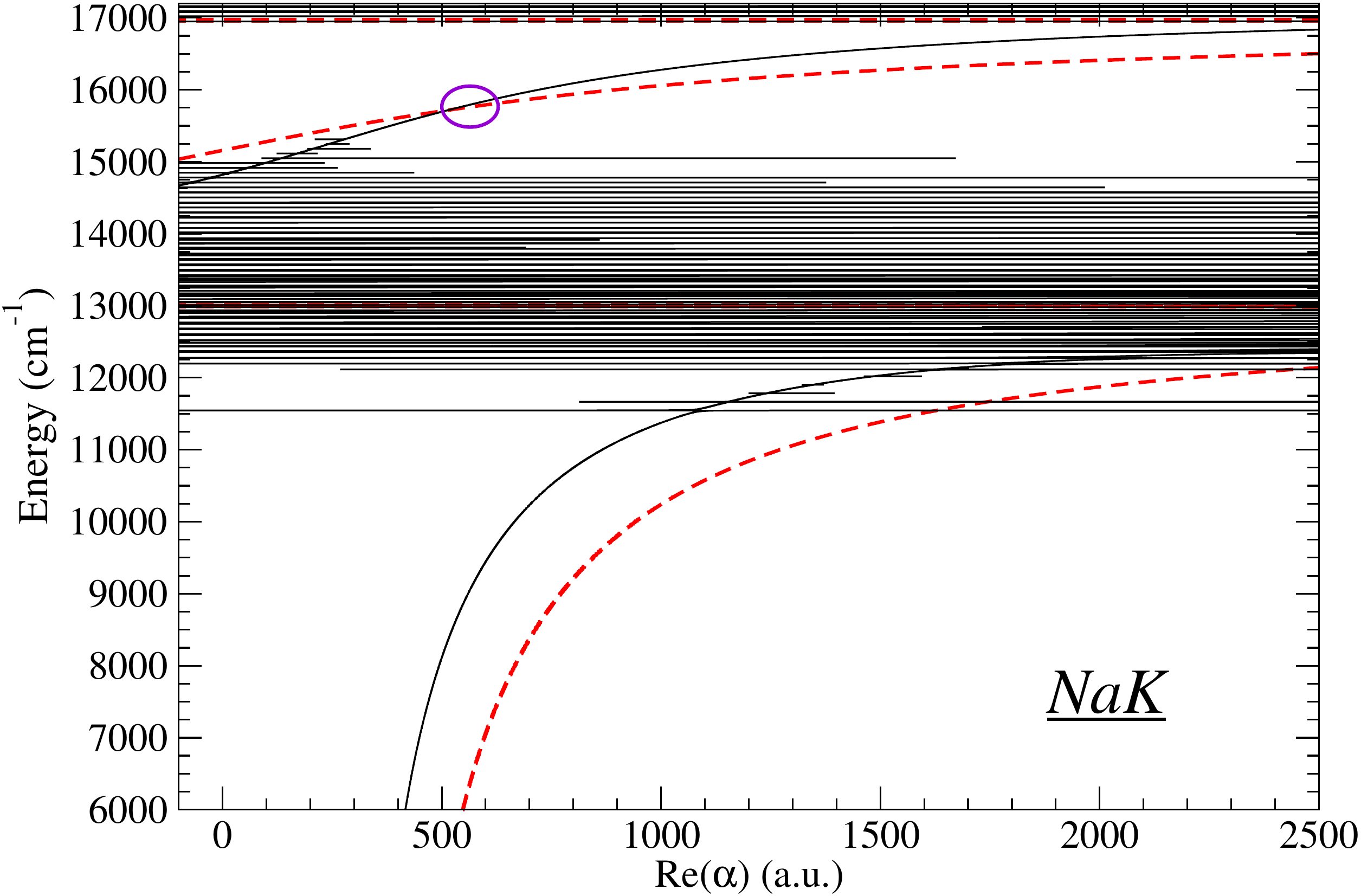}

    \caption{\label{fig:dynpol_nak} \small \underline{NaK :} (left Fig. ) Potential energy curves of NaK as a function of the internuclear distance (left panel). Real part of the polarizability of a $X^1\Sigma^+,v=0$ NaK molecule (middle panel) and imaginary part of the polarizability of a $X^1\Sigma^+,v=0$ NaK molecule (right panel) as a function of the energy of the laser. (Right Fig. ) zoom on the real part of the polarizability and comparaison with the sum of the atomic polarizability Na and K (red dashed curve). Purple circle point out the magic frequencies.}
  \end{center}
\end{figure*}

 \begin{figure*}[htbp]
  \begin{center}  

        \includegraphics[width=0.9\textwidth]{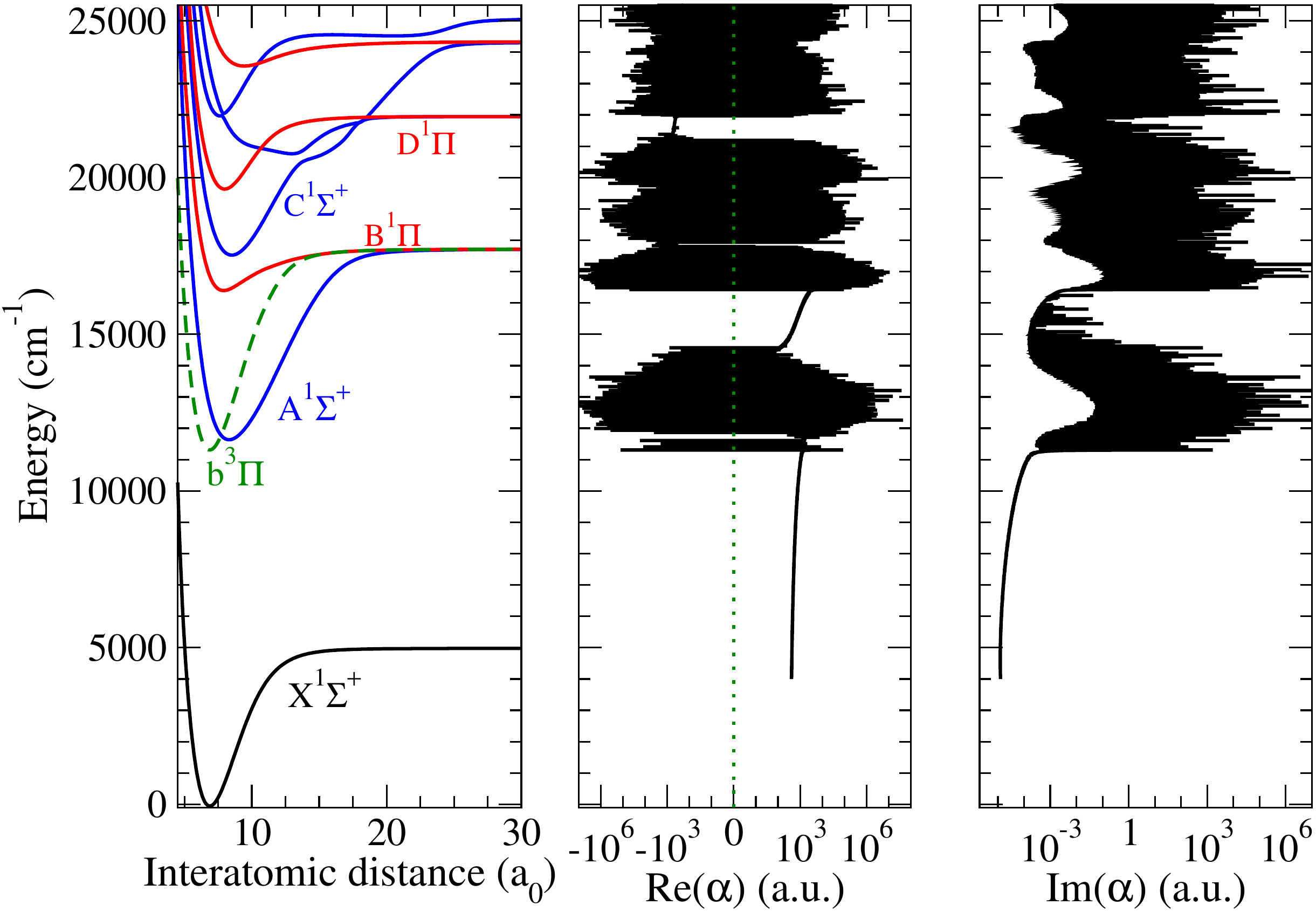}

        \includegraphics[width=0.9\textwidth]{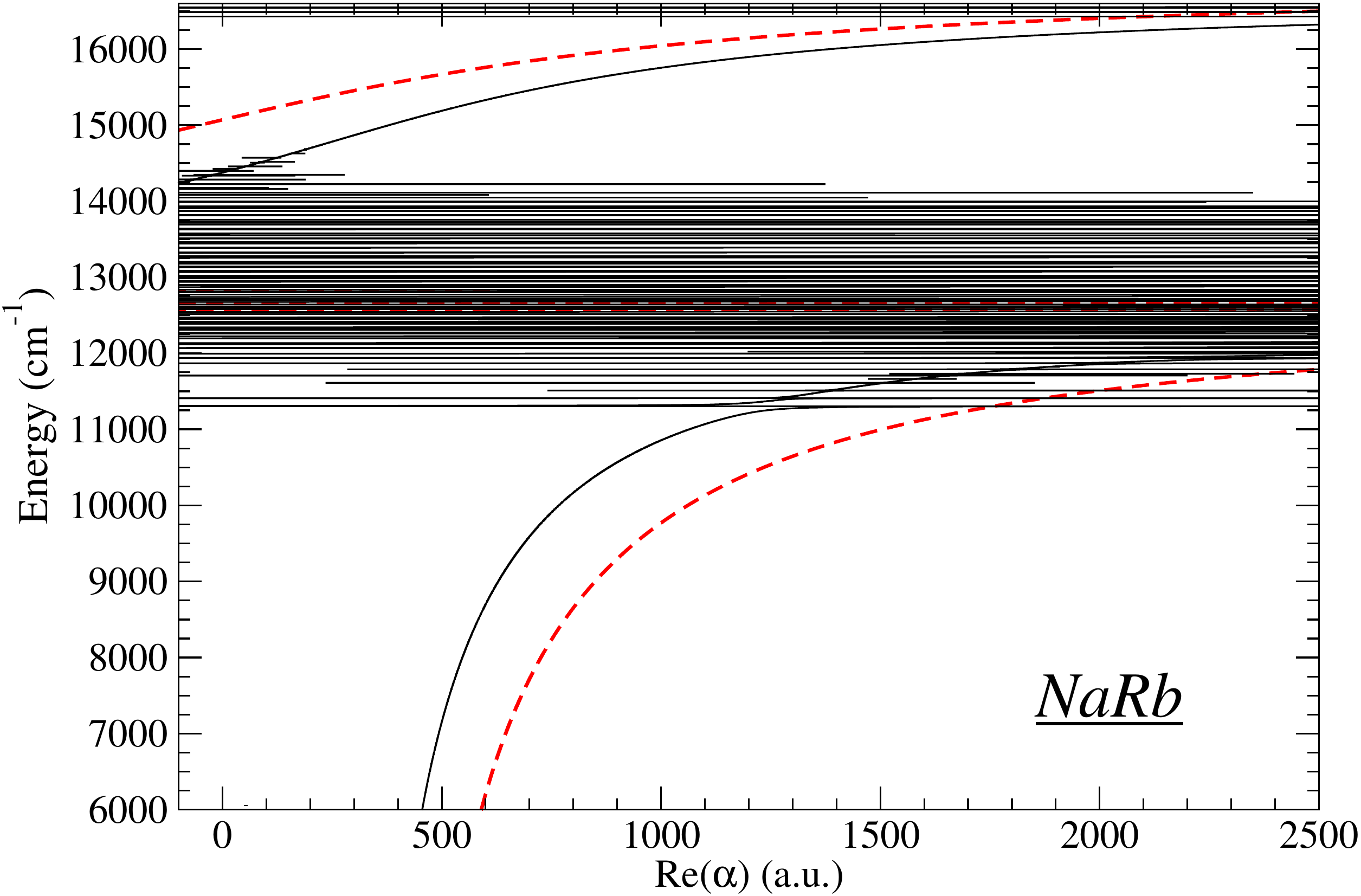}

    \caption{\label{fig:dynpol_narb} \small \underline{NaRb :} (left Fig. ) Potential energy curves of NaRb as a function of the internuclear distance (left panel). Real part of the polarizability of a $X^1\Sigma^+,v=0$ NaRb molecule (middle panel) and imaginary part of the polarizability of a $X^1\Sigma^+,v=0$ NaRb molecule (right panel) as a function of the energy of the laser. (Right Fig. ) zoom on the real part of the polarizability and comparaison with the sum of the atomic polarizability Na and Rb (red dashed curve). Purple circle point out the magic frequencies.}
  \end{center}
\end{figure*}
 

 \begin{figure*}[htbp]
  \begin{center}

        \includegraphics[width=0.9\textwidth]{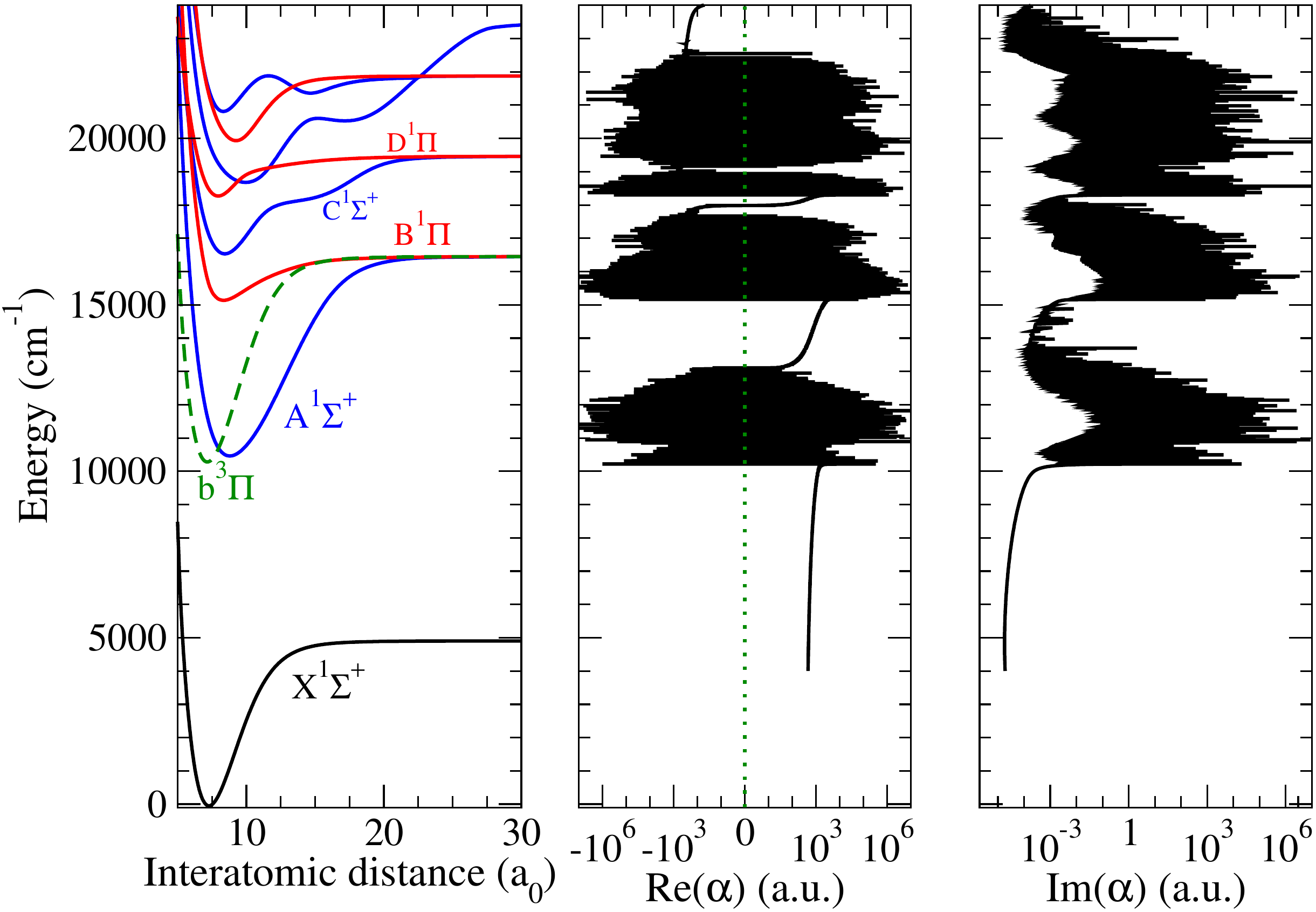}

        \includegraphics[width=0.9\textwidth]{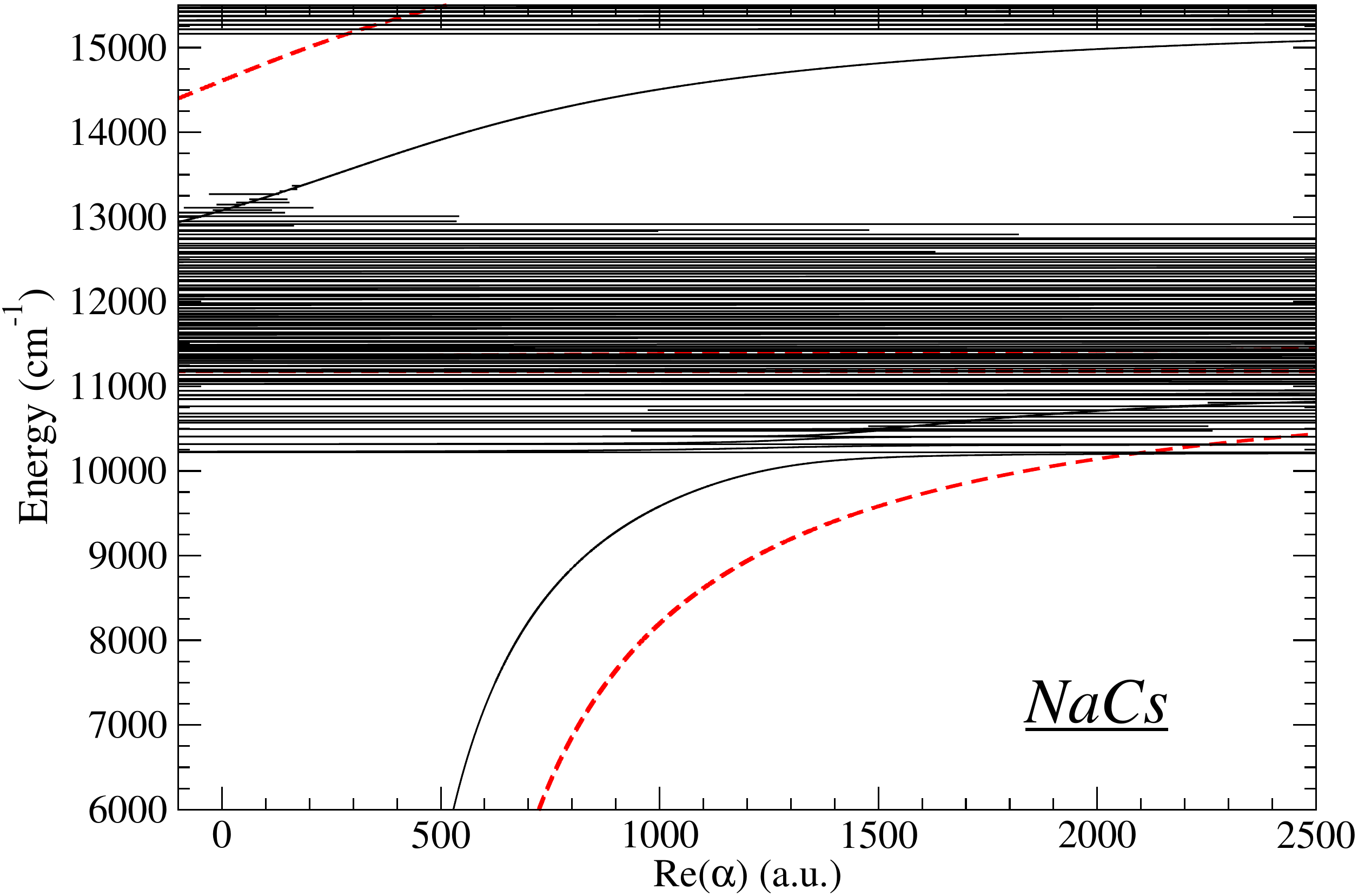}

    \caption{\label{fig:dynpol_nacs} \small \underline{NaCs :} (left Fig. ) Potential energy curves of NaCs as a function of the internuclear distance (left panel). Real part of the polarizability of a $X^1\Sigma^+,v=0$ NaCs molecule (middle panel) and imaginary part of the polarizability of a $X^1\Sigma^+,v=0$ NaCs molecule (right panel) as a function of the energy of the laser. (Right Fig. ) zoom on the real part of the polarizability and comparaison with the sum of the atomic polarizability Na and Cs (red dashed curve). Purple circle point out the magic frequencies.}
  \end{center}
\end{figure*}

 \begin{figure*}[htbp]
  \begin{center}  

        \includegraphics[width=0.9\textwidth]{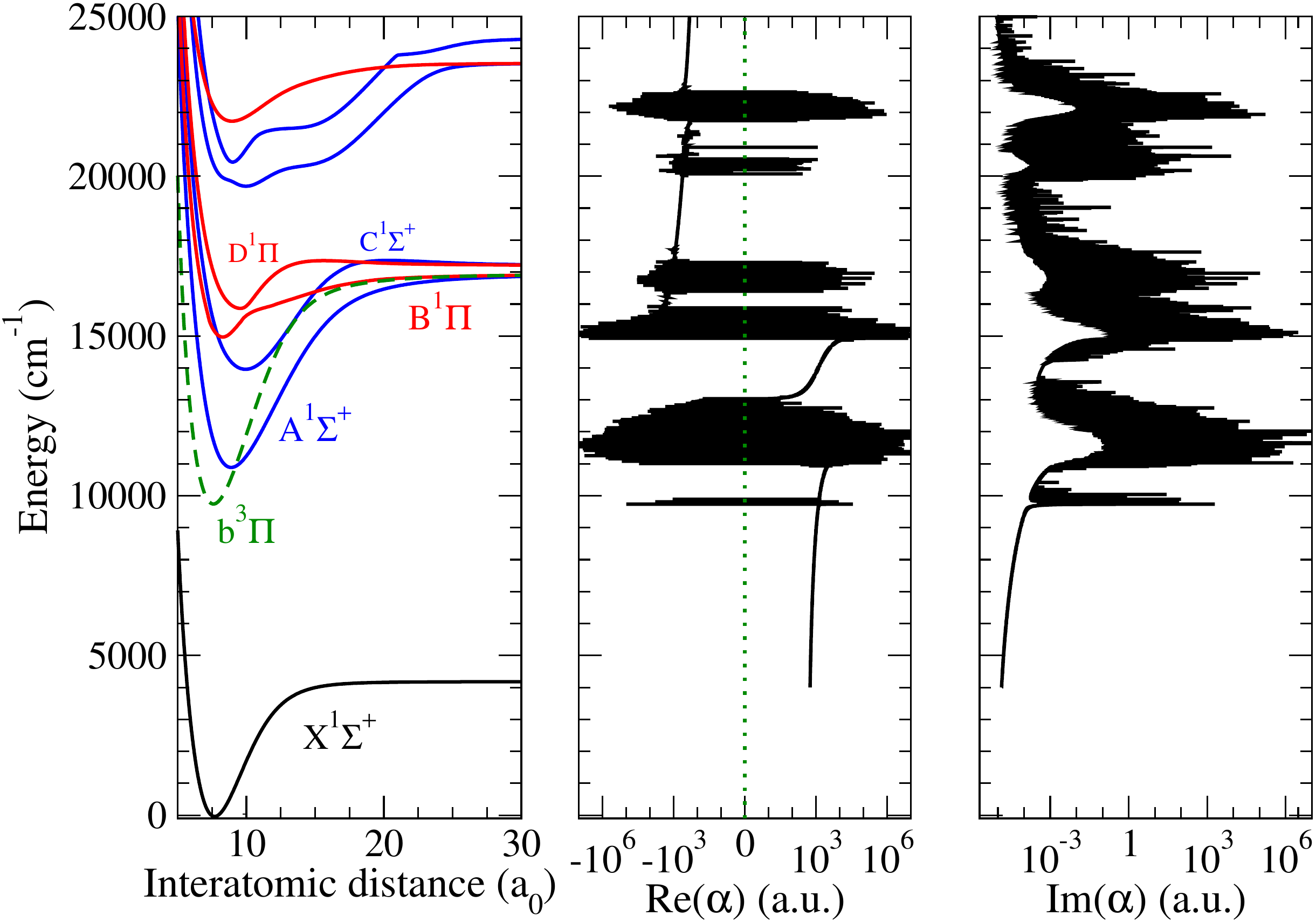}

        \includegraphics[width=0.9\textwidth]{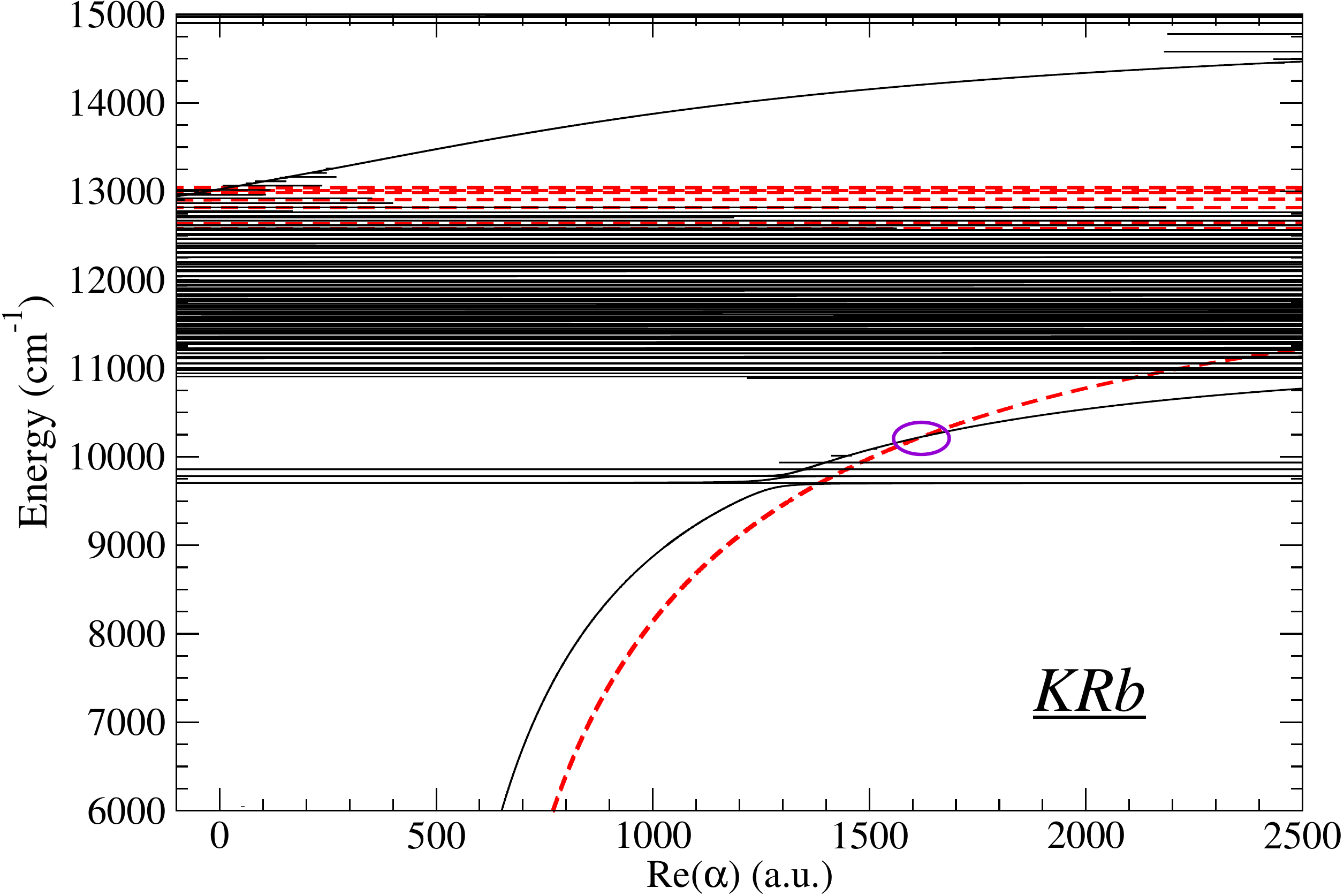}

    \caption{\label{fig:dynpol_krb} \small \underline{KRb :} (left Fig. ) Potential energy curves of KRb as a function of the internuclear distance (left panel). Real part of the polarizability of a $X^1\Sigma^+,v=0$ KRb molecule (middle panel) and imaginary part of the polarizability of a $X^1\Sigma^+,v=0$ KRb molecule (right panel) as a function of the energy of the laser. (Right Fig. ) zoom on the real part of the polarizability and comparaison with the sum of the atomic polarizability K and Rb (red dashed curve). Purple circle point out the magic frequencies.}
  \end{center}
\end{figure*}


\begin{figure*}[htbp]
  \begin{center}

        \includegraphics[width=0.9\textwidth]{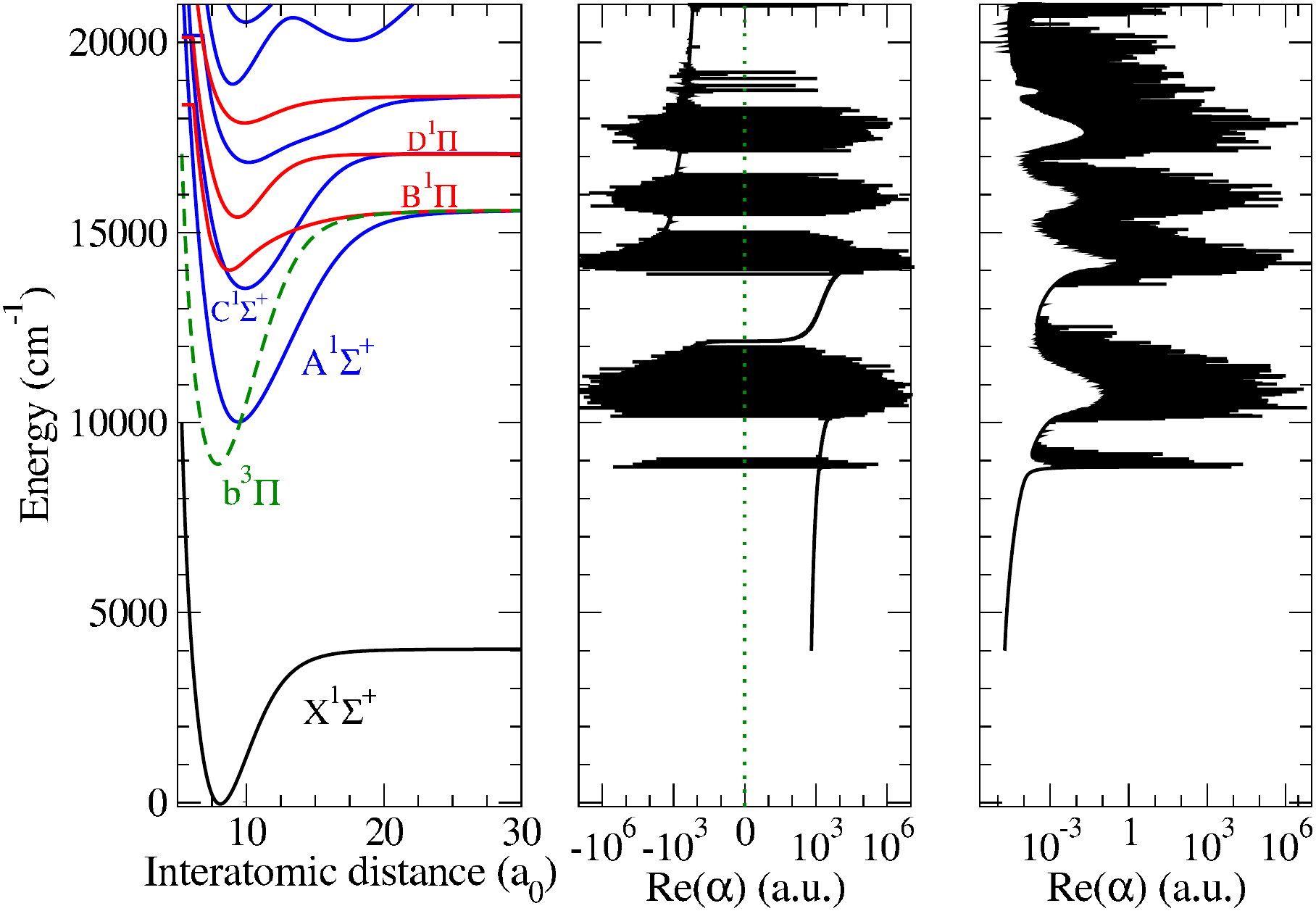}

        \includegraphics[width=0.9\textwidth]{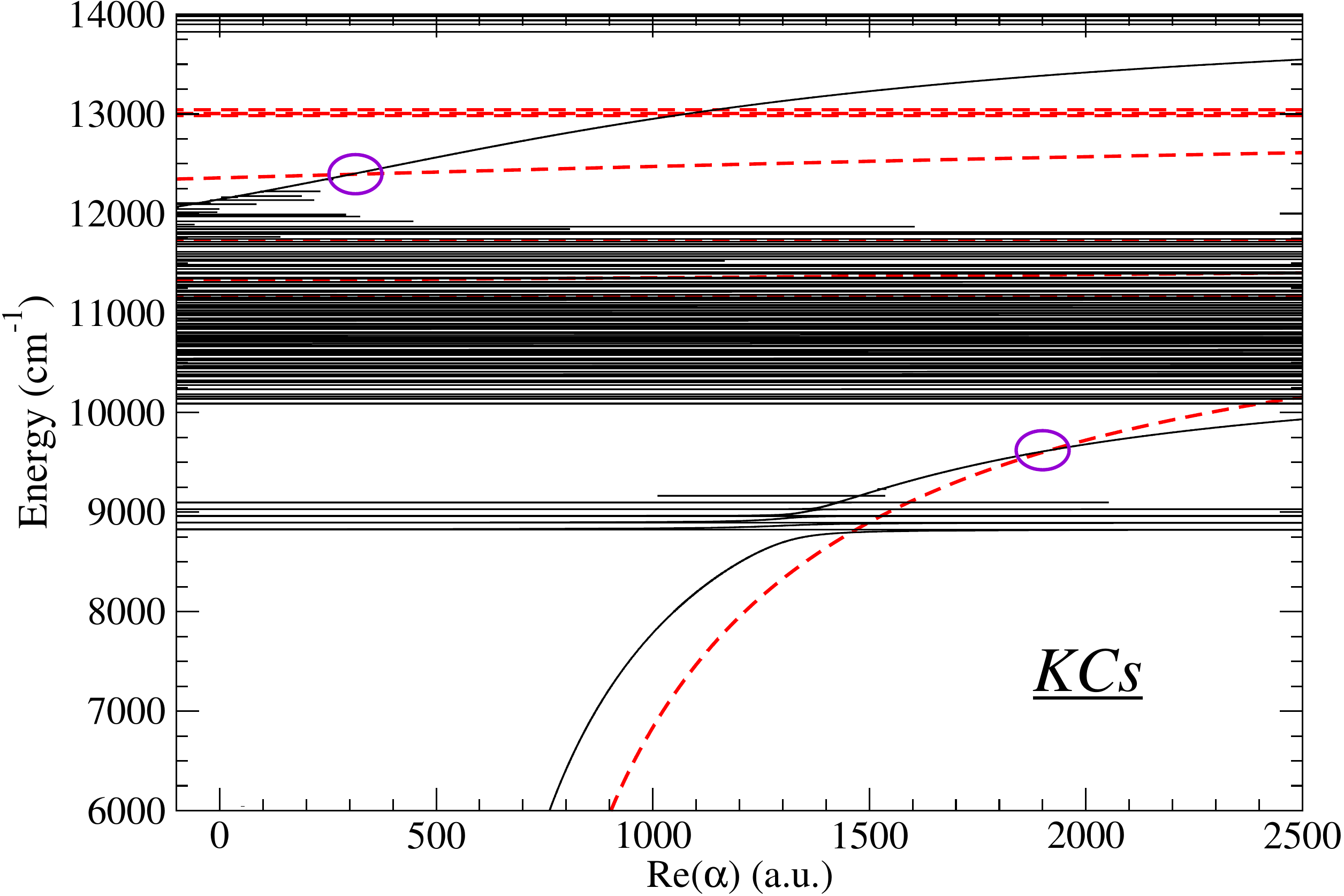}

    \caption{\label{fig:dynpol_kcs} \small \underline{KCs :} (left Fig. ) Potential energy curves of KCs as a function of the internuclear distance (left panel). Real part of the polarizability of a $X^1\Sigma^+,v=0$ KCs molecule (middle panel) and imaginary part of the polarizability of a $X^1\Sigma^+,v=0$ KCs molecule (right panel) as a function of the energy of the laser. (Right Fig. ) zoom on the real part of the polarizability and comparaison with the sum of the atomic polarizability K and Cs (red dashed curve). Purple circle point out the magic frequencies.}
  \end{center}
\end{figure*}

 \begin{figure*}[htbp]
 \begin{center}
 \includegraphics[width=\linewidth]{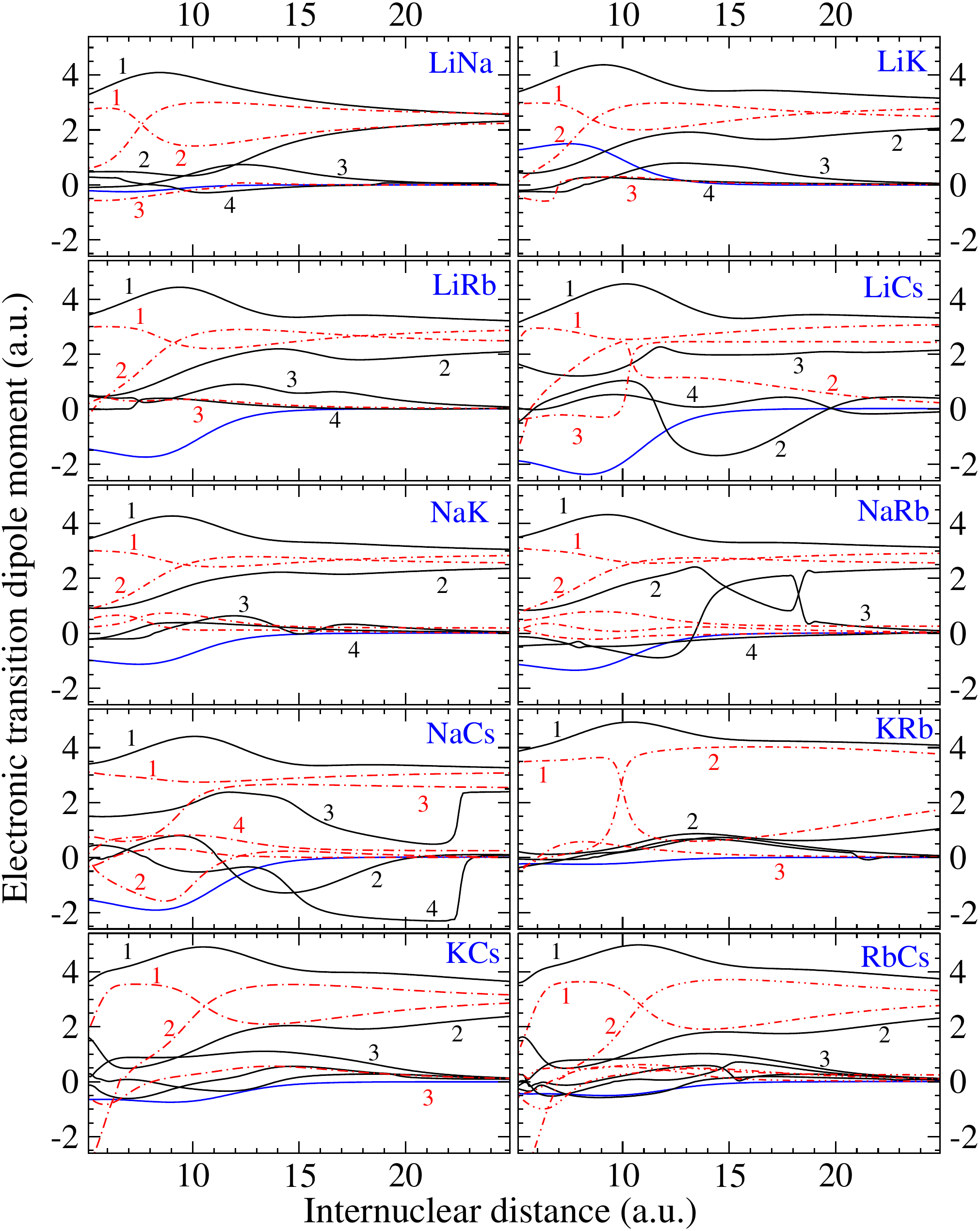}
 \caption{\label{fig:etdm} \small Electronic permanent and transition dipole moments from the ground state $X^1\Sigma^+$ of the ten heteronuclear molecules included in the calculation. For each molecule the permanent dipole moment (blue solid line), and the transition dipole moments towards $\Sigma^+$ states (black solid lines) and towards $\Pi$ states (dashed red lines) are shown. All these data can be found in a text format in the supplementary material [Insert ref. to supplementary material].}
 \end{center}
 \end{figure*}
 
 \begin{figure*}[htbp]
 \begin{center}
 \includegraphics[width=\linewidth]{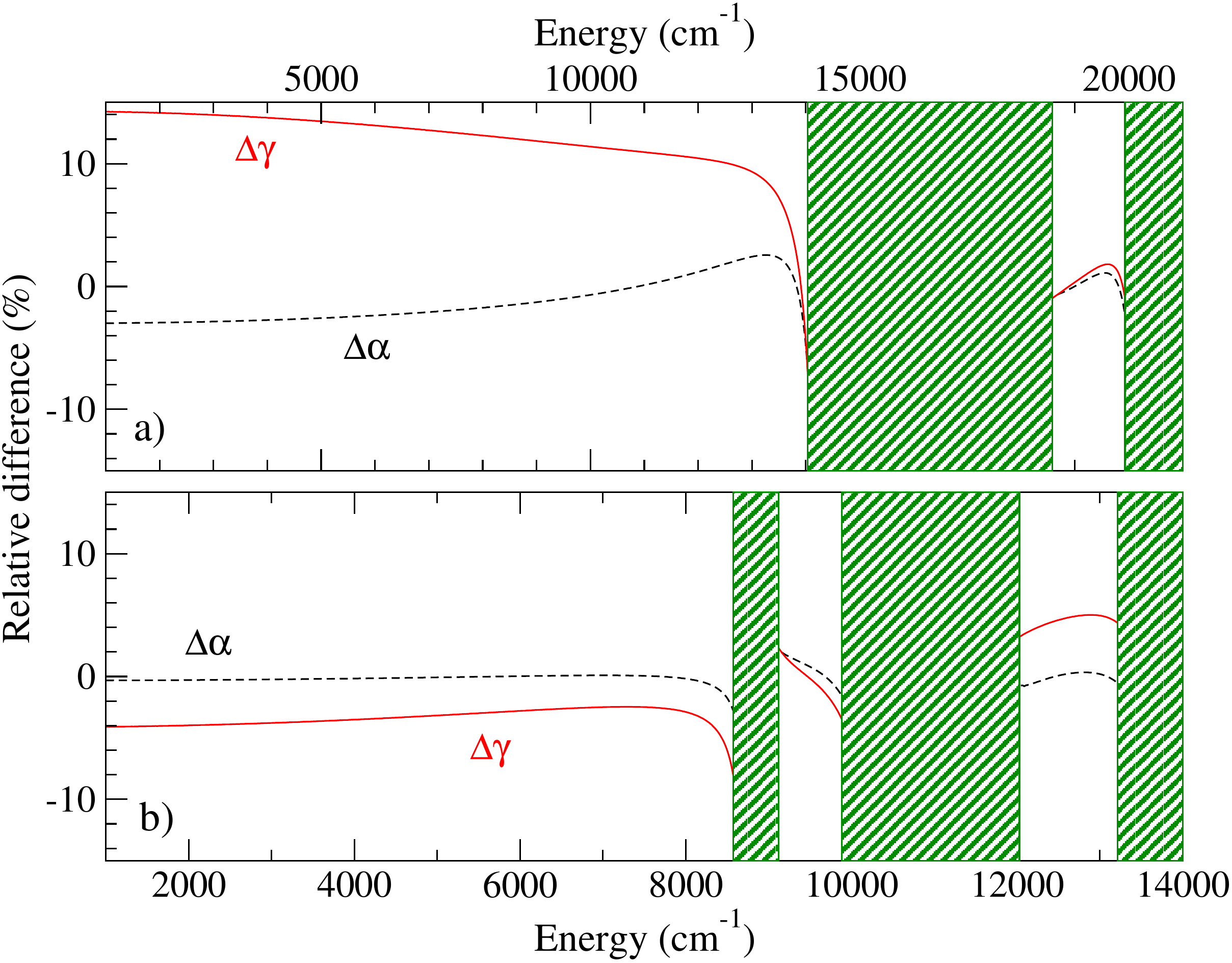}
 \caption{\label{fig:anis} \small Ratio of the numerically computed over effective isotropic DDP (resp. DDP anisotropy) $\Delta \alpha$ (resp. $\Delta \gamma$). Upper panel: for the LiNa ground state molecule in $v=0, J=0$. Lower panel: Idem for RbCs. The range of energies excluded from the model are reported in dashed areas (see Table \ref{tab:effective}). }
 \end{center}
 \end{figure*}

  \clearpage

\bibliography{bibliocold} 

\end{document}